%% file: ms_v7.1.tex
\documentclass[11pt,preprint]{aastex}
\tightenlines
\slugcomment{Accepted for publication in the Astrophysical Journal}

\def	\Angstrom	{\,{\rm \AA}}

\def	\beq	{\begin{equation}}
\def	\beqa	{\begin{eqnarray}}
\def    \Cabs   {C_{\rm abs}}
\def	\cm	{\,{\rm cm}}

\def	\eeq	{\end{equation}}
\def	\eeqa	{\end{eqnarray}}
\def	\erg	{\,{\rm ergs}}

\def	\eV	{\,{\rm eV}}

\def	\g	{\,{\rm g}}
\def	\gtsim	{\gtrsim}		 

\def	\K		{\,{\rm K}}

\def	\ltsim	{\lesssim}		 
\def    \Mdust  {M_{\rm dust}}

\def	\Msol	{M_{\odot}}

\def    \qpah   {q_{\rm PAH}}
\def	\s	{\,{\rm s}}
\def	\sr	{\,{\rm sr}}
\def    \UFDS   {U_{\rm FDS}}
\def    \Umax   {U_{\rm max}}
\def    \Umin   {U_{\rm min}}

\def    \um     {$\mu${\rm m}}
\def    \mum    {\,\mu {\rm m}}
\def    \Teff   {T_{\rm eff}}

\newlength{\figwidthhex}
\newlength{\figwidthtrack}

\newcommand{\newtext}[1]{#1}

\countdef\decade=200
\advance\decade by \year
\countdef\hours=201
\advance\hours by \time
\divide\hours by 60
\countdef\mins=202
\advance\mins by \hours
\multiply\mins by 60
\multiply\hours by 100
\countdef\miltime=203
\advance\miltime by \hours
\advance\miltime by \time
\advance\miltime by -\mins

\begin{document}

\title{%
	Infrared Emission from Interstellar Dust.
	IV. The Silicate-Graphite-PAH Model 
            in the Post-Spitzer Era
	}

\author{B.T. Draine}
\affil{Princeton University Observatory, Peyton Hall, Princeton,
       NJ 08544; {\tt draine@astro.princeton.edu}}
\and
\author{Aigen Li}
\affil{Department of Physics and Astronomy, University of Missouri,
       Columbia, MO 65211; {\tt LiA@missouri.edu}}

\begin{abstract}
Infrared (IR) emission spectra are calculated for dust heated
by starlight, for mixtures of amorphous silicate and graphitic grains,
including varying amounts of polycyclic aromatic hydrocarbon (PAH) particles.
The models are constrained to reproduce the average Milky Way 
extinction curve.

The calculations include the effects of single-photon heating.
Updated IR absorption properties for the PAHs 
are presented, that are consistent with observed emission spectra, 
including those newly obtained by {\it Spitzer Space Telescope}.
We find a size distribution for the PAHs 
giving emission band ratios consistent with the observed spectra  
of the Milky Way and other galaxies.

Emission spectra are presented for a wide range
of starlight intensities.
We calculate how the efficiency of emission into 
different IR bands depends on PAH size; 
the strong 7.7$\mum$ emission feature is produced mainly 
by PAH particles containing $<10^3$ C atoms.
We also calculate how the emission spectrum depends on $U$, the
starlight intensity relative to the local interstellar radiation field.
The submm and far-infrared emission is compared to the observed emission
from the local interstellar medium.

Using a simple distribution function,
we calculate the emission spectrum for dust heated by
a distribution of starlight intensities, such as occurs within galaxies.
The models are parameterized 
by the PAH mass fraction $\qpah$, 
the lower cutoff $\Umin$ of the starlight intensity distribution,
and the fraction $\gamma$ of the dust heated by starlight
with $U>\Umin$.
We present graphical procedures using Spitzer IRAC and MIPS 
photometry to estimate the parameters $\qpah$, $\Umin$, 
and $\gamma$, 
the fraction $f_{\rm PDR}$ of the dust luminosity coming from
photodissociation regions with $U>100$, and the total dust mass $\Mdust$.
\end{abstract}


\section{Introduction
	\label{sec:intro}}

Interstellar dust in galaxies absorbs energy from starlight;
this absorbed energy is then
reradiated at infrared (IR) and far-IR wavelengths.  
The unprecedented sensitivity of Spitzer Space Telescope 
\citep{Werner+Roellig+Low_etal_2004}
allows this 
IR emission to be measured for a wide range of
galaxy sizes and morphologies.
The spectral properties
of the IR emission from dust
allow one to infer the composition of the dust,
the size distribution of the dust particles, 
the intensity of the starlight that is heating
the dust, and the total mass of dust.

Deducing the dust properties and starlight intensity 
is by no means direct or straightforward.
To start, one needs to assume a provisional dust model with the physical 
nature of the dust (composition, geometry, and size
distribution) fully specified. One then tries to constrain 
the grain properties by comparing model predictions with 
observations (e.g. interstellar extinction, scattering,
polarization, IR and microwave emission, and interstellar 
depletions). 

Various models have been proposed for interstellar 
dust. The models fall into three broad 
categories: the silicate-graphite model 
\citep{Mathis+Rumpl+Nordsieck_1977,
       Draine+Lee_1984,
       Kim+Martin+Hendry_1994}
and its natural extension -- 
the silicate-graphite-PAH model 
\citep{Siebenmorgen+Krugel_1992,
       Li+Draine_2001b,
       Weingartner+Draine_2001a};
the silicate core-carbonaceous mantle model
\citep{Desert+Boulanger+Puget_1990,
       Jones+Duley+Williams_1990,
       Li+Greenberg_1997},
and the composite model which assumes the dust to 
be low-density aggregates of small silicates and carbonaceous particles
\citep{Mathis+Whiffen_1989,
       Mathis_1996,
       Zubko+Dwek+Arendt_2004}.
The core-mantle model is challenged by 
the nondetection of the 3.4$\mum$ C--H aliphatic hydrocarbon
polarization on sightlines where the 9.7$\mum$ Si--O
silicate band is observed to be polarized 
\citep{Chiar+Adamson+Whittet_etal_2006,
       Mason+Wright+Adamson+Pendleton_2006}
[see also
\citet{Adamson+Whittet+Chrysostomou_etal_1999,
       Li+Greenberg_2002}].
The original composite model proposed by 
\citet{Mathis+Whiffen_1989},
with $\sim$\,80\%  
of the grain volume consisting of vacuum,
may have too flat a far-IR emissivity to 
explain the observational data 
\citep{Draine_1994}.
The updated version of the composite model of 
\citet{Mathis_1996}
assumes a vacuum fraction $\sim$\,45\%
in order to make economical use of the available carbon 
to account for the observed extinction while satisfying 
the ``subsolar'' interstellar abundance budget suggested by 
\citet{Snow+Witt_1996}.
However, 
\citet{Dwek_1997} 
argues that this dust model
emits too much in the far-IR in 
comparison with observations.
The latest version of the composite model put forward by 
\citet{Zubko+Dwek+Arendt_2004}
aims at reproducing the observed
interstellar extinction and IR emission within a ``subsolar''
abundance budget. 
The ability of composite grain models to
reproduce the observed extinction within
``subsolar'' abundance constraints was
recently challenged by
\citet{Li_2005} 
based on the Kramers-Kronig relations.\footnote{%
  \newtext{
  Total (gas~+~dust) interstellar abundances
  remain uncertain.
  The most recent determination of the solar oxygen abundance
  $({\rm O/H})_\odot=10^{-3.34\pm0.05}$
  \citep{Asplund+Grevesse+Sauval_etal_2004}
  is close to the value for
  B stars in the Orion association ${\rm O/H}=10^{-3.31\pm0.03}$
  \citep{Cunha+Hubeny+Lanz_2006}.
  The warm interstellar medium gas has ${\rm O/H}=10^{-3.41\pm0.01}$
  \citep{Cartledge+Lauroesch+Meyer+Sofia_2006}, consistent
  with $\sim$20\% of the oxygen in silicate grains.
  For ``refractory'' elements
  such as Mg, Si, and Fe, however, 
  partial separation
  of dust and gas by processes such as radiation pressure,
  ambipolar diffusion, and gravitational sedimentation in the
  process of star formation could result in stellar abundances differing
  from interstellar abundances
  \citep{Snow_2000, Draine_2004a}.
  Photospheric abundances may therefore not be a reliable
  indicator of interstellar abundances for the elements that form refractory
  grains.
  }
  }

In this paper we calculate the IR emission 
expected from a specific physical model of
interstellar dust -- the silicate-graphite-PAH model.
In this model, 
the dust is assumed to consist of a mixture of 
carbonaceous grains and amorphous silicate grains, 
with size distributions that are consistent with
the observed wavelength-dependent extinction in 
the local Milky Way 
\citep{Weingartner+Draine_2001a}, 
including different amounts of 
polycyclic aromatic hydrocarbon (PAH) material.

\newtext{Spectroscopic observations with {\it ISO} revealed the
rich spectrum of PAHs (see, e.g., 
\cite{Beintema+vandenAncker+Molster_etal_1996};
\cite{Tielens+Hony+vanKerckhoven+Peeters_1999};
\cite{Joblin+Abergel+Bregman_etal_2000}).
}
The model \newtext{presented here} posits 
\newtext{``astro}-PAH\newtext{''} 
absorption properties that are 
consistent with spectroscopic observations of PAH emission 
from dust in nearby galaxies 
\citep{Smith+Draine+Dale_etal_2006}.
The model calculations
show how the IR emission depends on the PAH abundance, 
and also on the intensity of starlight heating the dust.
Comparison of models to infrared photometry obtained
with {\it Spitzer Space Telescope} allows one to estimate
PAH abundances, starlight intensities, and total dust masses.

The plan of this paper is as follows:

With new laboratory data 
[particularly the near-IR absorption spectra of PAH ions 
measured by 
\citet{Mattioda+Hudgins+Allamandola_2005}]
and Spitzer spectroscopic data 
[with new PAH features discovered 
\citep{Smith+Dale+Armus+Draine_etal_2004,
       Werner+Uchida+Sellgren_etal_2004,
       Smith+Draine+Dale_etal_2006}
and the familiar 7.7, 8.6, and 11.3$\mum$ features 
resolved into several subfeatures 
\citep{Smith+Draine+Dale_etal_2006}
we update in \S\ref{sec:PAH Cross Sections}
the cross sections $\Cabs(\lambda)$ that we previously assumed  
for the PAH particles\footnote{
   Note that ``PAH particles'', ``PAH grains'',
   and ``PAH molecules'' are synonymous.
   }
\citep{Li+Draine_2001b}.
We distinguish the PAHs by 2 charge states: 
either neutral (PAH$^{\rm o}$) or ionized (PAH$^{\pm}$).
We assume that multiply-charged (both positively
and negatively) PAHs all have the same cross sections
as those of singly-charged cations. 
\newtext{The resulting ``astro-PAH'' absorption cross sections do not
represent any specific material, but should approximate the
actual absorption properties of the PAH mixture in interstellar space.}

In \S\ref{sec:Heating of Dust by Starlight}
we show examples of temperature distribution functions 
for both neutral and ionized PAHs heated by starlight.
These temperature distribution functions are then used to calculate
emission spectra for individual particles in 
\S\ref{sec:Single-Grain Emission Spectra}, which are in turn used to
prepare a plot (Figure \ref{fig:band efficiencies})
showing how the efficiency of emitting into different
emission features depends on the PAH size.

When we observe emission from a region in the Milky Way or another
galaxy, there will always be a mixture of dust types and sizes present.
We consider the emission from a specific set of dust mixtures,
as described in \S\ref{sec:Dust Mixtures}.
All the dust mixtures considered here are
consistent with the observed ``average'' extinction curve for
diffuse regions in the local Milky Way, but they differ from one another in
the assumed abundance of small PAH particles.
In \S\ref{sec:fir and submm emission} we show that 
the far-IR and submm emission calculated for the model is consistent with the
COBE-FIRAS observations of emission from dust in the local Milky Way.
In \S\ref{sec:irac band ratios}
we show that 
the calculated IRAC band ratios appear to be 
consistent with the spectrum of diffuse emission from the Milky Way,
as determined by 
\citet{Flagey+Boulanger+Verstraete_etal_2006} 
from Spitzer Space Telescope observations.

The long wavelength emission from the dust model depends on the intensity
of the starlight heating the dust.  It will often be the case
that the region observed (e.g., an entire star-forming galaxy)
will include dust heated by a wide range of starlight intensities.
In \S\ref{sec:alpha-gamma dust models}
we describe a simple parametric model for the distribution of the
dust mass between regions with starlight intensities ranging from
low to very high.
We show in \S\ref{sec:dust mass estimation}
how observations in the 3 MIPS bands 
(24$\mum$, 71$\mum$, and 160$\mum$)
plus the 3.6$\mum$ and 7.9$\mum$ IRAC bands\footnote{%
  \newtext{
  The IRAC and MIPS bands are often referred to using nominal wavelengths
  3.6, 4.5, 5.8, 8.0, 24, 70, and 160$\mum$.
  However, to two significant digits, the effective wavelengths of
  IRAC bands 3 and 4 are $5.7$ and $7.9\mum$
  (IRAC Data Handbook Version 3.0), and that of
  MIPS band 2 is $71\mum$ (MIPS Data Handbook Version 3.2) and we
  will use these wavelengths to refer to the bands.
  }
  }
can be used to estimate the parameters describing 
the distribution of starlight intensities 
as well as the fraction $\qpah$ of the total dust mass 
that is in PAHs, thereby allowing estimation of the
total dust mass $\Mdust$ in the emitting region.
A reader interested primarily in applying the results of this
paper to interpretation of IRAC and MIPS observations may
wish to proceed directly to \S\ref{sec:dust mass estimation}.

The results of the paper are discussed in \S\ref{sec:discussion}
and summarized in \S\ref{sec:summary}.

\section{PAH Cross Sections: Post-Spitzer Era
         \label{sec:PAH Cross Sections}}

\begin{figure}[h]
\begin{center}
\includegraphics[width=12cm]{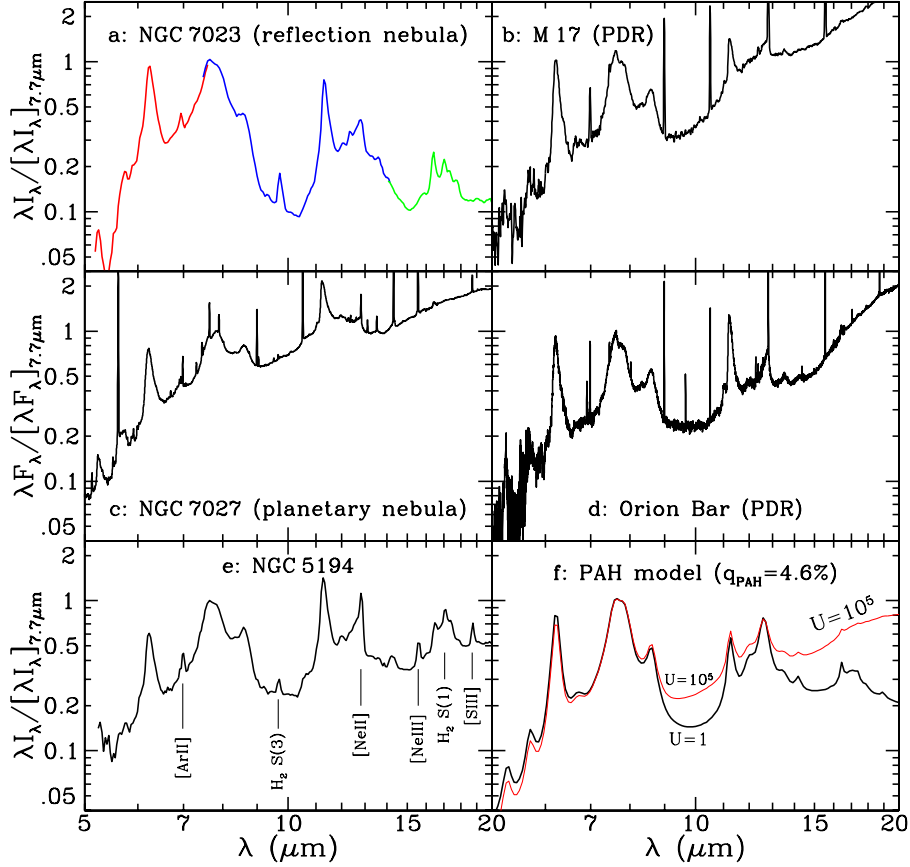}
\caption{\label{fig:observed spectra} \footnotesize
         Observed 5--20\um\ spectra for: 
	 (a) Reflection nebula NGC~7023
             \citep{Werner+Uchida+Sellgren_etal_2004};
	 (b) Orion Bar PDR
             \citep{Verstraete+Pech+Moutou_etal_2001};
	 (c) M17 PDR
             \citep{Peeters+Tielens+Boogert_etal_2005};
	 (d) Planetary nebula NGC~7027
	     \citep{vanDiedenhoven+Peeters+vanKerckhoven_etal_2004};
	 (e) Seyfert Galaxy NGC~5194
             \citep{Smith+Draine+Dale_etal_2006}.
	 Also shown (f) is the emission calculated for
	 the present dust model with $\qpah=4.6\%$, illuminated
	 by the local diffuse starlight with $U=1$ \newtext{and $10^5$} 
	 (see Fig.\ \ref{fig:variousU_spec}).
	 }
\end{center}
\end{figure}
\begin{table}[h]
\footnotesize
\begin{center}
\caption{\footnotesize
         \label{tab:drude}
         PAH Resonance Parameters.
}
\begin{tabular}{cccccl}
    &             &            
                  & \multicolumn{2}{c}{$\sigma_{{\rm int},j}\equiv\int \sigma_{{\rm abs},j} d\lambda^{-1}$}\\
$j$ & $\lambda_j$ & $\gamma_j$ &  neutral           &   ionized  \\
    & (\um)       &            &  ($10^{-20}$cm/C)  &   ($10^{-20}$cm/C)   
    & \newtext{tentative identification} \\
\hline
1   & 0.0722  & 0.195   & $7.97\times10^{7}$      & $7.97\times10^{7}$ 
    & \newtext{$\sigma\rightarrow\sigma^*$ transition in aromatic C} \\
2   & 0.2175  & 0.217   & $1.23\times10^{7}$      & $1.23\times10^{7}$ 
    & \newtext{$\pi\rightarrow\pi^*$ transition in aromatic C} \\
3   & 1.050   & 0.055   & $0$                     & $2.0\times10^{4}$
    & \newtext{weak electronic transition(s) in PAH cations} \\
4   & 1.260   & 0.11    & $0$                     & $0.078$  
    & \newtext{weak electronic transition(s) in PAH cations} \\
5   & 1.905   & 0.09    & $0$                     & $-146.5$ 
    & \newtext{?} \\
6   & 3.300   & 0.012   & $394$(H/C)              & $89.4$(H/C)
    & \newtext{aromatic C--H stretch} \\
\newtext{7}   & \newtext{5.270}    & \newtext{0.034}   & $2.5$                   & $20.$  
    & \newtext{C-H bend + C-H stretch combination mode} \\
8   & 5.700   & \newtext{0.035} & \newtext{4} & \newtext{32.} 
    & \newtext{C-H bend + C-H stretch combination mode} \\
9   & 6.220   & \newtext{0.030}  & $29.4$                  & $235.$            
    & \newtext{aromatic C-C stretch (in-plane)} \\
10  & 6.690   & 0.070   & $7.35$                  & $59.$ 
    & \newtext{?} \\
11  & 7.417   & 0.126   & $20.8$                  & $181.$ 
    & \newtext{aromatic C-C stretch} \\
12  & 7.598   & 0.044   & $18.1$                  & $163.$ 
    & \newtext{aromatic C-C stretch} \\
13  & 7.850   & 0.053   & $21.9$                  & $197.$ 
    & \newtext{C-C stretch + C-H bending} \\
14  & 8.330   & 0.052   & $6.94$(H/C)             & $48.$(H/C)      
    & \newtext{C-C stretch + C-H bending?} \\
15  & 8.610   & 0.039   & $27.8$(H/C)             & $194.$(H/C) 
    & \newtext{C-H in-plane bending} \\
\newtext{16}  & \newtext{10.68} & \newtext{0.020} & \newtext{0.3(H/C)} & \newtext{0.3(H/C)} 
    & \newtext{C-H out-of-plane bending, solo?} \\
17  & 11.23   & \newtext{0.012}  & $18.9$(H/C)             & $17.7$(H/C)       
    & \newtext{C-H out-of-plane bending, solo} \\
18  & \newtext{11.33}   & \newtext{0.032}   & $52.$(H/C)             & $49.$(H/C)      
    & \newtext{C-H out-of-plane bending, solo} \\
19  & 11.99   & \newtext{0.045}   & $24.2$(H/C)             & $20.5$(H/C)      
    & \newtext{C-H out-of-plane bending, duo} \\
20  & \newtext{12.62}   & \newtext{0.042}   & \newtext{35(H/C)} & \newtext{31(H/C)}                
    & \newtext{C-H out-of-plane bending, trio} \\
21  & \newtext{12.69}   & \newtext{0.013}   & \newtext{1.3(H/C)}                & \newtext{1.3(H/C)}                 
    & \newtext{C-H out-of-plane bending, trio} \\
22  & \newtext{13.48}   & \newtext{0.040}   & \newtext{8.0(H/C)} & \newtext{8.0(H/C)} 
    & \newtext{C-H out-of-plane bending, quartet?} \\
23  & 14.19   & 0.025   
   & $0.45$                   & $0.45$ 
    & \newtext{C-H out-of-plane bending, quartet?} \\
24  & 15.90   & 0.020   
   & $0.04$                   & $0.04$ 
    & \newtext{?} \\
25  & 16.45  & 0.014   
   & $0.5$                  & $0.5$ 
    & \newtext{C-C-C bending?} \\
26  & 17.04  & 0.065   
   & $2.22$                  & $2.22$ 
    & \newtext{C-C-C bending?} \\
27  & 17.375  & 0.012   
   & $0.11$                  & $0.11$  
    & \newtext{C-C-C bending?} \\
28  & 17.87  & 0.016   
   & $0.067$                  & $0.067$ 
    & \newtext{C-C-C bending?} \\
\newtext{29}  & \newtext{18.92}  
   & \newtext{0.10}   & \newtext{0.10}                    & \newtext{0.17} 
    & \newtext{C-C-C bending?} \\
30  & 15      & 0.8     & $50.$                   & $50.$ 
    & \newtext{} \\
\hline
\end{tabular}
\end{center}
\end{table}

The strong and ubiquitous interstellar emission features 
observed at 3.3$\mum$, 6.2$\mum$, 7.7$\mum$, 8.6$\mum$,
and 11.3$\mum$ almost certainly arise from 
vibrational modes of polycyclic aromatic hydrocarbon 
(PAH) material, with 
C--H stretching modes producing the 3.3$\mum$ feature, 
C--H bending modes producing the in-plane 8.6$\mum$ 
and out-of-plane 11.3$\mum$ features,
and C--C stretching and bending modes 
producing emission features at 6.2$\mum$, 7.7$\mum$, 
and longer wavelengths
\citep{Leger+Puget_1984,Allamandola+Tielens+Barker_1985}.\footnote{%
  \newtext{
  Other carriers have also been proposed, 
  including hydrogenated amorphous carbon (HAC)
  \citep{Jones+Duley+Williams_1990},
  quenched carbonaceous composite (QCC)
  \citep{Sakata+Wada+Onaka+Tokunaga_1987},
  coal 
  \citep{Papoular+Ellis+Guillois_etal_1993}, 
  hydrogenated fullerenes 
  \citep{Webster_1993}, 
  and nanodiamonds 
  \citep{Jones+dHendecourt_2000}.
  The HAC, QCC, and coal hypotheses assume that the emission arises following
  photon absorption in
  small thermally-isolated aromatic units within or attached to these
  bulk materials \citep{Duley+Williams_1981}.
  However, it does not appear
  possible to confine the absorbed photon energy within these aromatic
  ``islands'' for the time $\gtsim10^{-3}\s$ required for the thermal
  emission process (see \cite{Li+Draine_2002a}).
  Free-flying fullerenes and nanodiamonds have the required small 
  heat capacity, but: (1) There are strong upper limits on the abundance of
  C$_{60}$ and C$_{60}^+$ \citep{Moutou+Sellgren+Verstraete+Leger_1999};
  (2) Although not ruled out, there is little spectroscopic
  evidence for hydrogenated nanodiamond in the ISM.
  }
  }

To reproduce this emission, a dust model must include a substantial
population of ultrasmall grains or large molecules with the vibrational
properties of PAH material and with 
sizes such that single-photon heating can excite
the observed vibrational emission.
Because the exact composition of the interstellar PAH material is
unknown, and also because laboratory knowledge of the optical properties
of PAHs is very limited, it is necessary to make assumptions regarding the
absorption cross sections of the PAH particles.

\begin{figure}[h]
\begin{center}
\includegraphics[width=9.5cm,angle=0]{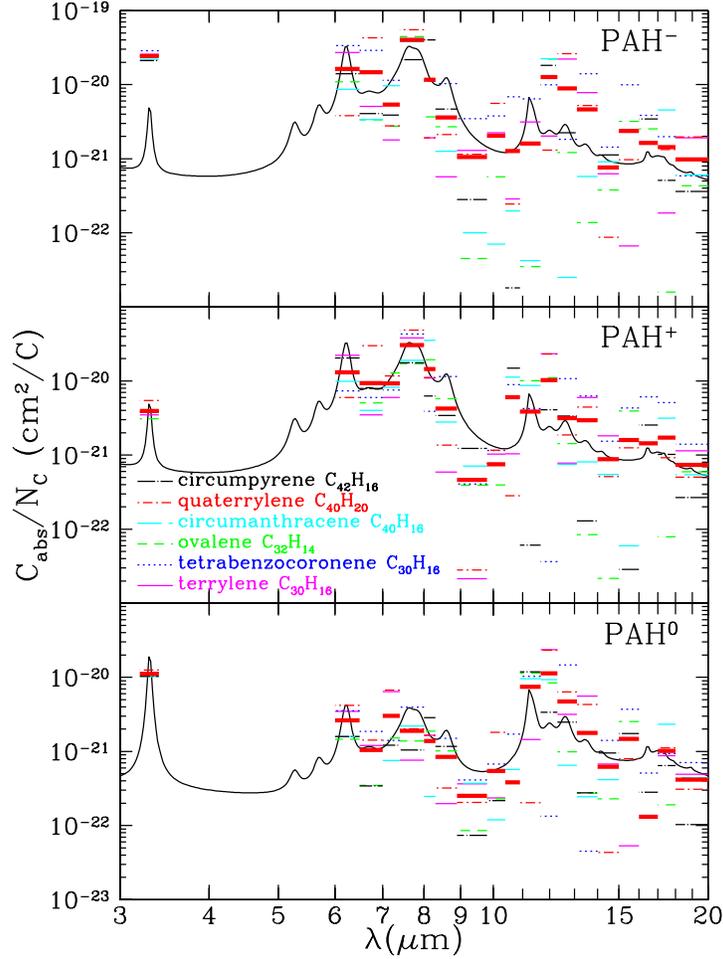}
\caption{\label{fig:Cabs_vs_theory}
         \footnotesize
         Solid curve: adopted absorption cross section per C 
         from eq.\ (\ref{eq:adopted Cabs})
	 with C/H$\approx$3.2 (e.g., C$_{64}$H$_{20}$). 
	 For the neutrals, anions, and cations listed in the figure legends, 
	 the 
	 horizontal line segments indicate the average absorption
	 over that frequency intervals, taken from theoretical
	 calculations by \citet{Malloci+Joblin+Mulas_2006}. 
	 The heavy solid line segment is the average for the 6 species shown.
	 }
\end{center}
\end{figure}

The approach taken here to modeling the PAHs is to try to
find optical properties for ``astro-PAH'' material that appear to be physically
reasonable (i.e., band strengths within the 
range measured for PAHs in the laboratory),
in order to estimate the sizes and abundance of interstellar PAHs
that would be consistent with the observed emission spectra.
To this end, we adopt feature profiles that are based on
astronomical observations.
At wavelengths $\lambda > 5.5\micron$, 
the spectra for the central regions of galaxies in the
SINGS galaxy sample
\citep{Smith+Draine+Dale_etal_2006} 
provide observed profiles for
the spectral features that the present model attempts to mimic.
We follow 
\citet{Li+Draine_2001b} (herafter LD01) 
in describing the PAH vibrational resonances by Drude profiles
(components $j=6$--26 in Table \ref{tab:drude}),\footnote{%
  \newtext{\citet{Boulanger+Boissel+Cesarsky+Ryter_1998} 
  have shown that Lorentz profiles 
  provide good fits to the PAH emission
  features from the NGC 7023 reflection nebula 
  and the $\rho$ Oph molecular cloud.
  The Drude profile closely resembles a Lorentz profile (both have
  more extended wings
  than a Gaussian profile).
  We favor the Drude profile as it is 
  expected for classical damped harmonic oscillators.
  }
  }

\beq
\Delta \Cabs(\lambda) = N_{\rm C}\sum_{j=1}^{25}
\frac{2}{\pi} 
\frac{\gamma_j \lambda_j \sigma_{{\rm int},j}}
     {(\lambda/\lambda_j-\lambda_j/\lambda)^2+\gamma_j^2}~~~,
\eeq
but with new profile
parameters, adjusted to
\newtext{closely} 
resemble the observed profiles. 
The profile parameters are given
in Table \ref{tab:drude},
where $\lambda_j$, $\gamma_j\lambda_j$,
and $\sigma_{\rm int}\equiv\int \sigma_{\rm abs} d\lambda^{-1}$ 
are, respectively, the peak wavelength, the FWHM, 
and the integrated strength per C atom of the $j$-th Drude resonance.
\newtext{%
The features in Table \ref{tab:drude}, and their identifications,
have been discussed previously [see, e.g., \citet{Tielens_2005}],
with \citet{Smith+Draine+Dale_etal_2006} providing a comprehensive
study of the 5--35\um\ emission from galaxies.}
We assume the PAH particles to have absorption cross sections 
per carbon atom $\Cabs(\lambda)/N_{\rm C}$ 
equal to those adopted by 
\citet{Li+Draine_2001b} 
[see eq.\ 5--11 of that paper] 
except for the following changes:


\begin{enumerate}
\item For PAH ions,
we add additional absorption in the near-IR as recommended by
\citet{Mattioda+Allamandola+Hudgins_2005}.
This consists of a ``continuum'' term
\beq
\frac{\Delta\Cabs(\lambda)}{N_{\rm C}}
=
3.5\times10^{-19 -1.45/x}
\exp\left(-0.1x^2\right)\cm^2 ~~~{\rm for~ions},~~ 
x\equiv\left(\lambda/\mu{\rm m}\right)^{-1}
~~~.
\eeq
Because PAH ions were already assumed to absorb strongly at 
$\lambda \ltsim 0.8\micron$, 
this additional absorption is numerically insignificant for 
$\lambda \ltsim 0.8\micron$; 
the factor $\exp\left(-0.1x^2\right)$ has been added 
simply to force this term to go smoothly to zero 
as $\lambda\rightarrow 0$.

\item For PAH ions, we 
add near-IR resonances at wavelengths $1.05\micron$ and $1.26\micron$
plus a negative ``resonance'' term at $1.905\micron$ to
suppress aborption in the 1.8--2.0$\mum$ region,
as recommended by 
\citet{Mattioda+Allamandola+Hudgins_2005};
the three features are
represented by Drude profiles, 
with parameters as given in Table \ref{tab:drude}.
The negative term at $\lambda$\,=\,1.905$\mum$ ($j$\,=\,5)
was suggested by the recent laboratory data of
\citet{Mattioda+Hudgins+Allamandola_2005}.
Removal of this term
has a negligible effect on the heating or cooling rates of PAHs 
except in regions illuminated by very cool stars ($T_{\rm eff} \ltsim 1500\K$)
or in regions where the PAHs are excited to unusually
high temperatures ($T\gtsim 1500\K$).

\item Small changes have been made to central wavelengths and
feature widths (e.g., 
$\lambda_j=6.20\micron\rightarrow6.22\micron$
and $\gamma_j=0.032\rightarrow0.0284$ for the feature near 6.2$\mum$,
$\lambda_j=11.9\micron\rightarrow11.99\micron$ and 
$\gamma_j=.025\rightarrow0.050$ for the feature near 12$\mum$,
and $\lambda_j=12.7\micron\rightarrow12.61\micron$ and 
$\gamma_j=.024\rightarrow0.0435$ for the feature near 12.7$\mum$)
guided by spectra obtained recently
by {\it Spitzer Space Telescope} 
\citep{Smith+Draine+Dale_etal_2006}.

\item The integrated strength 
$\sigma_{\rm int}\equiv\int \sigma_{\rm abs} d\lambda$ 
of the 3.3$\mum$ feature has been increased by a factor 1.5 for neutrals,
and a factor 2 for ions, to better agree with the range of values calculated
for a number of PAHs (see Figure \ref{fig:Cabs_vs_theory}).

\item $\sigma_{\rm int}$ for the 6.22\um\ feature is 50\% of the value
in LD01.
 
\item The 7.7$\mum$ complex is now
composed of three components, at 7.417, 7.598, and 7.850$\mum$, 
with 
$\sigma_{\rm int}$ equal to 50\% of the 7.7$\mum$ feature in LD01.

\item $\sigma_{\rm int}$ of the 8.6$\mum$ feature in LD01 is
now shared by features at 8.330 and 8.610$\mum$,
with $\sigma_{\rm int}$ equal to 50\% of the 8.6\um\ feature in LD01.

\item The 11.3$\mum$ feature is now composed of
features at 11.23 and 11.30$\mum$, with 
$\sigma_{\rm int}$ equal to 50\% of the 11.3$\mum$ feature in LD01.

\item The integrated strength $\sigma_{\rm int}$ of 
the 12.7$\mum$ feature has been multiplied by 0.63
for both neutrals and ions relative to LD01.

\item Weak features have been added at 5.70$\mum$, 6.69$\mum$, 
13.60$\mum$, 14.19$\mum$, 15.90$\mum$, \newtext{and 18.92$\mum$,} 
as seen in spectra of
star-forming galaxies in the SINGS survey
\citep{Smith+Draine+Dale_etal_2006}.  The 5.70\um\ feature has
previously been seen in planetary nebulae 
\citep{Allamandola+Bregman+Sandford_etal_1989} and
PDRs
\citep{Verstraete+Puget+Falgarone_etal_1996,
       Peeters+Allamandola+Bauschlicher_etal_2004},
and is presumed to be due to combination and overtone bands 
involving C-H out-of-plane bending modes.

\item A weak feature at 5.25\um\,
seen in spectra of the M17 PDR and the Orion Bar 
\citep{Verstraete+Puget+Falgarone_etal_1996,
       Peeters+Allamandola+Bauschlicher_etal_2004}
and presumed to be due to C-H out-of-plane combination and overtone modes
\citep{Allamandola+Bregman+Sandford_etal_1989},
has been added.

\item The strength of the 16.4$\mum$ feature has been multiplied
by 0.14 relative to LD01.

\item A new emission complex near 17$\mum$ has been added, 
composed of features at
17.038, 17.377, and 17.873$\mum$ 
\citep{Smith+Dale+Armus+Draine_etal_2004, 
       Werner+Uchida+Sellgren_etal_2004,
       Smith+Draine+Dale_etal_2006}.

\item Emission features at 21.2 and 23.1$\mum$ were seen
in some laboratory samples 
\citep{Moutou+Leger+DHendecourt_1996},
and were therefore included by LD01 as examples 
of features that might be observed 
at $\lambda \gtsim 20\micron$.
However, the SINGS spectra 
\citep{Smith+Draine+Dale_etal_2006}
do not show any features at $\lambda > 19\micron$. The
21.2 and 23.1$\mum$ features have therefore 
been eliminated in the new model.

\item LD01 included a broad absorption component with $\lambda_j=26\micron$,
$\gamma_j=0.69$, and $\sigma_{\rm int}=18\times10^{-20}\cm/{\rm C}$.  
This has been replaced by a broad absorption 
component with $\lambda_{27}=15\micron$, $\gamma_{27}=0.8$, and
$\sigma_{\rm int}=50\times10^{-20}\cm/{\rm C}$ to provide continuum
emission from 13\um\ longward.

\end{enumerate}

In general, PAHs and larger grains will not be spherical, 
but we will characterize a grain of mass $M$ by
the effective radius $a$, defined to be the radius of an equal volume sphere:
$a \equiv (3M/4\pi\rho)^{1/3}$, 
where amorphous silicate is assumed to have a mass density
$\rho=3.5\g\cm^{-3}$,
and carbonaceous grains are assumed to have a mass density due to
graphitic carbon alone of
$\rho=2.2\g\cm^{-3}$.

Thus the number of carbon atoms in a carbonaceous grain is
\beq
N_{\rm C}=460 \left(\frac{a}{10\Angstrom}\right)^3
~~~.
\eeq
The smallest PAH considered in this paper
has $N_{\rm C} = 20$ C atoms 
(corresponding to $a\approx$\,3.55\,\AA)
since smaller PAHs are 
photolytically unstable 
\citep{Allamandola+Tielens+Barker_1989}.
As in LD01, we assume the number of H atoms per C atom to depend on the
size of the PAH:
\beq
\begin{array}{l c l c l}
{\rm H/C} &=& 0.5 & {\rm for} & N_{\rm C} \leq 25
\\
    &=& 0.5 \left(25/N_{\rm C}\right)^{1/2} && 25 \leq N_{\rm C} \leq 100
\\
          &=& 0.25                       && N_{\rm C}\geq 100
\end{array}
\eeq

Figure \ref{fig:Cabs_vs_theory} shows the adopted $C_{\rm abs}(\lambda)$
in the infrared for a PAH molecule with H/C=5/16 (e.g., C$_{64}$H$_{20}$, with
$a=5.18\Angstrom$).
Also shown are values of $C_{\rm abs}$ per C atom, averaged over
wavelength intervals, for a number of molecules for which $C_{\rm abs}$
has been calculated theoretically 
\citep{Malloci+Joblin+Mulas_2006}
for selected PAH molecules, cations, and anions.
The first thing to note is the wide range of absorption 
cross sections per C.  
For example, in the case of the 3.3$\mum$ C--H stretch, 
the integrated absorption cross section
per C varies by a factor of 25 among the PAH cations.
The absorption averaged over the 7.5--8.0$\mum$ range
varies by a factor of 30 among the neutral PAHs.
Similar large variations in absorption cross sections are also seen
at other wavelengths.
Our adopted cross section falls within the range found for the sample
of molecules shown in Fig.\ \ref{fig:Cabs_vs_theory}.

\begin{figure}[h]
\begin{center}
\includegraphics[width=12cm,angle=270]{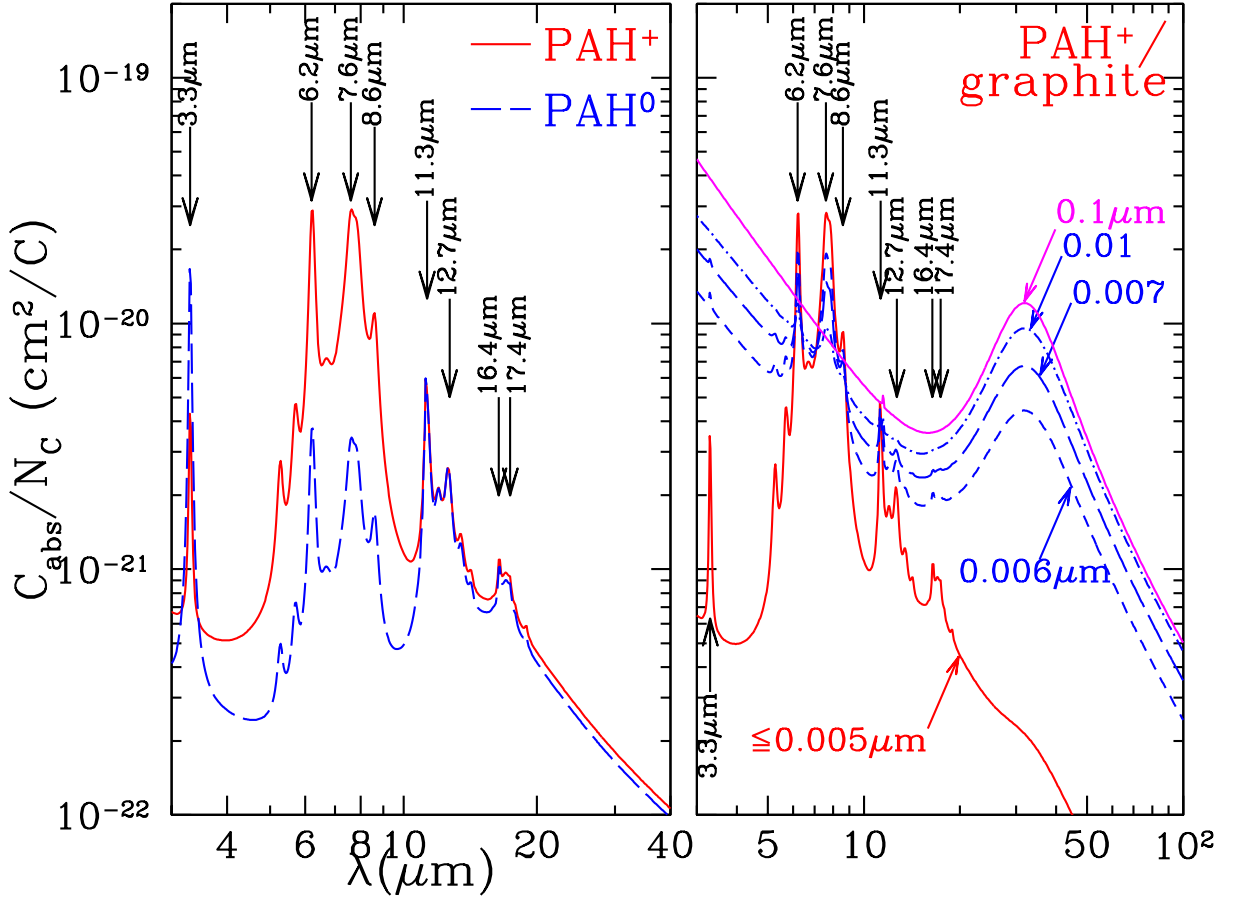}
\caption{\label{fig:Cabs for carbonaceous grains}
         \footnotesize
         Absorption cross section per C atom for (a) neutral and ionized PAHs,
         and (b) for ionized carbonaceous grains, with properties of
	 PAHs for $6\Angstrom<a<50\Angstrom$, 
	 and properties of graphite spheres for $a\gtsim 100\Angstrom$.
         See \S\ref{sec:PAH Cross Sections} for details.
         }
\end{center}
\end{figure}

Figure \ref{fig:Cabs for carbonaceous grains} compares our adopted 
absorption cross sections for both neutral and ionized PAHs.

As in LD01, as the number $N_{\rm C}$
of carbon atoms in the grain increases,
we assume a continuous change in optical properties from those of
PAH material when $N_{\rm C}$ is small, to those of graphite when
$N_{\rm C}$ is large.
The transition from PAH to graphite is entirely ad-hoc: we take
\beq \label{eq:adopted Cabs}
\Cabs(\lambda) 
= \left(1-\xi_{\rm gra}\right) \Cabs({\rm PAH},N_{\rm C}) 
+ \xi_{\rm gra}\,\Cabs({\rm graphite},a)
\eeq
where we take the graphite ``weight'' $\xi_{\rm gra}$ to be
\beqa
\xi_{\rm gra} 
&=& 0.01 ~{\rm for}~ a \leq 50\Angstrom ~~~ (N_{\rm C} \leq 5.75\times10^4)
\\
&=& 0.01+0.99\left[1-\left(\frac{50\Angstrom}{a}\right)^3\right] ~~~{\rm for}~ a \geq 50\Angstrom  ~~~ ({\rm i.e.}~N_{\rm C} \geq 5.75\times10^4) ~~.
\eeqa
The rationale for this is as follows: 
in addition to the
C--H stretching mode emission at 3.3$\mum$,
there appears to be 2--5$\mum$
continuum emission from the interstellar medium 
\citep{Lu+Helou+Werner+Dinerstein+Dale_2003,
       Helou+Roussel+Appleton+Frayer+Stolovy_etal_2004}
and we therefore need a source
of continuum opacity in the 2--5$\mum$ region
that is not provided by the C--H and C--C stretching and bending
modes.
Here we assume that every small PAH has a small amount of ``continuum''
opacity, equal to 1\% of what would have been calculated with the
optical properties of bulk graphite (i.e. $\xi_{\rm gra}$\,=\,0.01).
When the carbonaceous particles have $N_{\rm C}\gtsim 10^5$, they
are assumed to behave like bulk graphite.

Graphite is highly anisotropic, with different
dielectric functions $\epsilon_\perp$ and $\epsilon_\parallel$ for
electric fields perpendicular and parallel to the ``{\it c}-axis'' 
(the {\it c}-axis is normal to the ``basal plane'' of graphite).  
For $1\micron < \lambda < 20\micron$,
absorption in small randomly-oriented graphite spheres is primarily due
to the free electrons in graphite moving in the basal plane.
However, the basal plane conductivity is large enough that at wavelengths
$\lambda > 30\micron$ the absorption
is primarily due to the weak but nonzero conductivity parallel to the
{\it c}-axis.
The contribution of ``free electrons'' to $\epsilon_\parallel$ results in
an absorption peak near 30$\mum$ \citep{Draine+Lee_1984}.
This peak is seen for $a\geq60\Angstrom$ 
in Figure \ref{fig:Cabs for carbonaceous grains}b.
This absorption peak results in a broad emission feature near 30$\mum$ when the
$a\gtsim60\Angstrom$ graphite particles are heated to $T\gtsim100\K$.

The peak in $\Cabs$ near 30\um\ in 
Figure \ref{fig:Cabs for carbonaceous grains}b
is a consequence of our adopted dielectric
function for graphite, which is based on a simple ``free electron'' model that
may not apply to realistic carbonaceous
grains.  For example, the measured
absorption in amorphous carbon grains
\citep{Tanabe+Nakada+Kamijo+Sakata_1983,
       Mennella+Brucato+Colangeli+Palumbo_1999}
does not appear to show a peak near 30\um.
This is further discussed in \S\ref{sec:others}

\section{Heating of Dust by Starlight
         \label{sec:Heating of Dust by Starlight}}

We consider heating of grains by radiation with 
energy density per unit frequency
\beq
u_\nu = U \times u_\nu^{\rm MMP83}
\eeq
where $U$ is a dimensionless scaling factor and
$u_\nu^{\rm MMP83}$ is
interstellar radiation field (ISRF) estimated by 
\citet{Mathis+Mezger+Panagia_1983}[hereafter MMP83]
for the solar neighborhood.

\begin{figure}[h]
\begin{center}
\includegraphics[width=10cm,angle=270]{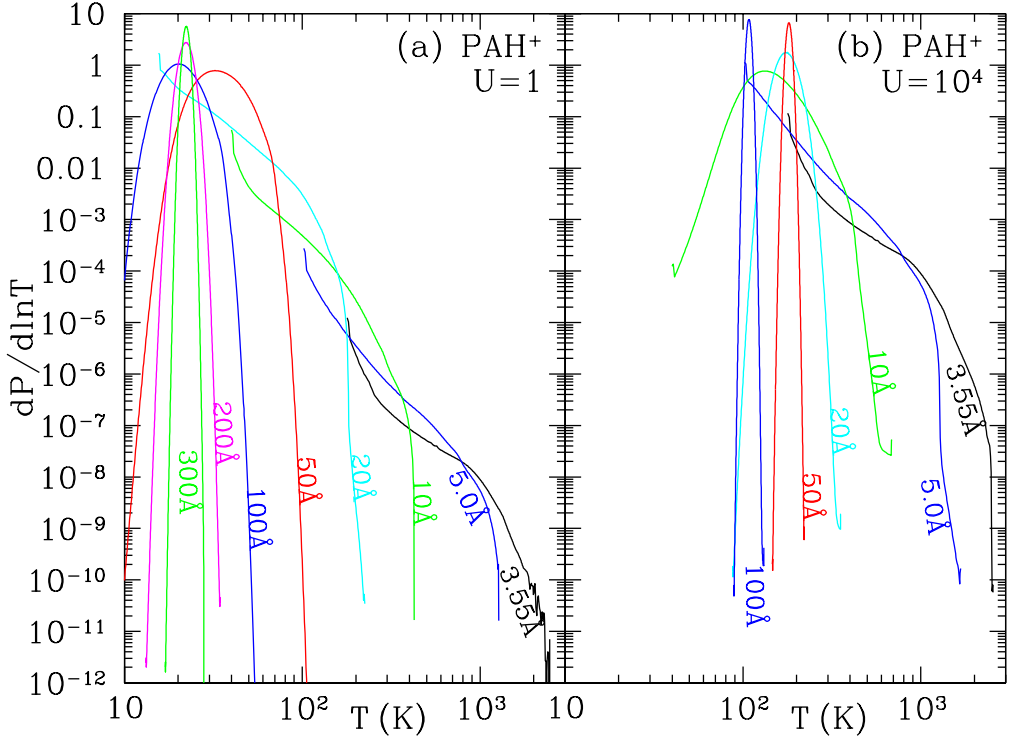}
\caption{\label{fig:dPdlnT for PAH+}
         \footnotesize
         Temperature probability distribution $dP/d\ln T$ 
         for selected carbonaceous grains
	 heated by starlight with $U=1$ and $U=10^4$.
	 }
\end{center}
\end{figure}

For each grain composition and radius $a$, we use a
detailed model for the heat capacity 
\citep{Draine+Li_2001} 
to calculate the function
$\bar{E}(T)$,
the expectation value for the vibrational energy of the grain when in
equilibrium with a heat reservoir at temperature $T$.
For a grain with vibrational energy $E$,
the grain ``temperature'' $T(E)$
is taken to be the temperature at which the expectation value for
the vibrational energy would be $E$: $\bar{E}(T)=E$;
we use this temperature estimate for all values of the vibrational
energy $E$, in this respect departing from
\citet{Draine+Li_2001}, 
who used a different estimate for the temperature
when dealing with the first 20 vibrational excited states.
This does not appreciably affect the emission spectrum since almost all of 
the absorbed photon energy is reradiated while the grain is at high 
temperatures.

For each grain composition, radius $a$, 
and radiation intensity scale factor $U$, we determine
the probability distribution function $dP/dT$, 
where $dP$ is the probability of finding the
grain with temperature in $[T,T+dT]$.  
For large grains, $dP/dT$ is approximated by a 
delta function $dP/dT\approx\delta(T-T_{\rm ss})$, 
where the ``steady-state'' temperature
$T_{\rm ss}(a)$ is the temperature at which the radiated power is equal to
the time-averaged heating rate for a grain of radius $a$.
For small grains, we find the steady-state solution $dP/dT$ for grains
subject to stochastic heating by photon absorption, and cooling by
emission of infrared photons, 
using the ``thermal-discrete'' approximation
\citep{Draine+Li_2001}, 
where the downward transition probabilities for 
a grain with vibrational energy $E$ are estimated
using a thermal approximation.
For each grain size $a$ and radiation intensity $U$, we divide the
energy range $[E_{\rm min},E_{\rm max}]$ into 500 bins.  
$E_{\rm min}$ and $E_{\rm max}$ are found iteratively, 
with the requirement that the probability of the grain 
being outside the range $[E_{\rm min},E_{\rm max}]$ be negligible.

Figure \ref{fig:dPdlnT for PAH+} shows $dP/d\ln T$ for PAH$^+$/graphite
grains for selected
grain sizes, for $U=1$ and $U=10^4$.
In Figure \ref{fig:dPdlnT for PAH+}a one
sees that small grains undergo extreme temperature excursions 
(the $a=3.55\Angstrom$ PAH occasionally reaches $T> 2000\K$), whereas
larger grains (e.g., $a=300\Angstrom$) have temperature distribution functions
that are very strongly-peaked, corresponding to only small excursions
around a ``steady-state'' temperature $T_{\rm ss}$
(the temperature for which the
rate of radiative cooling would equal the time-averaged rate of energy
absorption).

Figure \ref{fig:dPdlnT for PAH+}b shows $dP/d\ln T$ for $U=10^4$.
It is apparent that when the rate of starlight heating is increased,
the ``steady-state'' temperature approximation becomes valid for smaller
grains.  For example, for $U=10^4$ one could approximate an $a=50\Angstrom$ 
grain as having a steady temperature $T_{\rm ss}\approx150\K$
whereas for $U=1$ the temperature excursions are very important for
this grain.
The radius below which single photon heating is important, and above which
the grain temperature can be approximated as being constant,
is the size for which the time between photon absorptions becomes equal
to the radiative cooling time
\citep{Draine+Li_2001}.
Or (equivalently), it is the size for which the thermal energy content
of the grain when at $T_{\rm ss}$ is equal to the energy of the
most energetic photons heating the grain.

\section{Single-Grain Emission Spectra
         \label{sec:Single-Grain Emission Spectra}}

\begin{figure}[h]
\begin{center}
\includegraphics[width=10cm,angle=270]{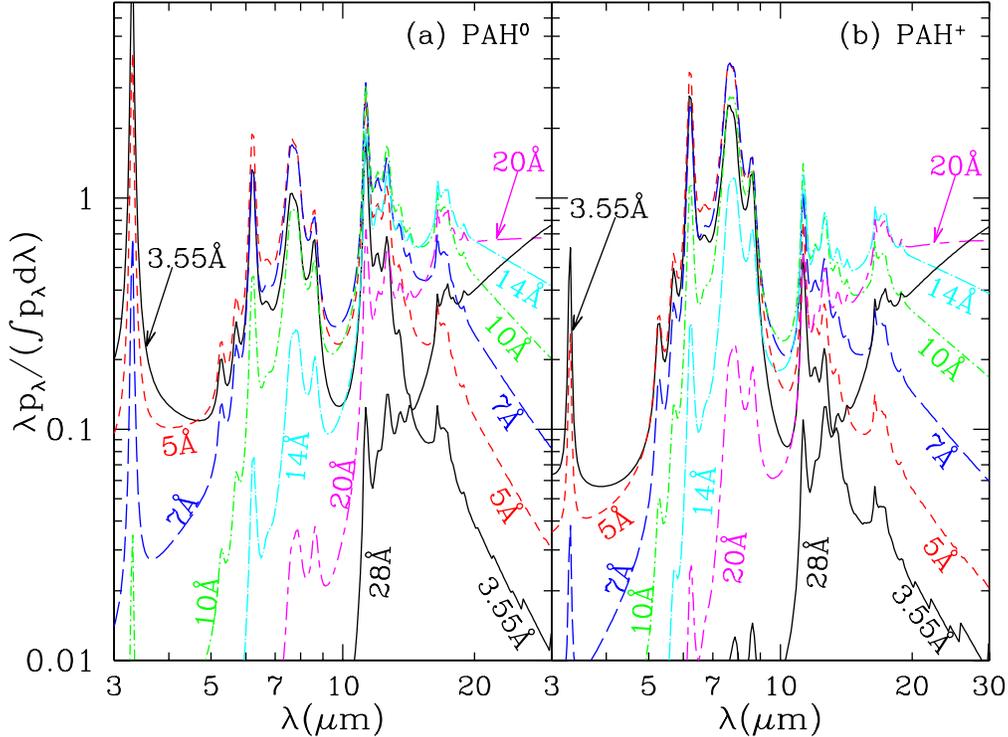}
\caption{\label{fig:PAHi and PAHn emission spectra}
         \footnotesize
         Normalized time-averaged emission spectra for $U<10^4$
         for (a) neutral and (b) ionized PAHs of various sizes (see text).
	 }
\end{center}
\end{figure}

From the probability distributions $dP/dT$, we calculate the
time-averaged
emission spectra for individual particles,
\beq
p_\lambda = \int 4\pi \Cabs(\lambda) B_\lambda(T) \frac{dP}{dT} dT
~~~,
\eeq
\beq
B_\lambda(T) \equiv 2hc^2\lambda^{-5}\left[\exp(hc/\lambda kT)-1\right]^{-1}
~~~.
\eeq
Figure \ref{fig:PAHi and PAHn emission spectra} shows
the 3--30$\mum$ emission from PAH ions and PAH neutrals heated
by $U=1$ starlight, with spectra shown for a number of different sizes.
As expected, the short wavelength emission (e.g., the 3.3$\mum$ feature) is
strong only for the smallest particles, which can be heated to $T\gtsim 10^3\K$
by absorption of a single ultraviolet photon 
(see Fig.\ \ref{fig:dPdlnT for PAH+}).
As the particle size becomes larger, the short wavelength emission falls off,
and an increasing fraction of the absorbed starlight energy is radiated in
the longer wavelength modes.
Emission in the 17$\mum$ complex, for example, is most efficient for
$a\approx 15\Angstrom$.

\begin{figure}[h]
\begin{center}
\includegraphics[width=10cm,angle=270]{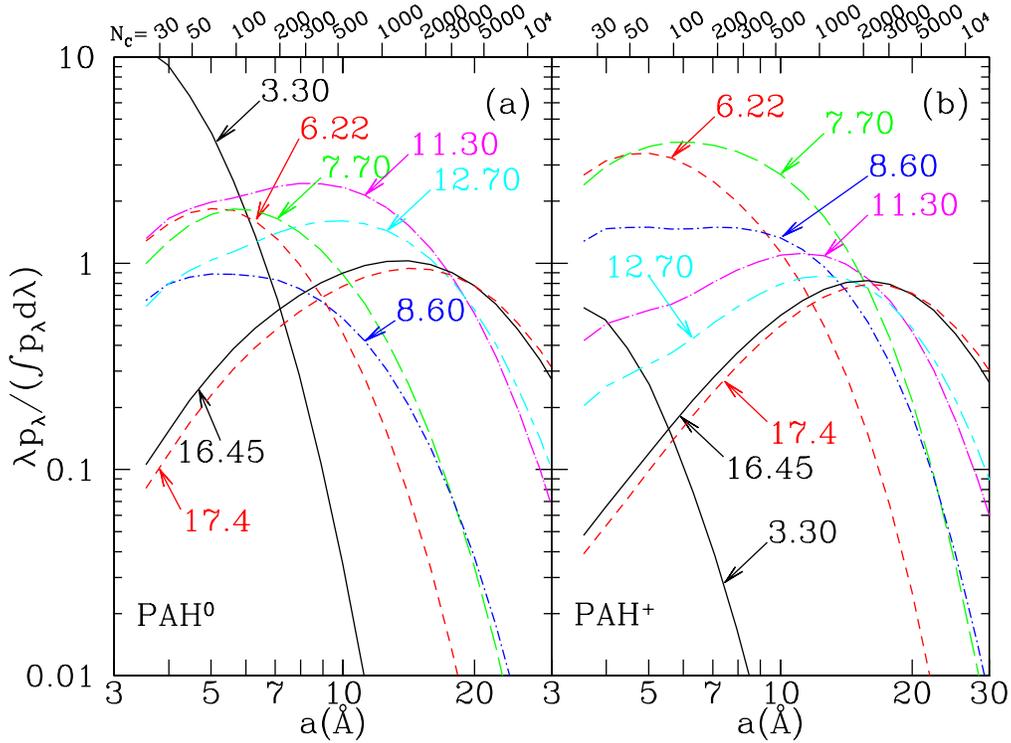}
\caption{\label{fig:band efficiencies}
         \footnotesize
         Efficiency for radiating in different emission bands,
         as function of size, for ionized and neutral PAHs (see text).
         }
\end{center}
\end{figure}

The dependence of the feature emission on grain size is shown in
Figure \ref{fig:band efficiencies}.
From this plot it is apparent that emission in the 3.30$\mum$ feature
will be almost entirely due to $a\ltsim 6\Angstrom$, or $N_{\rm C}\ltsim
10^2$, whereas grains with radii as large as 
$12\Angstrom$, or $N_{\rm C}\approx10^3$,
are efficient at converting absorbed starlight into 7.7$\mum$ emission.
The 11.3$\mum$ feature can be produced by particles as large as
$a\approx20\Angstrom$ or $N_{\rm C}\approx 4000$,
and the 16.45$\mum$ feature and 17.4$\mum$ complex are efficiently
produced by PAH particles in the 8--25$\Angstrom$ size range
($300\ltsim N_{\rm C}\ltsim 10^4$).

\begin{figure}[h]
\begin{center}
\includegraphics[width=10cm,angle=270]{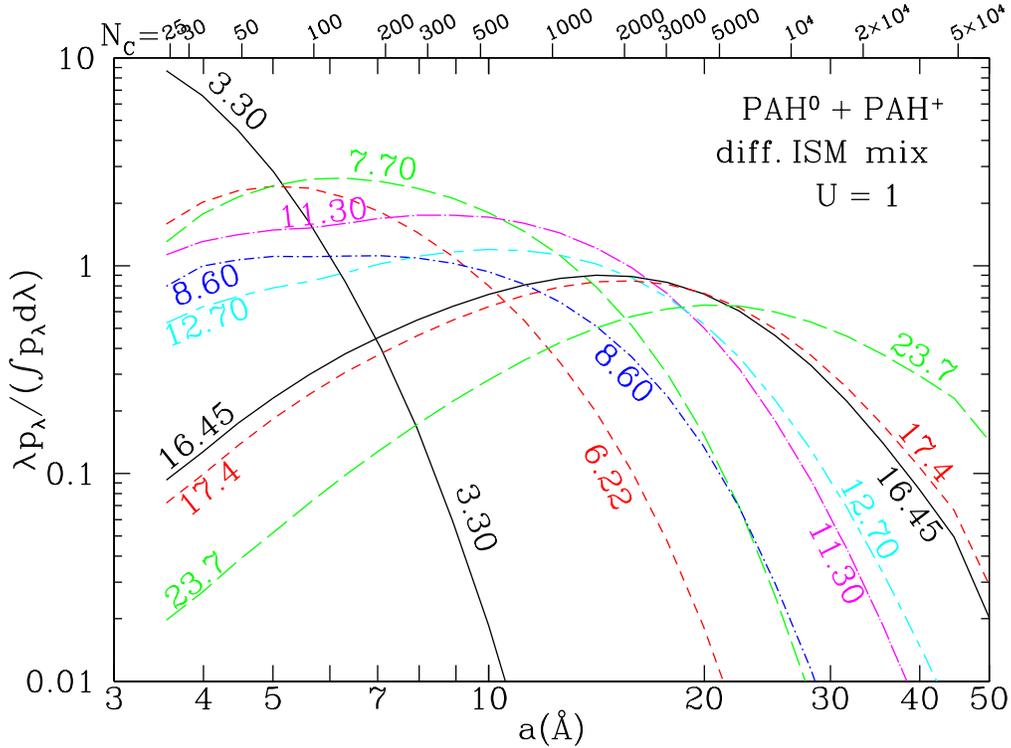}
\caption{\label{fig:band efficiencies, dism}
         \footnotesize
         $\lambda p_\lambda$ 
	 divided by the the total time-averaged power $p$ 
	 radiated by the grain,
         as a function of grain size $a$, for selected values of
	 wavelength $\lambda$; curves are labelled by
	 the value of $\lambda(\micron)$.
	 We assume a mix of neutral and ionized PAHs, with
	 ionization fraction as in Figure \ref{fig:xion for dism}.
	 24$\micron$ is the nominal wavelength of the MIPS band;
	 the other wavelengths are at peaks of emission 
	 features.
	 }
\end{center}
\end{figure}
\begin{figure}[h]
\begin{center}
\includegraphics[width=6cm,angle=0]{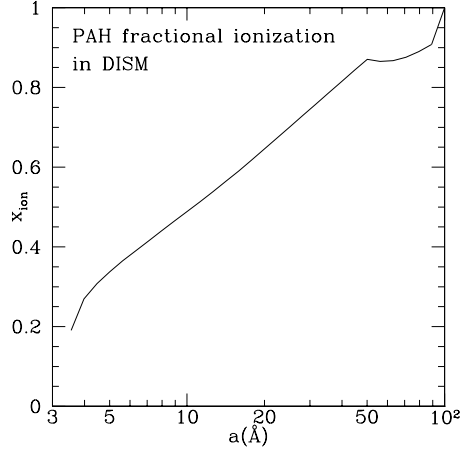}
\caption{\label{fig:xion for dism}
         \footnotesize
         Adopted fractional ionization in diffuse ISM
	 \citep{Li+Draine_2001b}.
	 }
\end{center}
\end{figure}
\begin{figure}[h]
\begin{center}
\includegraphics[width=12cm,angle=270]{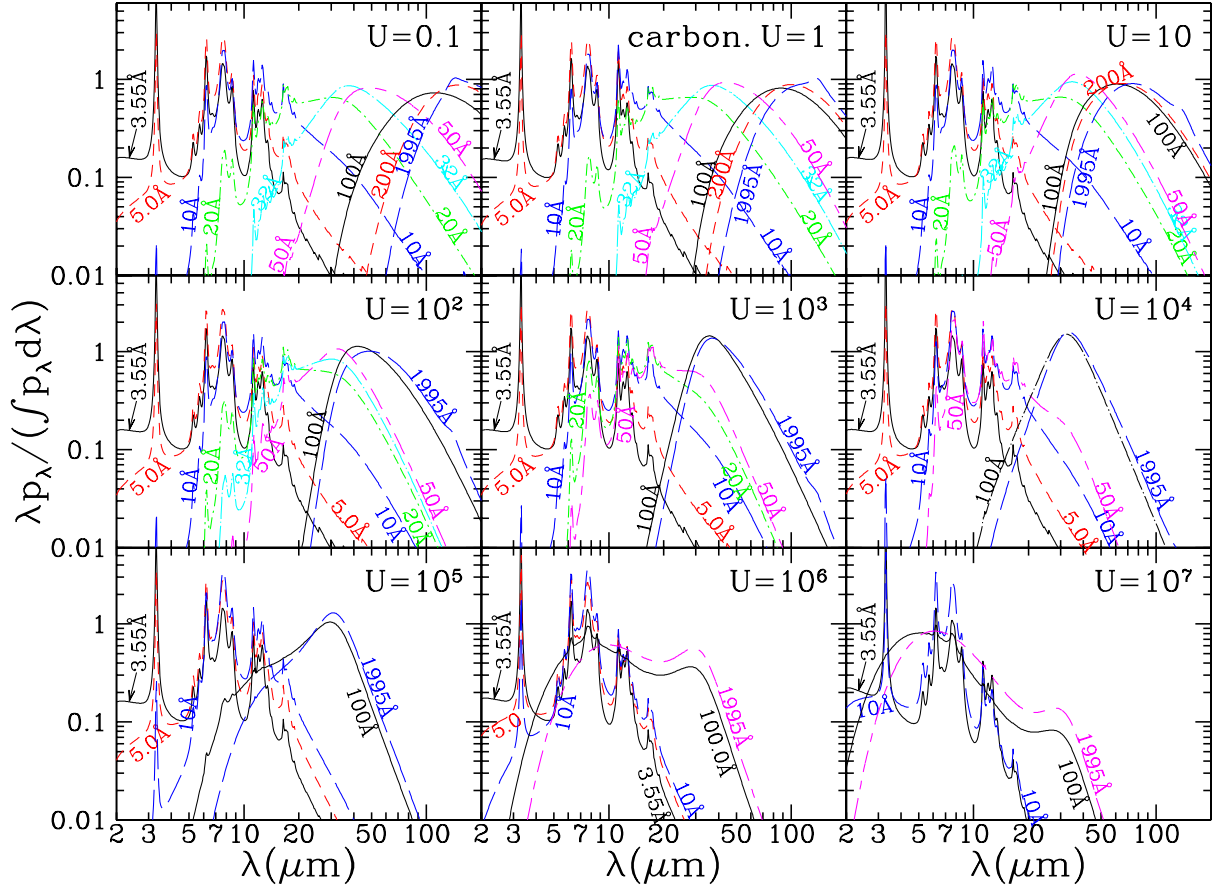}
\caption{\label{fig:carbon grain spectra for various U}
         \footnotesize
         Emission for selected sizes of carbonaceous grains 
	 for $U=0.1$ -- $10^7$.
	 For $a<100\Angstrom$ PAHs are assumed to have 
	 $x_{\rm ion}$ from Fig.\ \ref{fig:xion for dism} (see text).
	 }
\end{center}
\end{figure}
\begin{figure}[h]
\begin{center}
\includegraphics[width=12cm,angle=270]{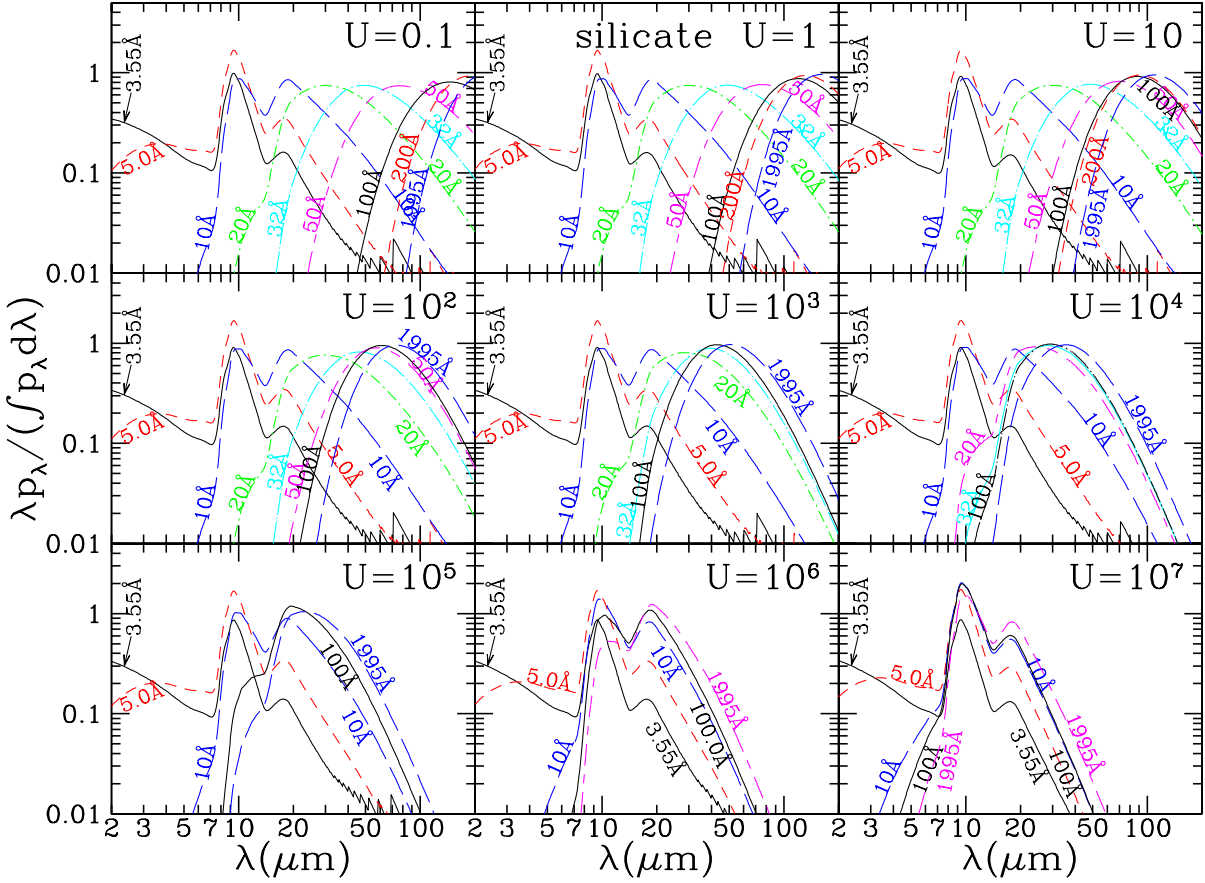}
\caption{\label{fig:silicate grain spectra for various U}
         \footnotesize
         Emission for amorphous silicate grains of various sizes
	 for $U=0.1$ -- $10^7$ (see text).
	 }
\end{center}
\end{figure}
\begin{figure}[h]
\begin{center}
\includegraphics[width=8cm,angle=270]{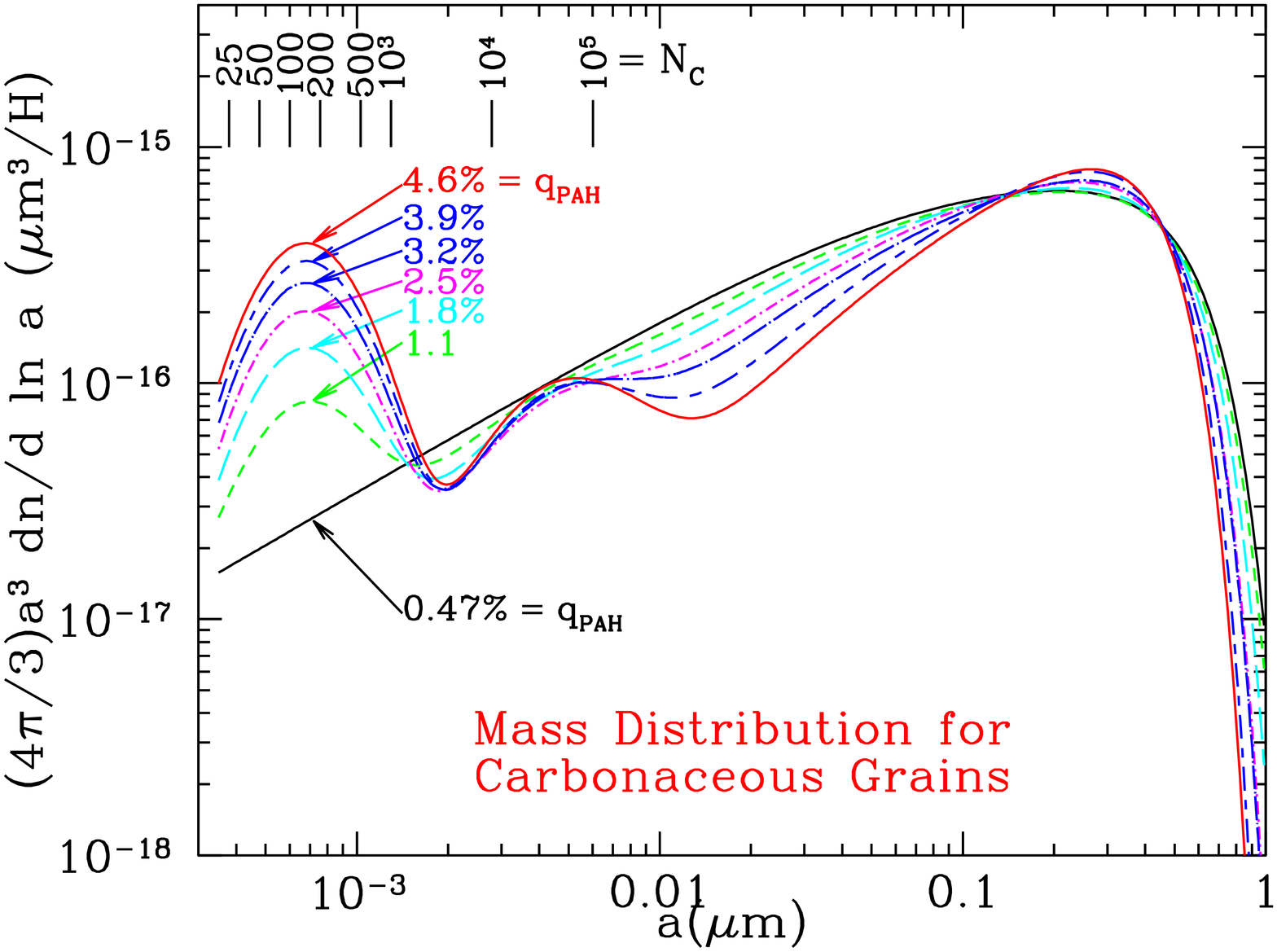}
\caption{\label{fig:dnda}
         \footnotesize
         Size distributions $j_M$=1--7 for carbonaceous grains.
	 Mass distributions of silicate grains are given by
	 Figure 2 of \citet{Weingartner+Draine_2001a}, multiplied by
	 0.92 (see text).
	 }
\end{center}
\end{figure}

The PAH emission spectra shown in Figure 
\ref{fig:PAHi and PAHn emission spectra} 
and the band emission efficiencies shown in Figure \ref{fig:band efficiencies}
are independent of the starlight
intensity $U$ in the single-photon-heating regime, $U\ltsim 10^4$ for
$\lambda \ltsim 30\micron$.
For larger grains, and at higher starlight intensities, the emission spectra
do depend on $U$.
If $\Cabs \propto \lambda^{-2}$ for $\lambda \gtsim30\micron$,
then the power radiated by the grain $\propto T^6$, the
steady-state temperature $T_{\rm ss}\propto U^{1/6}$, and the
infrared emission will peak at a wavelength 
$\lambda_p \propto 1/T_{\rm ss} \propto U^{-1/6}$.

The PAHs in the interstellar medium will consist of a mixture of
neutral and ionized particles.  For a given size PAH, 
the ionization balance will depend on
the gas temperature, the electron density, and the ultraviolet
radiation field \citep{Weingartner+Draine_2001c}.  Here we
adopt the ionization balance estimated by \citet{Li+Draine_2001b}
for the diffuse ISM in the Milky Way -- 
a weighted sum of the ionization fractions calculated for PAHs in the
CNM (cold neutral medium),
WNM (warm neutral medium) and WIM (warm ionized medium).
For this weighted sum of neutral and ionized PAHs,
$\lambda p_\lambda/p$ at selected wavelengths
is shown as a function of grain size in
Figure \ref{fig:band efficiencies, dism}.
Note that most of the selected wavelengths coincide with the positions
of PAH emission features, thus indicating how PAH size affects the
efficiency of converting starlight energy into IR emission features.
The adopted ionization fraction as a function of grain size is shown in
Figure \ref{fig:xion for dism}.

Figure \ref{fig:band efficiencies, dism} also shows $\lambda p_\lambda/p$
evaluated at the $23.7\mum$ wavelength of the MIPS~24\um\ band.
We see that PAHs in the 15--40\AA\ size range 
($2000\ltsim N_{\rm C} \ltsim 3\times 10^4$) 
are relatively efficient at converting starlight energy into
24\um\ continuum following single-photon heating
(Figure \ref{fig:dPdlnT for PAH+} shows that a single $h\nu<13.6\eV$
photon can heat a $20\Angstrom$ PAH to $\sim170\K$).
The term ``very small grains'' is sometimes used to describe small grains
that contribute continuum emission into the IRAS~25\um\ band or the
MIPS~24\um\ band -- here
we see that such grains must have effective radii in the 15--40\AA\
range to efficiently convert absorbed stellar energy into $\sim$24\um\
continuum emission.

Figures \ref{fig:carbon grain spectra for various U} and
\ref{fig:silicate grain spectra for various U} 
show 2--200$\mum$ emission spectra calculated for 
carbonaceous grains and amorphous silicate grains with
sizes extending from $a=3.55\Angstrom$ to $5000\Angstrom$.
As expected, the emission from $a\gtsim 60\Angstrom$ 
grains peaks at $\lambda\approx 100 U^{-1/6}\micron$.
However, for $U=10^6$ and $10^7$, $\lambda p_\lambda$
for $a\gtsim60\Angstrom$ carbonaceous grains has two peaks: one near the
peak in $\lambda B_\lambda$ near $100 U^{-1/6}\micron$, 
and the second near 30$\mum$,
the latter resulting from the peak in $\Cabs(\lambda)$ near 30$\mum$ 
(see Figure \ref{fig:Cabs for carbonaceous grains}b).
As discussed in \S\ref{sec:PAH Cross Sections} above,
the 30\um\ opacity peak in the present model produced by graphite in 
the present model may not apply to
the carbonaceous material in interstellar dust.

\begin{figure}[h]
\begin{center}
\includegraphics[width=8cm,angle=270]{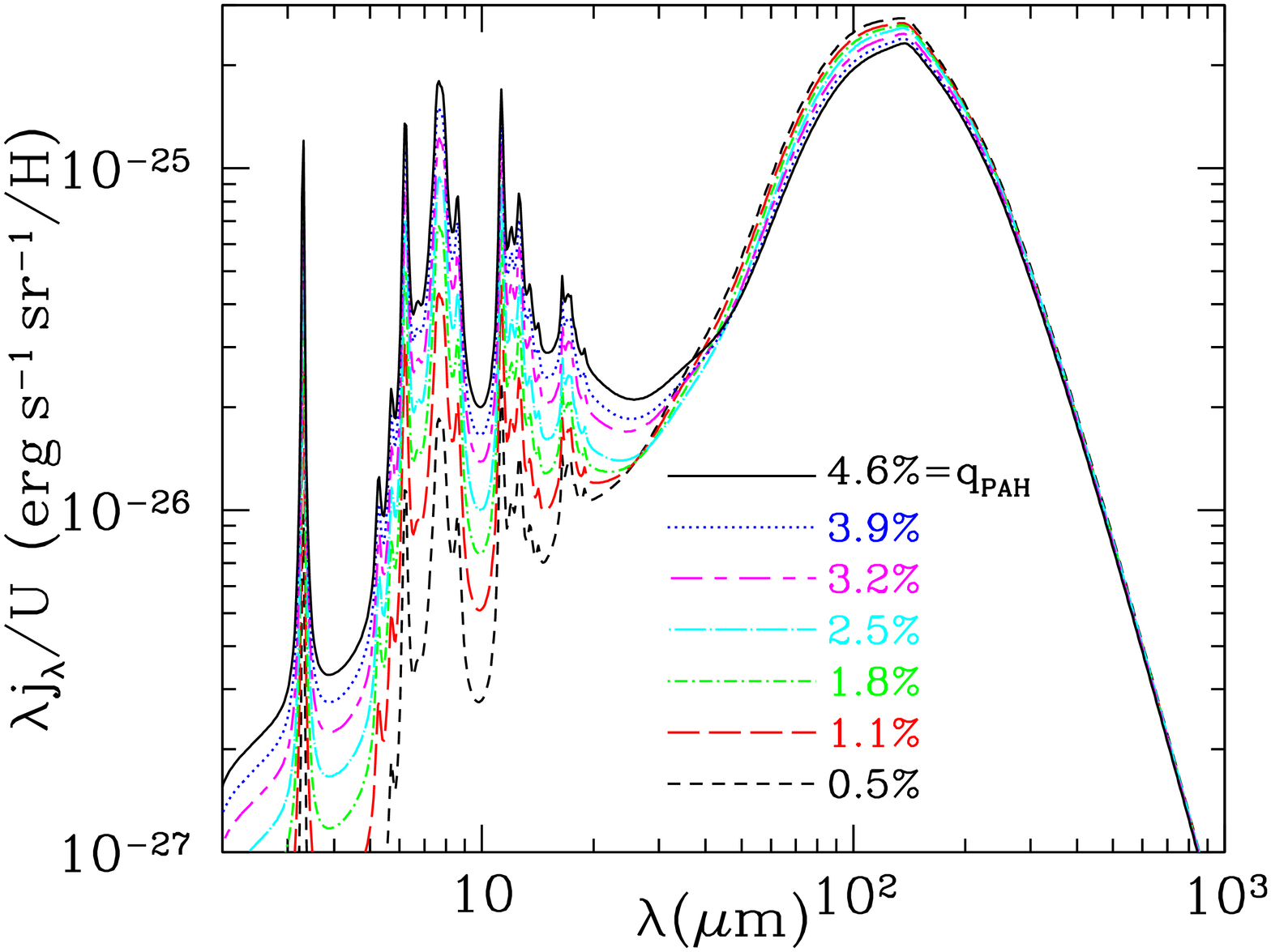}
\caption{\label{fig:variousqpah_spec}
         \footnotesize
	 Emission spectra for Milky Way dust models with
	 various $q_{\rm PAH}$ (dust models $j_M$=1--7), heated
	 by starlight with $U=1$.
	 Tabulated spectra for this and other cases will be available
	 at http://www.astro.princeton.edu/$\sim$draine/dust/irem.html.
	 }
\end{center}
\end{figure}
\begin{figure}[h]
\begin{center}
\includegraphics[width=10cm,angle=0]{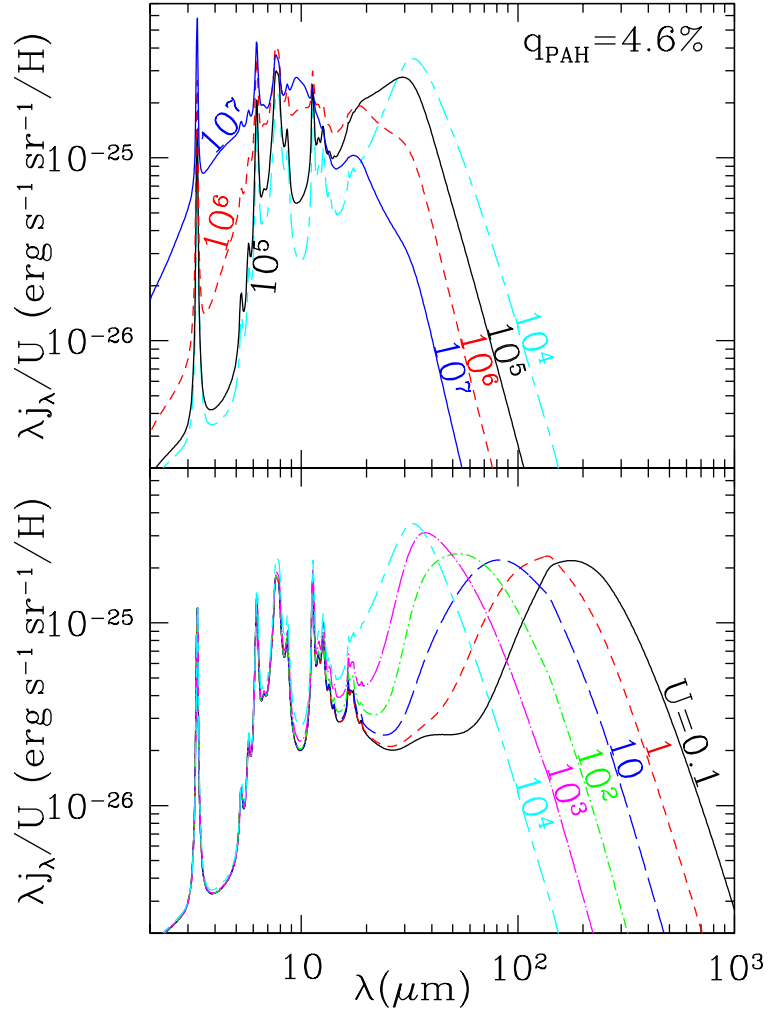}
\caption{\label{fig:variousU_spec}
         \footnotesize
         Emission spectra for size distribution $j_M=7$ for
         selected starlight intensity scale factors $U$.
         }
\end{center}
\end{figure}

\section{Dust Mixtures
         \label{sec:Dust Mixtures}}

\begin{table}[h]
\footnotesize
\begin{center}
\caption{\footnotesize
         \label{tab:lognormal parameters}
         PAH Size Distribution Parameters}
\begin{tabular}{l l c}
parameter & LD01 & present paper \\
\hline
$b_1/(b_1+b_2)$     & 0.75 & 0.75 \\
$b_2/(b_1+b_2)$     & 0.25 & 0.25 \\
$a_{01}(\Angstrom)$ & 3.5  & 4.0 \\
$a_{02}(\Angstrom)$ & 30.  & 20. \\
$\sigma_1$          & 0.4  & 0.4 \\
$\sigma_2$          & 0.4  & 0.55 \\
\hline
\end{tabular}
\end{center}
\end{table}
\begin{table}[h]
\begin{center}
{\footnotesize
\caption{\label{tab:dust models}
         \footnotesize
         {Physical Dust Models}}
\begin{tabular}{l l cccccccc }
 & & &\multicolumn{5}{c}{C/H (ppm) in PAHs with} \\
$j_M$ & Model & $q_{\rm PAH}$(\%) & ($M_{\rm dust}/M_{\rm H}$)$^a$ & $N_{\rm C}<100$ & $N_{\rm C}<200$ &
                                   $N_{\rm C}<500$ & $N_{\rm C}<10^3$ & 
                                   $N_{\rm C} < 10^4$ & $N_{\rm C}<10^5$ \\
\hline
1 & MW3.1\_00 & 0.47 & 0.0100  &  1.2 &  1.8 &  2.9 &  3.9 &  8.7 & 17.0 \\
2 & MW3.1\_10 & 1.12 & 0.0100  &  3.5 &  5.6 &  8.0 &  9.4 & 13.6 & 21.5 \\
3 & MW3.1\_20 & 1.77 & 0.0101  &  5.8 &  9.4 & 13.2 & 14.9 & 18.6 & 26.0 \\
4 & MW3.1\_30 & 2.50 & 0.0102  &  8.3 & 13.4 & 18.6 & 20.8 & 24.2 & 31.1 \\
5 & MW3.1\_40 & 3.19 & 0.0102  & 10.9 & 17.6 & 24.4 & 27.1 & 30.7 & 37.9 \\
6 & MW3.1\_50 & 3.90 & 0.0103  & 13.5 & 21.8 & 30.2 & 33.4 & 37.1 & 44.4 \\
7 & MW3.1\_60 & 4.58 & 0.0104  & 16.1 & 26.1 & 36.1 & 39.8 & 43.9 & 51.6 \\
8 & LMC2\_00  & 0.75 & 0.00343 &  0.9 &  1.2 &  1.8 &  2.2 &  3.5 &  5.0 \\
9 & LMC2\_05  & 1.49 & 0.00344 &  1.6 &  2.6 &  3.7 &  4.3 &  5.2 &  6.7 \\
10 & LMC2\_10 & 2.37 & 0.00359 &  2.6 &  4.4 &  6.3 &  7.1 &  7.9 &  9.6 \\
11 & SMC      & 0.10 & 0.00206 &  0.1 &  0.1 &  0.2 &  0.3 &  0.4 &  0.5 \\
\hline
\multicolumn{9}{l}{$^a$ $M_{\rm dust}/M_{\rm gas}=(1/1.36)(M_{\rm dust}/M_{\rm H})$}\\
\end{tabular}
}
\end{center}
\end{table}

Interstellar dust in the Milky Way and other galaxies includes
a wide range of grain sizes.  
Here we consider the size distributions
put forward by WD01 to reproduce the wavelength-dependent 
extinction on Milky Way
sightlines with extinction curves characterized by
$R_V\equiv A_V/(A_B-A_V) = 3.1$.
WD01 included a population of small ($a\ltsim 50\Angstrom$) 
carbonaceous particles with specified total mass.
\citet{Li+Draine_2001b} 
showed that if these small carbonaceous particles 
had the physical properties of PAHs and 
were distributed in two log-normal components,
\beq \label{eq:dnda}
\frac{dn}{da} = \sum_{j=1}^2\frac{n_{0j}}{a} 
                \exp\left[-\frac{(\ln(a/a_{0j}))^2}{2\sigma_j^2}
                    \right]
                + {\rm non\!-\!log\!-\!normal~contribution}~~~,
\eeq
the resulting infrared emission was approximately 
consistent with the diffuse emission observed by IRTS 
\citep{Onaka+Yamamura+Tanabe_etal_1996,
       Tanaka+Matsumoto+Murakami_etal_1996}.
The non-log-normal contribution to eq.\ (\ref{eq:dnda}) is
given by eq.\ (4) of WD01.
with parameter values taken from that paper
except for a reduction in the numbers of grains per H by a factor 0.92,
as recommended by \citet{Draine_2003a}.
Note that the non-log-normal term also extends continuously down to the smallest
sizes.

We continue to use eq.\ (\ref{eq:dnda}) for $dn/da$, 
but we have modified the values of the parameters 
$a_{0j}$ and $\sigma_j$, as given in 
Table \ref{tab:lognormal parameters}.
The factors $n_{0j}$ in eq.\ (\ref{eq:dnda}) are related to the numbers
$b_j$ of carbon atoms per total H in each of the log-normal components:
\beq
n_{0j} = \frac{3}{(2\pi)^{3/2}}
\frac{\exp(4.5\sigma_j^2)}{1+{\rm erf}(x_j)}
\frac{m_{\rm C}}{\rho_{\rm C} a_{Mj}^3\sigma_j}b_j
\eeq
\beq
x_j = \frac{\ln(a_{Mj}/a_{\rm min})}{\sigma_j\sqrt{2}}
\eeq
where $\rho_{\rm C}$ is the carbon mass density,
$m_{\rm C}$ is the mass of a carbon atom, and
\beq
a_{Mj}\equiv a_{0j} \exp\left(3\sigma_j^2\right)
\eeq
is the location of the peak in 
the mass distribution $\propto a^3 dn/d\ln a$.
Note that the non-log-normal term in eq.\ (\ref{eq:dnda}) also contributes
to the population of $N_{\rm C}<10^3$ PAHs.

These size distributions have been constructed for various amounts
of carbonaceous material in the very small PAH particles.
The size distributions
all reproduce the observed wavelength-dependent
extinction in the Milky Way, for sightlines with
$R_V\equiv A_V/E(B-V)\approx 3.1$. 
Because the abundance of ultrasmall grains is important for the infrared
emission as well as other applications, Table \ref{tab:dust models}
gives the amount of carbon present in grains containing
$<10^2$, $<200$, $<500$, $<10^3$, $<10^4$, and $<10^5$ C atoms, for
the 7 different grains models.
Models $j_M=1-7$ have $(b_1+b_2)=0.92\times 60\times 10^{-6}\times[(j_M-1)/6]$.
The mass distributions for carbonaceous grains are shown in
Fig.\ \ref{fig:dnda}.
[The silicate mass distribution is shown in Fig.\ 2 of
\citet{Weingartner+Draine_2001a}, except that the dust
abundances should be multiplied by 0.92.]

The emissivity per H nucleon for a dust mixture heated by starlight
intensity $U$ is
\beq
j_\nu(U) = \sum_j \int da \frac{dn_j}{da} 
\int \Cabs(j,a,\nu)B_\nu(T) \left(\frac{dP}{dT}\right)_{j,a,U} dT
~~~,
\eeq
\beq
B_\nu(T)\equiv \frac{2h\nu^3}{c^2} \frac{1}{\exp(h\nu/kT)-1} ~~~,
\eeq
where the sum is over compositions $j$, and the temperature
distribution function $dP/dT$ depends on composition $j$,
radius $a$, and starlight intensity $U$.

Taking $dP/dT$ calculated for the local starlight intensity
$U=1$,
summing over the grain size distribution, and over the
fractional ionization shown in Fig.\ \ref{fig:xion for dism},
we obtain the emission per H nucleon $j_\lambda$ shown in Figure 
\ref{fig:variousqpah_spec}.
The emission at $\lambda < 20\micron$ depends strongly on the
PAH abundance $\qpah$.

Figure \ref{fig:variousU_spec} shows the emission calculated for
the model with $\qpah=4.6\%$, but for different starlight intensities
$U$.
For $U\ltsim10^3$, the normalized 
emission at $\lambda < 20\micron$ is essentially
independent of $U$, because the emission is almost exclusively the result
of single-photon heating, with the PAH particles cooling off almost
completely between photon absorptions.
\newtext{This remarkable invariance of the PAH emission spectra
over orders of magnitude variations in starlight intensities
has been observed in a wide range of environments
[e.g. see \citet{Boulanger+Abergel+Cesarsky_etal_2000}, 
\citet{Kahanpaa+Mattila+Lehtinen_etal_2003},
and \citet{Sakon+Onaka+Isihara_etal_2004}].}
For $U\gtsim10^4$, however, the mean time between photon absorptions
becomes shorter than the radiative cooling time for a PAH with $\sim$1~eV
of internal energy, so that
the small grains do not cool completely between photon
absorptions.
Photon absorptions are then able to take them to
higher peak temperatures,
and the fraction of the power radiated at $\lambda < 20\micron$
increases with increasing $U$.

\section{\label{sec:fir and submm emission}
         Far-IR and Submm Emission}

The focus in the present dust model has been on the PAH features, but
the model is intended to reproduce the thermal dust emission at
far-infrared (FIR) and submm wavelengths as well.
The opacities of the amorphous silicate and graphite particles in the model are
calculated using the dielectric functions.  The graphite dielectric
function is based on laboratory measurements of graphite.
Because the nature of interstellar amorphous silicate material is
uncertain, the far infrared and submm behavior is not constrained by
laboratory data.  \citet{Draine+Lee_1984} made an estimate of the
imaginary part of the dielectric function of interstellar amorphous silicate.

\begin{figure}[h]
\begin{center}
\includegraphics[width=8cm,angle=0]{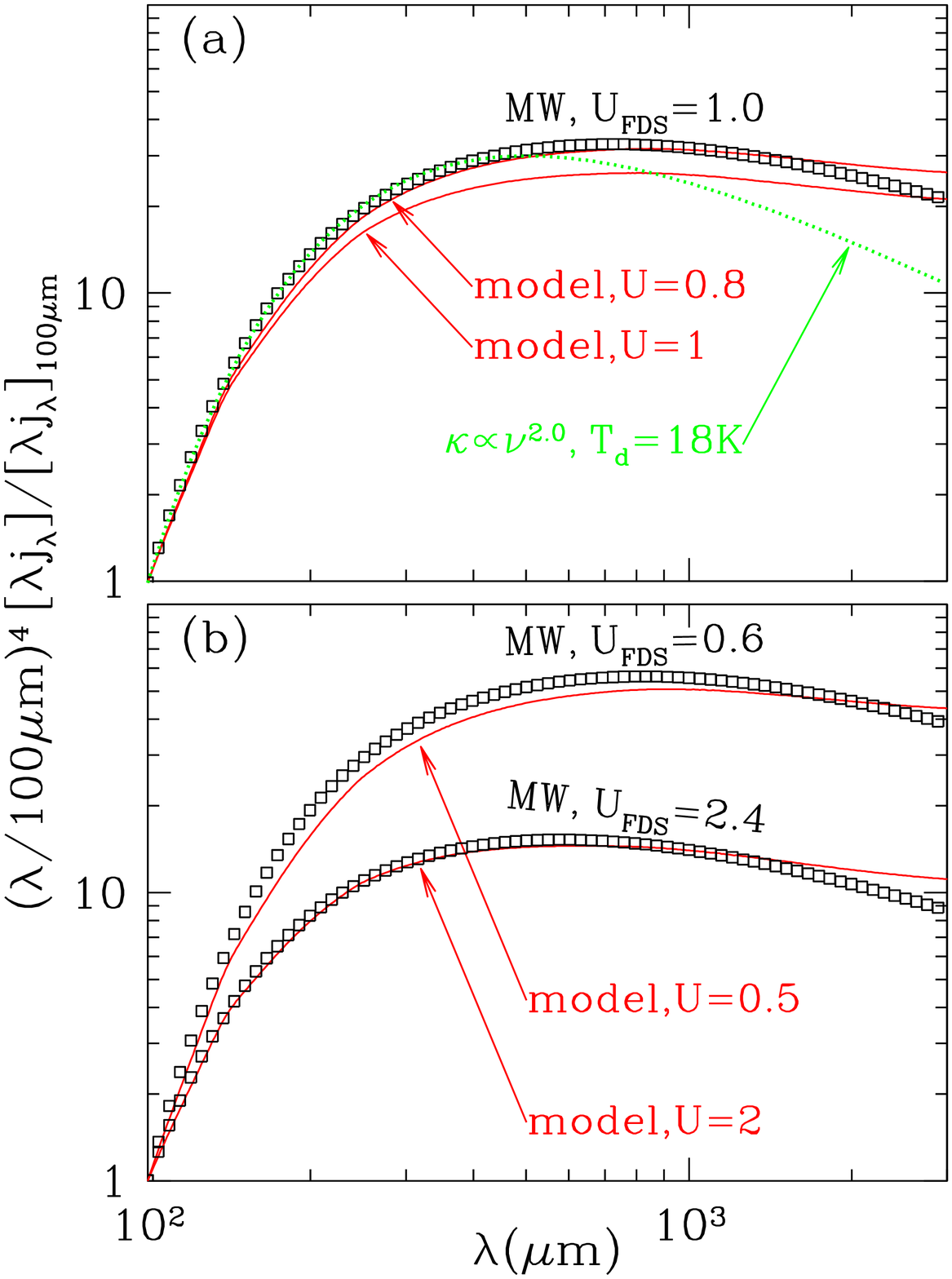}
\caption{\label{fig:firspec}\footnotesize
         (a) 
         Squares: observed spectrum, relative to the 100\um\ emission, of
         Milky Way dust from \citet{Finkbeiner+Davis+Schlegel_1999},
	 given by eq.\ (\ref{eq:firspec},\ref{eq:T2 vs U}) with $T_2=16.2$~K.
	 Solid curve: 100\um--3mm emission spectrum for
	 MW dust model with $\qpah=4.6\%$, for $U=0.8$ and 1.
	 The $U=0.8$ model approximately 
	 reproduces the observed emission for $T_2=16.2$~K, though having
	 excess emission at $\lambda > 1.5$~mm.
	 The good agreement between the $U=0.8$ model
	 and the observed spectrum with $\UFDS=1$ suggests a correspondence
	 $\UFDS\approx1.2U$.
	 Dotted curve: single-temperature dust model with opacity
	 $\kappa\propto \nu^2$ and $T=18\K$, for comparison.
	 A single modified blackbody produces insufficient emission for
	 $\lambda \gtsim 500\micron$.
	 (b)
	 Observed spectra for $\UFDS=0.6$ and 2.4 (squares)
	 and the present dust model for $U=0.5$ and 2 (solid curves).
	 }
\end{center}
\end{figure}
\citet[hereafter FDS]{Finkbeiner+Davis+Schlegel_1999} used
COBE-FIRAS observations of the sky at high galactic latitudes, after removal
of the cosmic background radiation and zodiacal emission, 
to characterize the emission from diffuse gas and dust in the Milky Way.
They excluded $|b|<7^\circ$, the Magellanic Clouds, 
HII regions in Orion and Ophiuchus, and an additional 16.3\% of the sky
where the data were of lower quality.  The final data set comprises
81\% of the $|b|>7^\circ$ sky.
For this region, FDS
find an empirical fit that quite accurately reproduces the 
observed $100\micron$--3mm spectra, 
using two parameters for each pixel: the 100\um\ intensity
$I_{\nu_0}$
and a temperature $T_2$ that determines the shape of the
$\lambda > 100\micron$ spectrum:
\beq \label{eq:firspec}
I_\nu = I_{\nu_0}
\frac{(\nu/\nu_0)^{2.70}B_\nu(T_2)+0.515(\nu/\nu_0)^{1.67}B_\nu(T_1)}
     {B_{\nu_0}(T_2)+0.515 B_{\nu_0}(T_1)}
~~~;~~~
T_1 = 9.4\K \left(\frac{T_2}{16.2\K}\right)^{1.182}
~~~,
\eeq
where $\nu_0=c/100\micron=3000$~GHz.
The mean value $\langle T_2\rangle=16.2\K$.
Variations in $T_2$ are presumably the result of variations in the
intensity of the starlight heating the grains; a simple model suggests
\beq \label{eq:T2 vs U}
T_2=16.2\K\,\UFDS^{1/6.70} ~~~,~~~ T_1=9.4\K\,\UFDS^{1/5.67}
~~~,
\eeq
where $\UFDS$ is the intensity
of the starlight heating the dust, relative to the average for the
region analyzed by FDS.

Based on the observed spectrum of eq.\ (\ref{eq:firspec}) with $T_2=16.2\K$,
\citet{Li+Draine_2001b} made small adjustments to the 
imaginary part of the dielectric
function for the amorphous silicate material at $\lambda>250\micron$
to improved agreement with the observed emission spectrum.
The real part of the dielectric function is obtained from the imaginary
part using the Kramers-Kronig relations \citep{Draine+Lee_1984}.
The resulting dielectric function is used to calculate absorption 
cross sections for amorphous silicate spheres in the present work.

The upper panel of Figure \ref{fig:firspec}a 
shows the observed far-infrared and
submm emission spectrum, as given by eq.\ (\ref{eq:firspec}) with
$T_2=16.2$~K (i.e., $\UFDS=1$).
Also shown are emission spectra calculated for the present model for
two values of the starlight intensity: $U=0.8$ and $U=1$.
Both spectra have shapes that are close to the observed spectrum, but the
$U=0.8$ spectrum agrees to within a few percent from 100\um\ to 1.5~mm.
Evidently $\UFDS\approx 1.2 U$ gives 
the correspondence between the starlight
intensity $U$ in the present model (normalized to the estimate of the
local interstellar radiation field by \citet{Mathis+Mezger+Panagia_1983})
and the starlight intensity $\UFDS$ relative to the ``average''
for the 71\% of the sky analyzed by FDS.

Fig.\ \ref{fig:firspec}b shows the emission spectra for
the model when the radiation field is lowered to $U=0.5$ and raised to
$U=2$; these spectra compare well with the FDS spectra
for $\UFDS=0.6$ and $\UFDS=2.4$, as expected from the correspondence
$\UFDS\approx 1.2U$ inferred from Fig.\ \ref{fig:firspec}a.

We conclude that the present dust model can successfully
reproduce the emission observed from dust in the diffuse interstellar medium
of the Milky Way (including dust in molecular clouds at $|b|>7\deg$ and away
from Orion and Ophiuchus)
out to wavelengths as long as 2mm without introduction of additional
emission components.

\section{\label{sec:irac band ratios}
         Spitzer IRAC and MIPS Band Ratios}

\begin{figure}[h]
\begin{center}
\includegraphics[width=14.0cm,angle=0]{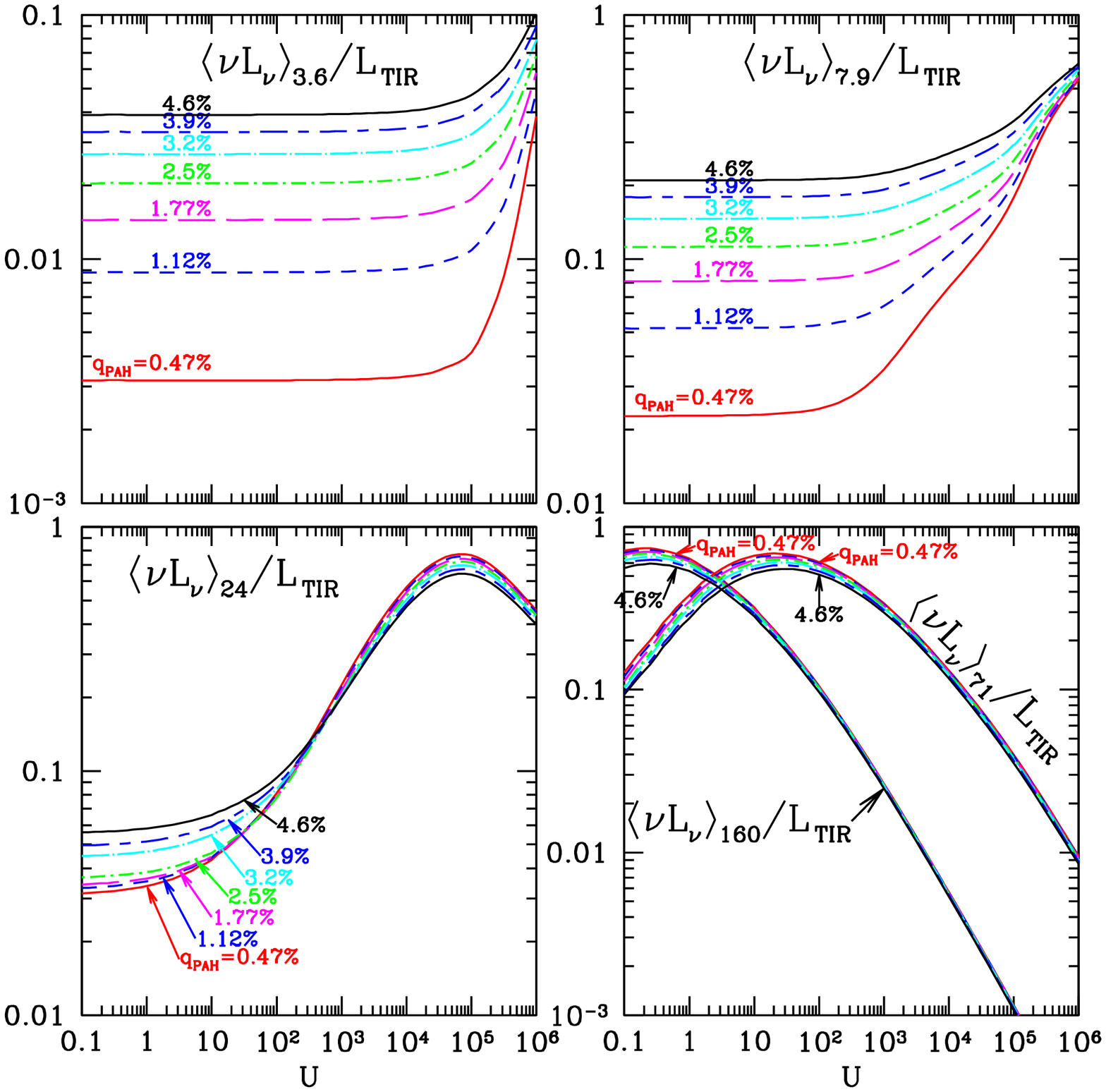}
\caption{\label{fig:spitzer band strengths vs U}
        \footnotesize
        IRAC~3.6$\mum$, IRAC~7.9$\mum$, MIPS~24$\mum$, MIPS~71$\mum$, and
	MIPS~160$\mum$ band strengths for dust illuminated by a single
	starlight intensity $U$, as a function of $U$,
	for dust models with 7 different values
	of $\qpah$.
	}
\end{center}
\end{figure}

\begin{figure}[h]
\begin{center}
\includegraphics[width=9.8cm,angle=0]{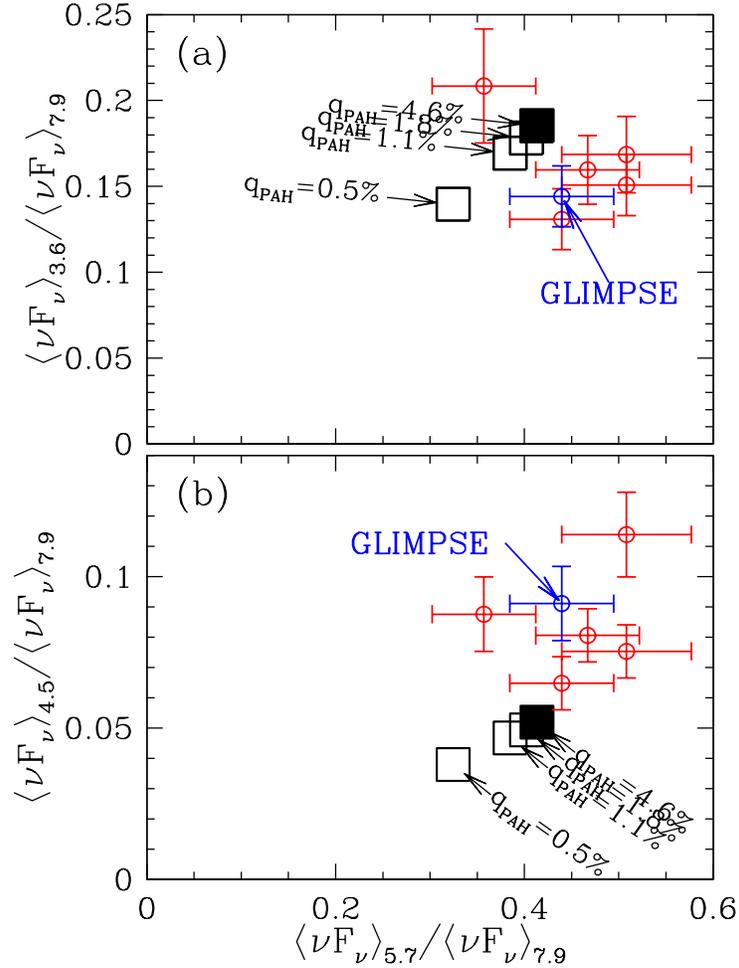}
\caption{\label{fig:irac band ratios}
         \footnotesize
         IRAC color-color plots for dust emission.
	 Symbols with error bars: diffuse emission extracted by
	 Flagey et al.\ (2006) for
	 5 GFLS fields and for one GLIMPSE field.
	 Squares: present models for Milky Way dust with different 
	 values of $\qpah=0.47$\%, 1.1\%, 1.8\%, and 4.6\%, 
	 for starlight intensities
	 $U \ltsim 10^3$ so that single-photon heating is dominant.
	 The dust model appears consistent with the observed 
	 diffuse emission in IRAC 3.6, 5.7, and 7.9\um\ bands,
	 but observed emission in the 4.5$\mum$ band 
         appears to be about 50\%
	 stronger than given by the model emissivities.
	 }
\end{center}
\end{figure}
\begin{figure}[h]
\begin{center}
\includegraphics[width=8.0cm,angle=0]{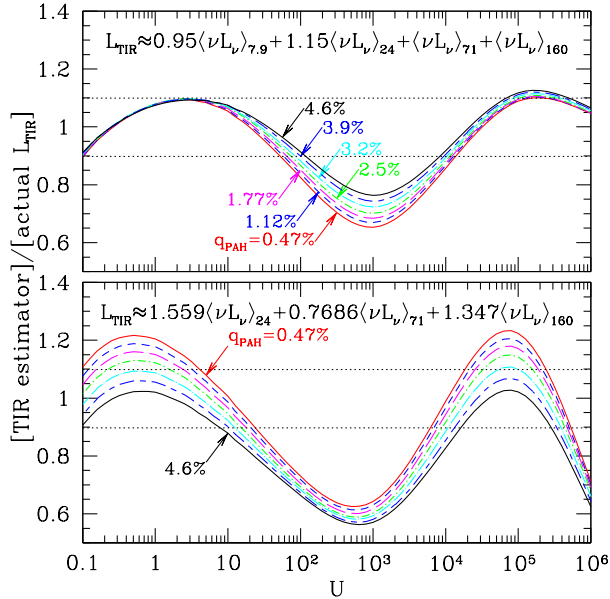}
\caption{\label{fig:TIR estimator}
         \footnotesize
         Upper panel: estimator (\ref{eq:TIR estimator})
	 for total infrared flux (TIR) divided by
	 actual TIR. The estimator (\ref{eq:TIR estimator}) is
	 within $\pm10\%$ of the actual
	 TIR (within the dotted lines) for $0.1 < U \ltsim 10^2$ and
	 $10^4 \ltsim U < 10^6$.
	 Lower panel: the TIR estimator from 
         Dale \& Helou (2002),
	 applied to the present dust models.
	 }
\end{center}
\end{figure}

For interpreting observations with the IRAC and MIPS cameras 
on Spitzer Space Telescope, the quantities of interest are 
the band-convolved emissivities and luminosities
\beq
\langle j_\nu\rangle_{\rm band} 
\equiv
\frac{\int R_{\rm band}(\nu) j_\nu d\nu}
     {\int (\nu/\nu_{\rm band})^\beta R_{\rm band}(\nu) d\nu}
~~~,~~~
\langle L_\nu\rangle_{\rm band} 
\equiv
\frac{\int R_{\rm band}(\nu) L_\nu d\nu}
     {\int (\nu/\nu_{\rm band})^\beta R_{\rm band}(\nu) d\nu}
\eeq
and
\beq
\langle \nu j_\nu\rangle_{\rm band} 
\equiv \nu_{\rm band}\langle j_\nu \rangle_{\rm band}
~~~,~~~
\langle \nu L_\nu\rangle_{\rm band} 
\equiv \nu_{\rm band}\langle L_\nu \rangle_{\rm band}
~~~,
\eeq
where $R_{\rm band}(\nu)$ is the relative response per unit power 
for the combination of optics, filter, and detector
and $\nu_{\rm band} \equiv c/\lambda_{\rm band}$, where $\lambda_{\rm band}$
the nominal wavelength of the band, is
given in Table \ref{tab:model emissivities}.

Table \ref{tab:model emissivities} gives
$\langle \nu j_\nu\rangle_{\rm band}$ for the four IRAC bands,\footnote{%
   $R_{\rm band}(\nu) \propto (1/h\nu)S_{\rm band}(\nu)$,
   where the relative 
   response per photon $S_{\rm band}(\nu)$ is obtained from
   {\tt http://ssc.spitzer.caltech.edu/irac/spectral\_response.html}.
   The IRAC calibration uses $\beta=-1$, 
   }
the 2 ``Peakup'' bands of the Infrared Spectrograph (IRS),\footnote{%
   \newtext{%
   $R_{\rm band}(\nu)\propto (1/h\nu)S_{\rm band}(\nu)$,
     where the relative quantum efficienty $S_{\rm band}(\nu)$ 
     is given in Fig. 6.1 of the Infrared Spectrograph Data Handbook
     Version 2.0.
   The IRS Peakup calibration uses $\beta=-1$.
   }}
the three MIPS bands,\footnote{%
   \newtext{%
   $R_{\rm band}(\nu)$ is obtained from
   {\tt http://ssc.spitzer.caltech.edu/mips/spectral\_response.html}.
   The MIPS calibration is for a $10^4\K$ blackbody, i.e., $\beta=2$.
   }}.
\newtext{
Table \ref{tab:model emissivities for Akari} gives
emissivities convolved with photometric bands of the AKARI satellite
\citep{Kawada+Shibai+Nakagawa_2004,Onaka+Fujishiro+Ihara_etal_2004},
and model emissivities convolved with the 
the three bands of the Herschel PACS instrument,\footnote{%
   \newtext{%
   A.\ Poglitsch has kindly provided provisional $R_{\rm band}(\nu)$ 
   for PACS.
   The calibration procedures for PACS have not yet
   been finalized.  Here we assume $\beta=2$.
   }}, 
and the three bands of the Herschel SPIRE instrument.\footnote{%
   \newtext{%
   M. Griffin has kindly provided provisional $R_{\rm band}(\nu)$ for
   SPIRE.
   The calibration procedures for SPIRE have not yet
   been finalized.  Here we assume $\beta=2$.
   }}
are given in Table \ref{tab:model emissivities for Herschel}.}
Emissivities $\langle \nu j_\nu\rangle_{\rm band}$ are given for
selected grain models with several different values of $\qpah$, and
for different values of the starlight intensity scale factor $U$.

Figure \ref{fig:spitzer band strengths vs U} shows
$\langle\nu L_\nu\rangle_{\rm band}$
relative to the total infrared (TIR) dust luminosity
\beq
L_{\rm TIR}\equiv \int_0^\infty L_\nu d\nu
\eeq
for
5 bands: IRAC~3.6$\mum$, IRAC~7.9$\mum$,
MIPS~24$\mum$, MIPS~71$\mum$, and MIPS~160$\mum$.
Results are given for grain models $j_M=1$--7, 
with the model results labelled by
the value of $\qpah$.
As expected, the 3.6$\mum$ emission and 7.9$\mum$ emission 
per unit total power is entirely
the result of single-photon heating at low values of $U$, 
and therefore 
$\langle\nu L_\nu\rangle_{7.9}/L_{\rm TIR}$ 
is independent of $U$ for $U\ltsim 10^3$,
and
$\langle\nu L_\nu\rangle_{3.6}/L_{\rm TIR}$ 
is independent of $U$ for $U\ltsim 10^5$.
$\langle\nu L_\nu\rangle_{3.6}/L_{\rm TIR}$ and 
$\langle\nu L_\nu\rangle_{7.9}/L_{\rm TIR}$ are 
also both approximately proportional to 
$\qpah$ in the low-$U$ limit, as expected.
For small values of $U$, 
the 24$\mum$ emission is also the result of 
single-photon heating, 
and $\langle\nu L_\nu\rangle_{24}/L_{\rm TIR}$ is 
independent of $U$ for $U\ltsim 10$.

\begin{deluxetable}{c@{}c@{}c@{}c@{}c@{}c@{}c@{}c@{}c@{}c@{}c}
\tabletypesize{\scriptsize}
\tablecaption{\label{tab:model emissivities}
              Model Emissivities
              $\langle \nu j_\nu\rangle_{\rm band} 
               ~~(\erg~\s^{-1}{\sr}^{-1}/{\rm H})^a$
	       \newtext{for Spitzer Space Telescope}
   }
\tablehead{
   \multicolumn{2}{c}{model}&
   \colhead{IRAC~1} &
   \colhead{IRAC~2} &
   \colhead{IRAC~3} &
   \colhead{IRAC~4} &
   \colhead{IRS~PU1}&
   \colhead{IRS~PU2}&
   \colhead{MIPS~1}&
   \colhead{MIPS~2}&
   \colhead{MIPS~3}\\
   \colhead{U} & \colhead{$\qpah$~(\%)} &
   \colhead{3.550\um} & 
   \colhead{4.493\um} & 
   \colhead{5.731\um} & 
   \colhead{7.872\um} & 
   \colhead{16\um} & 
   \colhead{22.\um} & 
   \colhead{23.68\um} & 
   \colhead{71.42\um} & 
   \colhead{155.9\um}
   }
\startdata
\input table4.dat
\enddata

$^a$~Additional models, with spectra, are available online
at http://www.astro.princeton.edu/$\sim$draine/dust/irem.html
\end{deluxetable}

\newtext{
\begin{deluxetable}{c@{~}c@{~}c@{~~}c@{~~} c@{~~} c@{~~} c@{~~} c@{~~} c@{~~} c@{~~} c@{~~} c@{~~} c@{~~} c}
\tabletypesize{\scriptsize}
\tablewidth{8.0in}
\rotate
\tablecaption{\label{tab:model emissivities for Akari}
              \newtext{Model Emissivities
              $\langle \nu j_\nu\rangle_{\rm band} 
               ~~(\erg~\s^{-1}{\sr}^{-1}/{\rm H})^a$
	       for Akari}
   }
\tablehead{
   \multicolumn{2}{c}{model}&
   \colhead{N3} &
   \colhead{N4} &
   \colhead{S7} &
   \colhead{S9W} &
   \colhead{S11}&
   \colhead{L15}&
   \colhead{L18W}&
   \colhead{L24}&
   \colhead{N60}&
   \colhead{WIDE-S}&
   \colhead{WIDE-L}&
   \colhead{N160}
\\
   \colhead{U} & 
   \colhead{$\qpah$~(\%)} &
   \colhead{3.2\um} & 
   \colhead{4.1\um} & 
   \colhead{7.0\um} & 
   \colhead{9.0\um} & 
   \colhead{11.0\um} & 
   \colhead{15.\um} &
   \colhead{18.\um} &
   \colhead{24.\um} &
   \colhead{65.\um} &
   \colhead{80.\um} &
   \colhead{140.\um} &
   \colhead{160.\um}
   }
\startdata
\input table5.dat
\enddata

$^a$~Additional models, with spectra, are available online
at http://www.astro.princeton.edu/$\sim$draine/dust/irem.html
\end{deluxetable}
\begin{deluxetable}{c c c c c c c c}
\tabletypesize{\footnotesize}
\tablewidth{5.5in}
\tablecaption{\label{tab:model emissivities for Herschel}
              \newtext{Model Emissivities 
              $\langle \nu j_\nu\rangle_{\rm band} 
               ~~(\erg~\s^{-1}{\sr}^{-1}/{\rm H})^a$
	      for Herschel}
   }
\tablehead{
   \multicolumn{2}{c}{model}&
   \colhead{PACS~1} &
   \colhead{PACS~2} &
   \colhead{PACS~3} &
   \colhead{SPIRE~1} &
   \colhead{SPIRE~2}&
   \colhead{SPIRE~3}\\
   \colhead{U} & \colhead{$\qpah$~(\%)} &
   \colhead{75\um} & 
   \colhead{110\um} & 
   \colhead{170\um} & 
   \colhead{250\um} & 
   \colhead{360\um} & 
   \colhead{520\um} 
   }
\startdata
\input table6.dat
\enddata
\\
$^a$~Additional models, with spectra, are available online
at http://www.astro.princeton.edu/$\sim$draine/dust/irem.html
\end{deluxetable}
}

\citet{Flagey+Boulanger+Verstraete_etal_2006} 
extracted the diffuse
ISM emission from IRAC imaging of various regions of the Milky Way 
obtained by the
Galactic First Look Survey (GFLS) and the 
Galactic Legacy Infrared MidPlane Survey Extraordinaire (GLIMPSE).
The model emission spectra, 
convolved with the response function for the IRAC bands,
can be compared to the observed colors of the diffuse emission.

Figure \ref{fig:irac band ratios}a shows 
the colors of the observed emission in the IRAC 
3.6$\mum$, 5.7$\mum$, and 7.9$\mum$ bands.  
The filled square show the color calculated for the present
dust model with the value of $\qpah\approx4.6\%$ 
that appears to be applicable to
the dust in the Milky Way and other
spiral galaxies with near-solar metallicity 
\citep{Draine+Dale+Bendo_etal_2006}.
The model with
$\qpah\approx 4.6\%$
appears to be in good agreement with the Milky Way observations
in Figure \ref{fig:irac band ratios}a.
First of all, 
$\langle\nu F_\nu\rangle_{5.7}/\langle\nu F_\nu\rangle_{7.9}$ 
is within the range
of observed values, indicating that the adopted $\Cabs(\lambda)$ has
approximately the correct shape over the 5.5--8.5$\mum$ wavelength range.
Second, the ratio of 3.6$\mum$ emission divided by 8.0$\mum$ emission
is close to the observed value.
The emission into the IRAC 3.6$\mum$ band is sensitive 
to the abundances of the very smallest PAHs,
with $N_{\rm C}\ltsim 75$ (see Figure \ref{fig:band efficiencies}), 
we can therefore conclude that
the size distribution adopted by the present model 
has approximately the correct
amount of PAH mass in the interval $25\ltsim N_{\rm C}\ltsim 75$, 
at least relative
to the PAH mass in the interval 
$25\ltsim N_{\rm C}\ltsim 10^3$ that radiates efficiently
into the IRAC 8$\mum$ band.
The ratio of the 3.6$\mum$ emission to 
the 8$\mum$ emission of course is sensitive to 
the assumed PAH charge state (see Fig.\ \ref{fig:band efficiencies}).
The present model assumes a mixture of 
ionization conditions that is intended to 
be representative of the local interstellar medium 
(see Figure \ref{fig:xion for dism}).
If the actual ionized fraction for the PAHs
is higher than in our model, 
then the actual mass fraction in the $25\ltsim N_{\rm C}\ltsim 75$
would have to be even higher than the adopted size distribution 
(shown in Figure \ref{fig:dnda}).
Conversely, if the PAH neutral fraction is actually higher 
than in our model, then the PAH mass fraction in 
the $25\ltsim N_{\rm C}\ltsim 75$ range could be reduced.

Observations of external galaxies by ISO 
and Spitzer Space Telescope
have been interpreted as
showing diffuse nonstellar emission 
in the 2.5--5$\mum$ wavelength range
\citep{Lu+Helou+Werner+Dinerstein+Dale_2003,
       Helou+Roussel+Appleton+Frayer+Stolovy_etal_2004},
consistent with thermal emission from dust 
with $Q_{\rm abs}\propto \nu^2$
and temperatures in the $\sim$\,750--1000$\K$ range.
The GFLS and GLIMPSE images also appear to contain diffuse emission in
the 4.5$\mum$ band (Flagey et al.\ 2006).
Figure \ref{fig:irac band ratios}b shows the observed ratio of
4.5\um\ emission relative to 7.9\um\ emission as well as the model
values.  
The observed 4.5$\mum$ emission is perhaps 50\% stronger (relative to
the 8.0$\mum$ emission) than the models.

In many cases it is desirable to try to 
estimate the total infrared luminosity from the
observed fluxes in the IRAC and MIPS bands.
The following weighted sum of IRAC~7.9\um\ and the 3 MIPS bands
is a reasonably accurate estimator for $L_{\rm TIR}$:
\beq \label{eq:TIR estimator}
L_{\rm TIR} \approx 
0.95\langle\nu L_\nu\rangle_{7.9} + 
1.15\langle\nu L_\nu\rangle_{24}+
\langle\nu L_\nu\rangle_{71}+\langle\nu L_\nu\rangle_{160} ~~~.
\eeq
Figure \ref{fig:TIR estimator} plots this estimator 
relative to the actual $L_{\rm TIR}$ for our dust models,
showing that eq.\ (\ref{eq:TIR estimator}) 
allows the total infrared luminosity to
be estimated to within $\sim10\%$ for our dust models 
heated by starlight with
$0.1 \ltsim U \ltsim 10^2$
or with $10^4 \ltsim U \ltsim 10^6$.
The accuracy becomes somewhat lower for 
$10^2 \ltsim U \ltsim 10^4$
(because of the factor of 3 gap in wavelength 
between the 71$\mum$ band and the 24$\mum$ band),
but the worst-case error is only $\sim 30\%$, occuring
for $U\approx10^3$ where the FIR emission peak falls
at $\sim40\micron$, halfway between the MIPS 24\um\ and 71\um\
filters, with the result that MIPS photometry underestimates the
actual power.
Also plotted in Figure \ref{fig:TIR estimator} is the TIR estimator
proposed by 
\citet{Dale+Helou_2002}
using only MIPS photometry,
applied to our models.  
The 
\citet{Dale+Helou_2002}
luminosity estimate
is accurate to within $\pm25\%$ for $U\ltsim10^2$.

\section{\label{sec:alpha-gamma dust models}
         Emission Spectra for Dust Models: 
         Distribution of Starlight Intensities}

The dust grains in a galaxy will be exposed to a wide range 
of starlight intensities.
The bulk of the dust in the diffuse interstellar medium 
will be heated by a general diffuse radiation field 
contributed by many stars.
However, some dust grains will happen to 
be located in regions close to luminous stars,
such as photodissociation regions near OB stars, where the starlight 
heating the dust will be much more intense than 
the diffuse starlight illuminating the bulk of the grains.

In principle, one could construct a model for 
the distribution of stars and dust in the galaxy, 
and solve the radiative transfer problem to
determine the heating rate for each dust grain.  
This, however, requires many uncertain assumptions, 
as well as heavy numerical calculations to
solve the radiative transfer problem
[see, e.g., 
\citet{Witt+Gordon_1996,
       Silva+Granato+Bressan+Danese_1998,
       Popescu+Misiriotis+Kylafis_etal_2000,
       Tuffs+Popescu+Volk_etal_2004,
       Piovan+Tantalo+Chiosi_2006}].
Here we take a much simpler approach, and assume a simple parametric form
for the fraction of the dust mass exposed to a distribution of
starlight intensities $U$ described by a delta function and a power-law
distribution for $\Umin < U < \Umax$:
\beq
\label{eq:dM_d/dU}
\frac{d \Mdust}{dU} = (1-\gamma) \Mdust \delta(U-\Umin)
                     + \gamma \Mdust 
                       \frac{(\alpha-1)}
		       {\left[\Umin^{1-\alpha}-\Umax^{1-\alpha}\right]}
                       U^{-\alpha}
~~~, ~~\alpha\neq 1,
\eeq
where $d\Mdust$ is the mass of dust heated by starlight intensities
in $[U,U+dU]$;
$\Mdust$ is the total mass of dust,
$(1-\gamma)$ is the fraction of the dust mass that is exposed to
starlight intensity $\Umin$, and $\alpha$ is a power-law index.
This functional form is similar to the power-law distribution used
by \citet{Dale+Helou+Contursi_etal_2001} and
\citet{Dale+Helou_2002}, except that we have added a delta function
component that contains most of the dust mass;
the delta-function term is intended to represent the dust in the
general diffuse interstellar medium.

\begin{figure}[h]
\begin{center}
\includegraphics[width=12cm,angle=270]{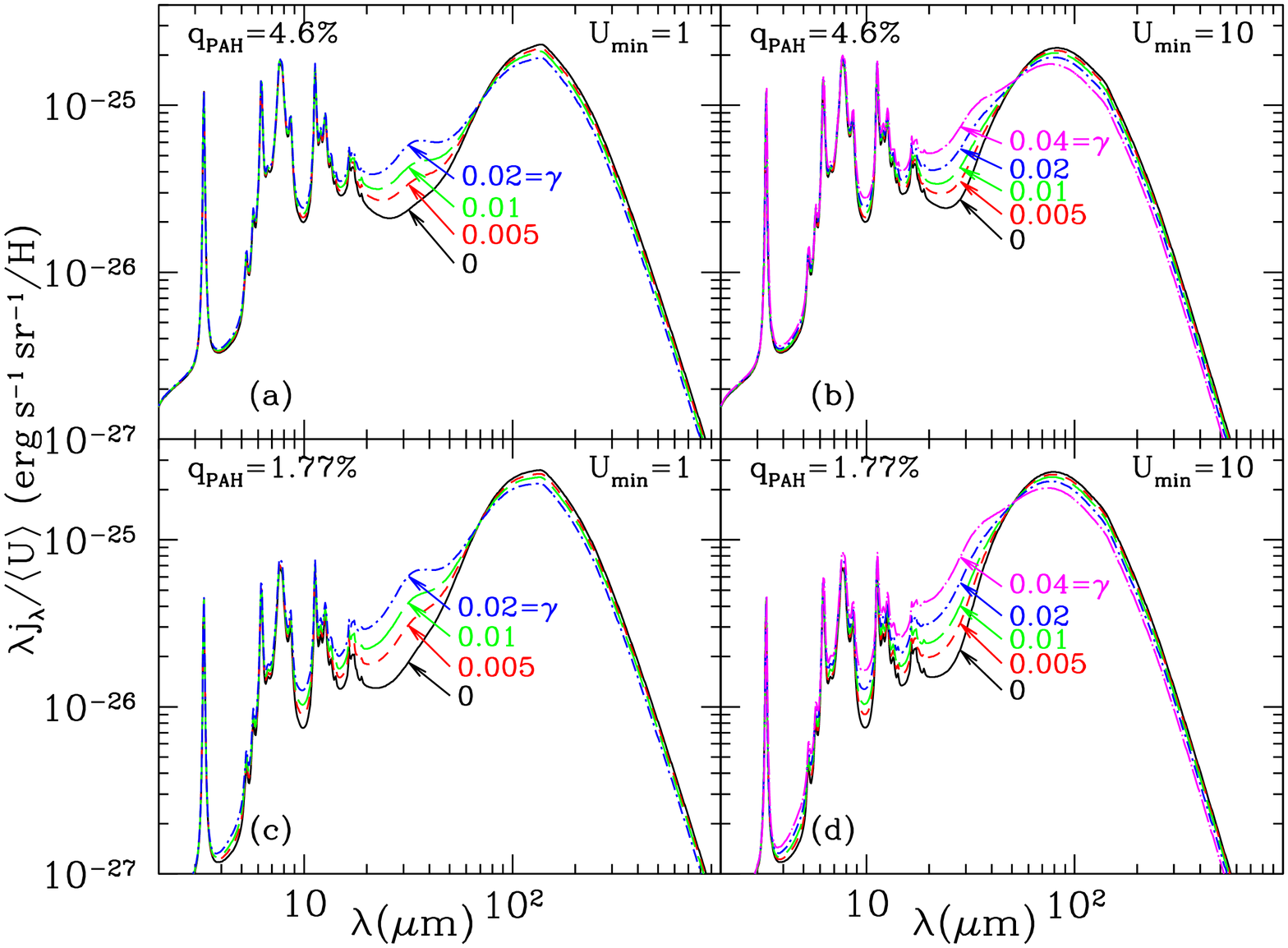}
\caption{\label{fig:spec for Udists}
         \footnotesize
         Emission spectra for models with MW dust with
	 (a) $\qpah=4.6\%$, $\Umin=1$,
	 (b) $\qpah=4.6\%$, $\Umin=10$,
	 (c) $\qpah=1.77\%$, $\Umin=1$,
	 (d) $\qpah=1.77\%$, $\Umin=10$.
	 All models with $\Umax=10^6$, $\alpha=2$.
	 Values of $\gamma$ are indicated.
	 }
\end{center}
\end{figure}
\begin{figure}[h]
\begin{center}
\includegraphics[width=14.0cm,angle=0]{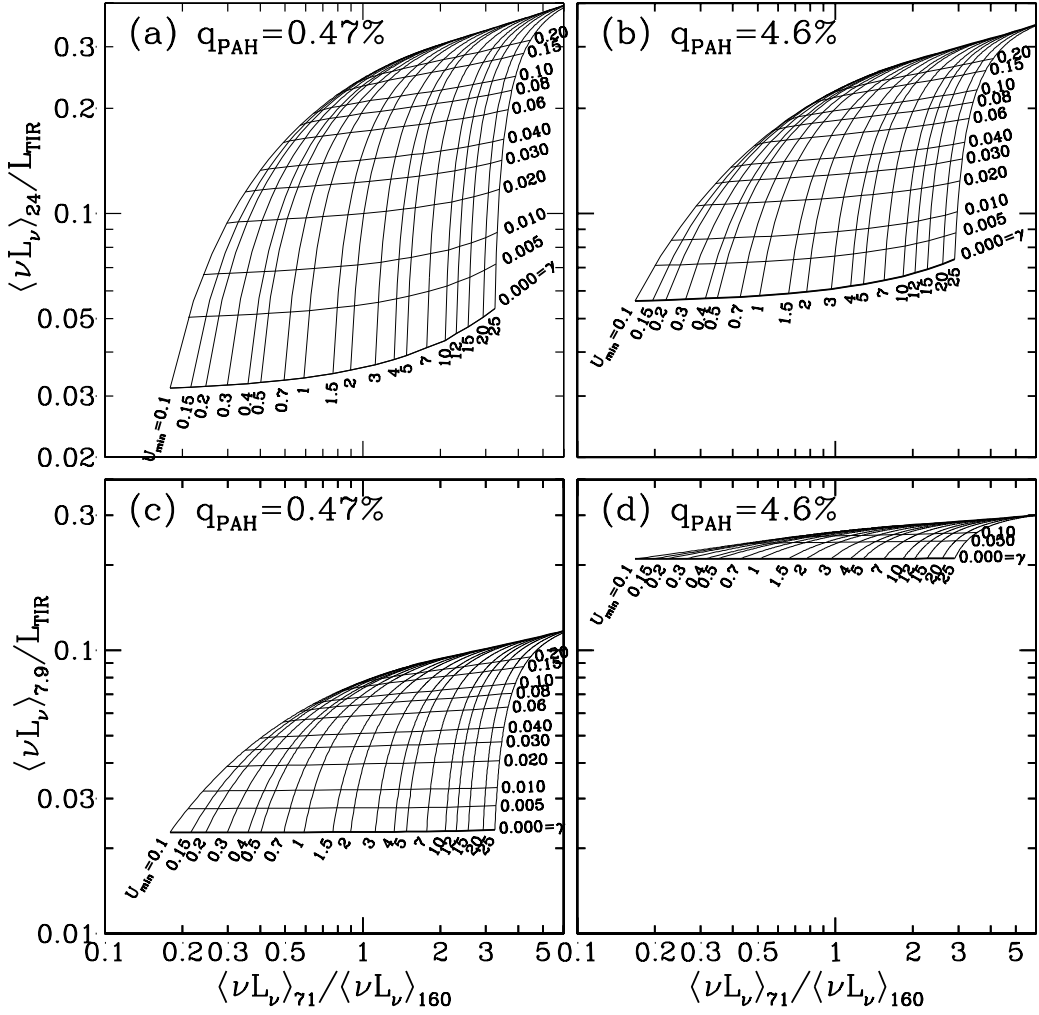}
\caption{\footnotesize
         \label{fig:band fluxes relative to TIR}
	 Ratios of $(\nu L_\nu)_{7.9}/L_{\rm TIR}$ and
	 $(\nu L_\nu)_{24}/L_{\rm TIR}$ for dust models
	 with $\qpah=0.47\%$ and 4.6\%, plotted against
	 the ratio $(\nu L_\nu)_{71}/(\nu L_\nu)_{160}$.
	 The model grids are labelled by the minimum
	 starlight intensity $\Umin$ and the
	 fraction $\gamma$ of the dust exposed to
	 starlight with $U>\Umin$ (see text).
	 }
\end{center}
\end{figure}

The starlight distribution function 
eq.\ (\ref{eq:dM_d/dU}) has parameters
$\Umin$, $\Umax$, $\alpha$, and $\gamma$.
However, \citet{Draine+Dale+Bendo_etal_2006} find that
the SEDs of galaxies in the SINGS survey 
appear to be satisfactorily reproduced with 
fixed $\alpha=2$ and $\Umax=10^6$, and we will therefore
use these fixed values for $\alpha$ and $\Umax$.
For $\alpha=2$ the fraction $\gamma$ of the dust mass 
that is exposed to starlight intensities 
$\Umin < U \leq \Umax$ has equal amounts 
of infrared power per unit log $U$.
Each of the grain models in Table \ref{tab:dust models} 
has a different value of $\qpah$, so the parameter 
$\qpah$ serves as a proxy for the dust model.
This leaves us with four free parameters: 
$\Mdust$, $\qpah$, $\Umin$, $\gamma$.
The shape of the dust emission spectrum is 
determined by only 3 free parameters: 
$\qpah$, $\Umin$, and $\gamma$.
Examples of emission spectra are shown 
in Figure \ref{fig:spec for Udists}.

\section{\label{sec:dust mass estimation}
         Estimating $\qpah$, $\Umin$, $\gamma$,
         $f_{\rm PDR}$, and $\Mdust$}

Photometry obtained with the IRAC and MIPS cameras on Spitzer Space Telescope
can be used to estimate the free parameters 
$\qpah$, $\Umin$, and $\gamma$,
the fraction $f_{\rm PDR}$ 
of the dust infrared emission radiated by dust grains in
photodissociation regions where $U>10^2$,
and the total dust mass $\Mdust$.
The flux measured in the IRAC 3.6\um\ band is almost entirely due
to starlight, and therefore can be used to remove 
the starlight contribution to the 7.9\um\ and 24\um\ bands, 
with the nonstellar flux densities
estimated to be 
\beqa \label{eq:fns8}
F_\nu^{\rm ns}(7.9\micron) &=& F_\nu(7.9\micron) - 0.232 F_\nu(3.6\micron)
\\ \label{eq:fns24}
F_\nu^{\rm ns}(24\micron) &=& F_\nu(24\micron) - 0.032 F_\nu(3.6\micron)
\eeqa
where the coefficients 0.232 and 0.032 are 
from Helou et al.\ (2004).
The stellar contribution to the 71$\mum$ and 160$\mum$ bands 
is negligible.

For a given dust type (i.e., a given value of $\qpah$) there
is a two dimensional family of emission models, parameterized
by $\Umin$ and $\gamma$.
Figures \ref{fig:band fluxes relative to TIR}a and b show the
flux into the MIPS 24\um\ band 
normalized by the total dust luminosity $L_{\rm TIR}$, plotted
against the ratio of the fluxes into the MIPS 71\um\ and 160\um\
bands.
The two-dimensional family of emission models is shown,
for $0.1\leq\Umin\leq20$, and $0\leq\gamma\leq 0.20$.
The models with $\gamma=0$ (the lower boundary of the model grid)
are models with dust heated by a single starlight intensity:
$\langle U\rangle=\Umin$.

For the $\gamma=0$ models, the ratio $(\nu L_\nu)_{24}/L_{\rm TIR}$
is essentially independent of $\Umin$ for $\Umin\ltsim 2$,
as already seen in Figure \ref{fig:spitzer band strengths vs U}: the
emission into the 24\um\ band is dominated by
single-photon heating of small grains.
For $\gamma>0$, some fraction of the dust is exposed to
starlight intensities $U\gtsim 10$, and the fraction of the total
power that enters the 24\um\ band increases.
For the particular size distributions that have been adopted for
these models (see Figure \ref{fig:dnda}), the models with
$\qpah=0.47\%$ and 4.6\% 
(Figures \ref{fig:band fluxes relative to TIR}a,b
differ by only a factor $\sim 1.5$ in $(\nu L_\nu)_{24}/L_{\rm TIR}$ in the
single-photon heating limit ($\Umin<1$, $\gamma=0$).

The emission into the 7.9\um\ band is a different story.
Figures \ref{fig:band fluxes relative to TIR}c,d
show that in the single-photon heating limit, the fraction of the
power radiated into the 7.9\um\ band is essentially proportional to
$\qpah$:
as $\qpah$ increases from 0.47\% to 4.6\%, 
$(\nu L_\nu)_{7.9}/L_{\rm TIR}$ increases from 0.02 to 0.2.
As $\gamma$ is increased, some additional 7.9\um\ emission is produced;
this results in a noticeable fractional increase
for the $\qpah=0.47\%$ model, but for the $\qpah=4.6\%$ model
the additional emission produces a very small fractional change
in $\langle \nu L_\nu\rangle_{7.9}/L_{\rm TIR}$, which remains
dominated by single-photon heating.

To determine $\qpah$, $\Umin$, $\gamma$, and $\Mdust$, 
the best procedure is to vary all the parameters 
to find the dust
model which comes closest to reproducing the photometry.
This procedure is used by Draine et al.\ (2006) 
in their study of the dust properties of the SINGS galaxy sample.

However, it is also possible to use a graphical procedure to 
find values of 
$\qpah$, $\Umin$, and $\gamma$ that are consistent with the observed data.
We assume that we have Spitzer photometry in 5 bands: the 3.6\um\ and 
7.9\um\ bands of IRAC, 
and the 24, 71, and 160\um\ bands of MIPS.
The 3.6\um\ band observations are used 
with eq.\ (\ref{eq:fns8}) and (\ref{eq:fns24}) to obtain 
$\langle F_\nu^{\rm ns}\rangle_{7.9}$
and 
$\langle F_\nu^{\rm ns}\rangle_{24}$.
We define three ratios of observables:
\beqa
P_{7.9} &\equiv& \frac{\langle\nu F_\nu^{\rm ns}\rangle_{7.9}}
                {\langle\nu F_\nu\rangle_{71}+
                \langle\nu F_\nu\rangle_{160}} ~~~,
\\
P_{24} &\equiv& \frac{\langle\nu F_\nu^{\rm ns}\rangle_{24}}
                {\langle\nu F_\nu\rangle_{71}+
                 \langle\nu F_\nu\rangle_{160}} ~~~,
\\
R_{71} &\equiv& \frac{\langle\nu F_\nu\rangle_{71}}
                     {\langle\nu F_\nu\rangle_{160}} ~~~.
\eeqa
For starlight intensities $0.1 \ltsim U\ltsim10^2$, the bulk of the power
radiated by dust emerges in the 50--200$\mum$ wavelength range, and therefore
the total dust luminosity is approximately proportional to
$[\langle\nu F_\nu\rangle_{71}+\langle\nu F_\nu\rangle_{160}]$.
The ratio $R_{71}$ is sensitive to the temperature of the $a\gtsim 0.01\micron$
grains that dominate the far-infrared emission; $R_{71}$ is
thefore an indicator for the intensity of the starlight heating the dust.

\subsection{\label{sec:determining qpah}
            Determining the PAH Fraction $\qpah$}

$P_{7.9}$ is proportional to the fraction of the dust power radiated in the
PAH features.
Because the 7.9\um\ emission is almost entirely the result of
single-photon heating, $P_{7.9}$ depends very weakly on the starlight intensity.

Figure \ref{fig:IRAC4/MIPS band ratios} shows $P_{7.9}$ vs. $R_{71}$ for
a sequence of grain models, each with a different PAH abundance $\qpah$.
The model with the lowest value of $\qpah$ naturally has very low values
of $P_{7.9}$.  
Because of the very low PAH abundance 
in Figure \ref{fig:IRAC4/MIPS band ratios}a, a very small amount of
dust exposed to high values of $U$ can enhance the 8$\mum$ emission,
and therefore $P_{7.9}$ increases notably as $\gamma$ increases from 0,
with increasing amounts of dust exposed to radiation intensities between
$\Umin$ and $\Umax=10^6$.
However, as $\qpah$ is increased, $P_{7.9}$ becomes less sensitive to
$\gamma$, because single-photon heating of the PAHs produces 7.9\um\
emission even when $U$ is small.

Based on modeling many spectra, we expect that the 7.9\um\ emission
will primarily arise from single-photon heating, i.e., $P_{7.9}$
will be close to the value calculated for $\gamma=0$.
The approximate value of $\qpah$ 
can therefore be determined by finding a dust model
among Figures \ref{fig:IRAC4/MIPS band ratios}a-g
for which the observed $(R_{71},P_{7.9})$ point falls just above the
$\gamma=0$ curve.  The location $(R_{71},P_{7.9})$ on the appropriate
plot also gives an estimate for $\Umin$ and $\gamma$, but 
these parameters are better determined using the 24\um\
emission, as described below.

\subsection{\label{sec:determining umin and gamma}
            Determining $\Umin$ and $\gamma$}

$P_{24}$ is proportional to the fraction of the dust power radiated
near 24$\mum$.
As seen in Figure \ref{fig:band efficiencies, dism}, 
24\um\ emission can be produced by single-photon heating of
$a\approx 15-40\Angstrom$ grains.
For a given dust model there is therefore a limiting value of $P_{24}$
that applies for $0.1\ltsim U \ltsim 10$ where the bulk of the dust power
is captured in the 71 and 160$\mum$ bands.
However, when some fraction of the grains are heated by starlight
with intensities $U\gtsim 20$ 
(see Figure \ref{fig:spitzer band strengths vs U}),
larger dust grains can be heated sufficiently
to add to the 24$\mum$ emission.  As a result $R_{24}$ is sensitive to the
value of $\gamma$.

\begin{figure}[h]
\begin{center}
\includegraphics[width=11.3cm,angle=0]{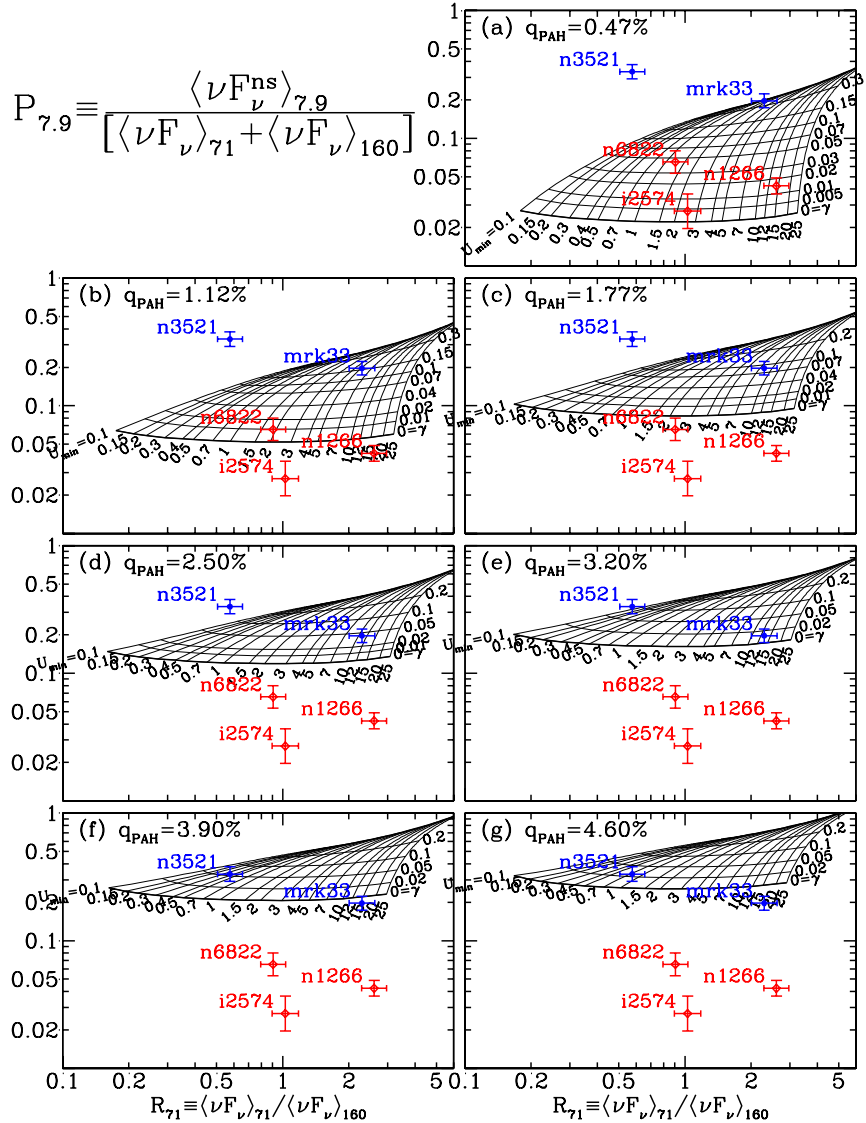}
\caption{\label{fig:IRAC4/MIPS band ratios}
         \footnotesize
         IRAC~7.9$\mum$ power relative to MIPS~71 and 160$\mum$
	 power versus MIPS 71/MIPS 160 band ratio.
	 }
\end{center}
\end{figure}

\begin{figure}[h]
\begin{center}
\includegraphics[width=11.3cm,angle=0]{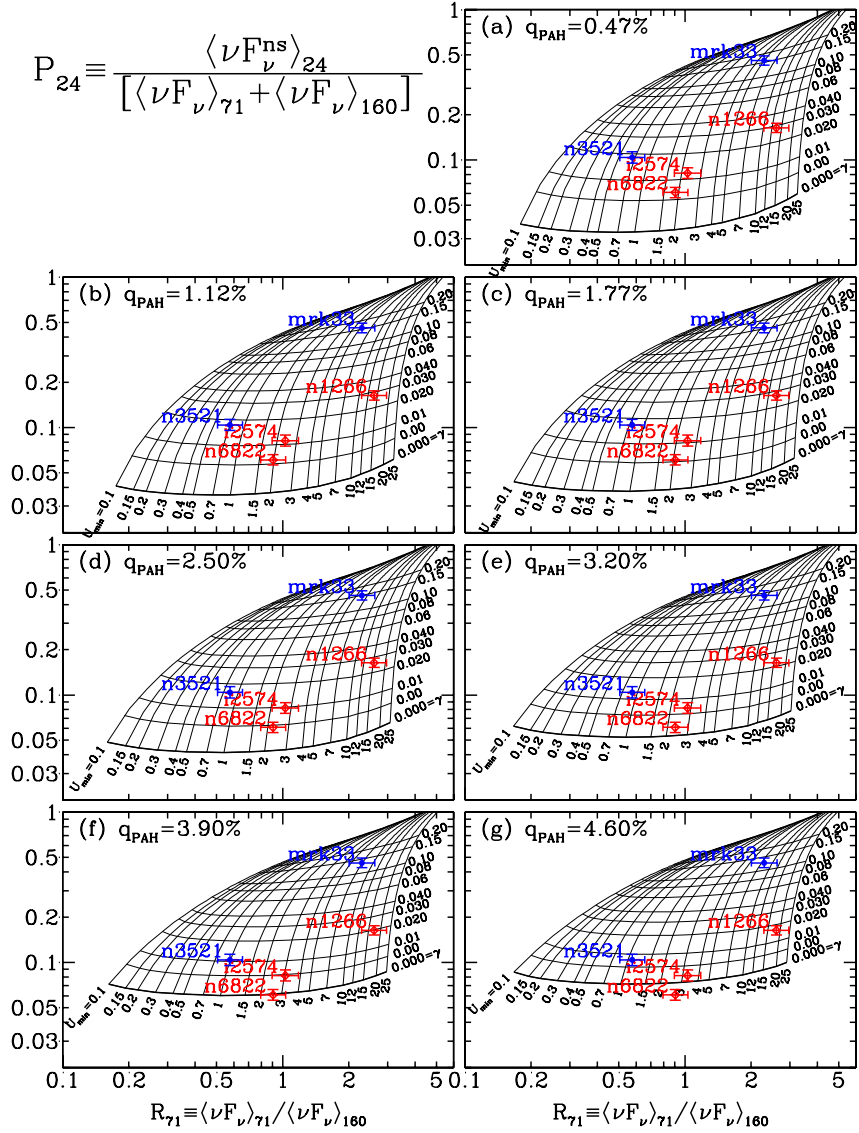}
\caption{\label{fig:MIPS band ratios}
         \footnotesize
         MIPS band ratios for models.
        }
\end{center}
\end{figure}

To estimate the value of $\gamma$, one looks for a dust model in Figure
\ref{fig:MIPS band ratios} with the value of $\qpah$ found from
Figure \ref{fig:IRAC4/MIPS band ratios}.
The observed location of $(R_{71},P_{24})$ on Fig.\ \ref{fig:MIPS band ratios} then allows
$\Umin$ and $\gamma$ to be determined.
It may happen that $(\Umin,\gamma)$ estimated from Figure
\ref{fig:MIPS band ratios} differs from $(\Umin,\gamma)$ indicated
by Figure \ref{fig:IRAC4/MIPS band ratios}.
When this occurs, it is an indication that a single dust model does not
perfectly reproduce the observed 7.9/24/71/160\um\ colors.
We recommend using the values of $(\Umin,\gamma)$ estimated from
Figure \ref{fig:MIPS band ratios} as they are less sensitive to
the adopted value of $\qpah$.

\begin{figure}[h]
\begin{center}
\includegraphics[width=10.0cm,angle=0]{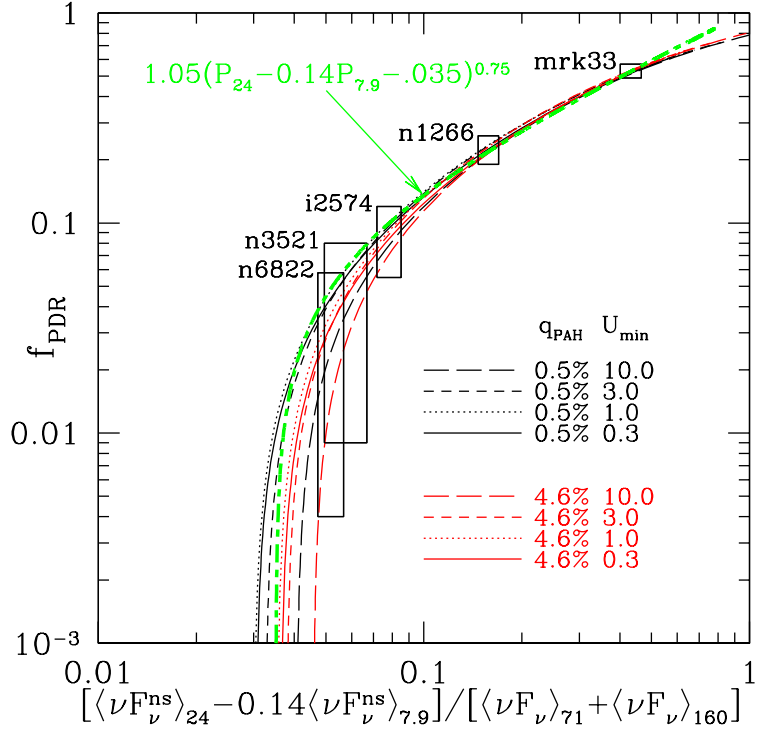}
\caption{\label{fig:fpdr}
         $f_{\rm PDR}$, the fraction of the dust infrared luminosity
         radiated by dust grains in regions where $U>10^2$, plotted
         as a function of the observable $(P_{24}-0.14P_{7.9})$.
         For the 5 example galaxies, the width of the rectangle
         corresponds to the $\pm1\sigma$ uncertainty in
         $P_{24}-0.14P_{7.9}$.
         The fitting function $1.05(P_{24}-0.14P_{7.9}-0.035)^{0.75}$
	 approximates the model results.
	 }
\end{center}
\end{figure}
\subsection{Determining $f_{\rm PDR}$}
Young stars tend to be in or near to dust clouds.
In star-forming galaxies such as the Milky Way, we expect that
a significant fraction of the starlight emitted by O and B stars
will be absorbed by dust that happens to be relatively close
to the star, so that the starlight intensity is significantly
above the average starlight intensity.
Photodissociation regions (PDRs) are typical examples of such environments.

For the power-law distribution of starlight intensities given by eq.\ (\ref{eq:dM_d/dU}),
we can calculate a quantity $f_{\rm PDR}$, which we define here as
the fraction of the total dust luminosity that is radiated by dust
grains in regions where $U > 10^2$:
\beq \label{eq:fpdr}
f_{\rm PDR} = 
\frac{\gamma\ln(\Umax/10^2)}
{(1-\gamma)(1-\Umin/\Umax)+\gamma\ln(\Umax/\Umin)}
\eeq
If $\Umin$ and $\gamma$ are known (e.g., either determined by finding
the best-fit $(\qpah,\Umin,\gamma)$ from the 3-dimensional model space,
or by following the graphical procedures described above), $f_{\rm PDR}$
can be calculated from eq.\ (\ref{eq:fpdr}).
Here we show how $f_{\rm PDR}$ can be estimated directly
from IRAC and MIPS photometry.

For the Milky Way dust models considered here, 
PAHs undergoing single-photon heating convert a small fraction of the
absorbed starlight power into 24\um\ emission, but
when a distribution of starlight intensities is present, there is
additional 24\um\ emission from high intensity regions.
The 24\um\ contribution from PAHs depends, of course, on the
PAH abundance.
We find that for our models
$f_{\rm PDR}$ can be closely related to the fraction of the power
radiated at 24\um\ after subtraction of the contribution of PAHs
to the 24\um\ power.  
We find that the combination $(P_{24}-0.14P_{7.9})$ gives a quantity
that is sensitive to $U$ and relatively insensitive to the value of
$\qpah$.

Figure \ref{fig:fpdr} shows $f_{\rm PDR}$ versus $P_{24}-0.14P_{7.9}$
for eight different dust models with different values of $\qpah$
and $\Umin$.
We see that the curves fall within a narrow band.
For these dust models, $f_{\rm PDR}$ can be estimated from the
observable $P_{24}-0.14P_{7.9}$ using Figure \ref{fig:fpdr}. 
\begin{figure}[h]
\begin{center}
\includegraphics[width=11.3cm,angle=0]{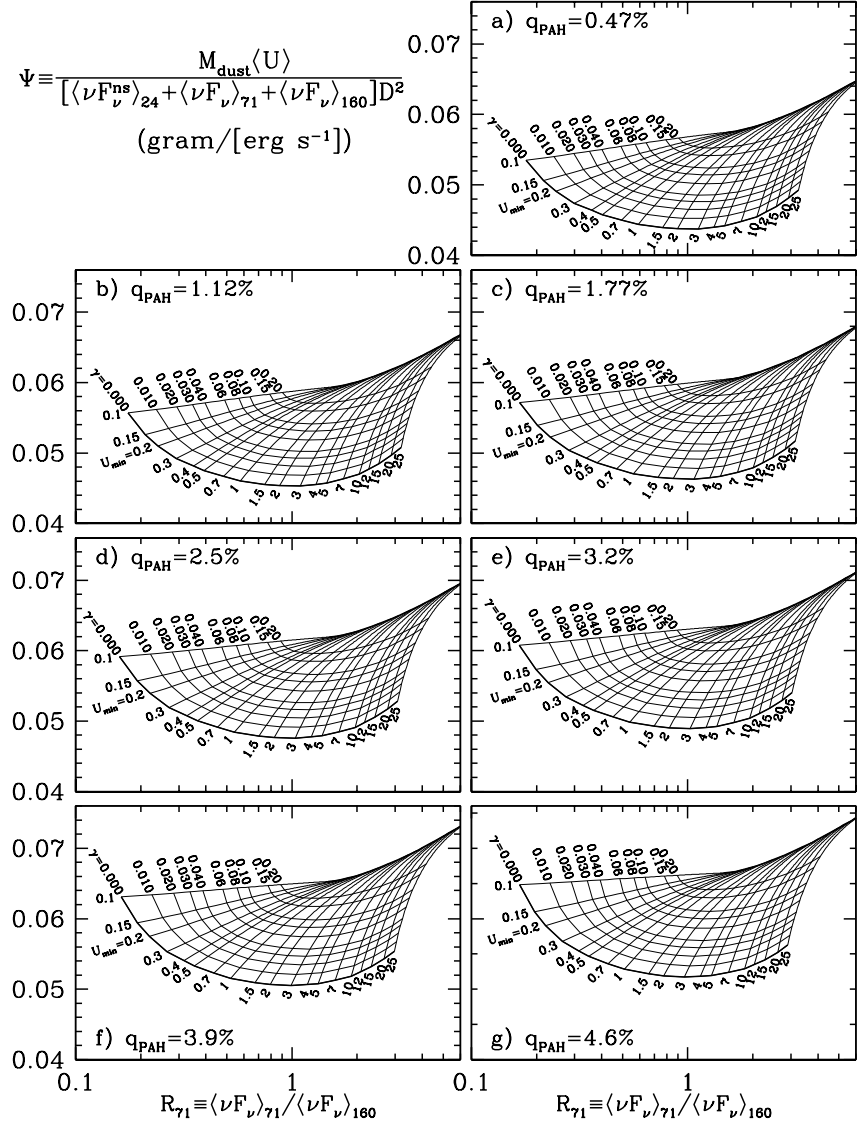}
\caption{\label{fig:Psi grid}
         \footnotesize
         Value of $\Psi$ for dust mass estimation (see text).
        }
\end{center}
\end{figure}

\subsection{Examples}

\begin{table}[h]
\footnotesize
\begin{center}
\caption{\footnotesize
         \label{tab:examples}
         Galaxy Examples.
}
\begin{tabular}{cccccc}
        & IC2574 & Mark 33 & NGC~1266 & NGC~3521 & NGC~6822 \\
\hline
$D$ (Mpc) & 4.02 & 21.7    & 31.3     & 9.0      & 0.49 \\
$\langle F_\nu\rangle_{3.6}$ (Jy)$^a$
                & $0.156\pm0.016$ & $0.027\pm0.003$ & $0.056\pm0.006$ & $2.12\pm0.21$ & $2.20\pm0.22$ \\
$\langle F_\nu\rangle_{7.9}$ (Jy)$^a$
                & $0.066\pm0.007$ & $0.125\pm0.013$ & $0.087\pm0.009$ & $6.23\pm0.62$ & $1.41\pm0.14$ \\
$\langle F_\nu\rangle_{24}$ (Jy)$^a$
                & $0.278\pm0.013$ & $0.836\pm0.034$ & $0.861\pm0.035$ & $5.47\pm0.22$ & $2.59\pm0.10$ \\
$\langle F_\nu\rangle_{71}$ (Jy)$^a$
                & $5.10\pm0.37$   & $3.82\pm0.28$   & $11.45\pm0.80$  & $57.2\pm4.0$ & $59.2\pm4.2$ \\
$\langle F_\nu\rangle_{160}$ (Jy)$^a$
                & $10.8\pm1.4$    & $3.63\pm0.46$   & $9.6\pm1.2$     & $217.\pm26$ & $143\pm17$ \\
 
$P_{7.9}$       & .027 & 0.196  & 0.0435 & 0.529 & 0.065 \\
$P_{24}$        & .0810 & 0.460  & 0.168  & 0.166 & 0.061 \\
$R_{71}$        & 1.08  & 2.33   & 2.55   & 0.582 & 0.944 \\
\hline
$\qpah$ from Fig.\ \ref{fig:IRAC4/MIPS band ratios}
         & 0.5\% & 3.2\%  & 0.5\%  & 4.6\% & 1.1\% \\
$\Umin$ from Fig.\ \ref{fig:MIPS band ratios}       
         & 2.3   & 7      & 12.5     & 1.1     & 2\\
$\gamma$ from Fig.\ \ref{fig:MIPS band ratios}
         &0.012  & 0.14   & 0.029   & 0.007 & 0.006 \\
$\langle U\rangle$ (from eq.\ \ref{eq:<U>})
         &2.6    & 18.   & 16.   & 1.2  & 2.1 \\
$\Psi~~(\g/(\erg\s^{-1}))$
         &0.046  & 0.061  & .049   & 0.054 & 0.0465 \\
$\log_{10}(\Mdust/\Msol)$
         &5.79   & 6.41   & 7.04   & 8.10  & 5.15 \\
\hline
\multicolumn{6}{l}{$^a$ \citet{Dale+GildePaz+Gordon_etal_2006}}
\end{tabular}

\end{center}
\end{table}
To illustrate this, five galaxies are indicated in 
Figures \ref{fig:IRAC4/MIPS band ratios} and \ref{fig:MIPS band ratios}: 
NGC~1266, NGC~3521, NGC~6822, IC~2574, and Markarian 33.
Table \ref{tab:examples} gives the measured global fluxes for these
galaxies 
\citep{Dale+Bendo+Engelbracht_etal_2005},
the derived ratios $P_{7.9}$, $P_{24}$, and $R_{71}$,
and the values of $\qpah$, $\Umin$, and $\gamma$ obtained from the
procedure described in 
\S\S\ref{sec:determining qpah},\ref{sec:determining umin and gamma}.
Below we comment on the individual cases.

IC~2574:  In Figure \ref{fig:IRAC4/MIPS band ratios}, 
          the very low value of $P_{7.9}=0.027$ only falls on the model grid
	  in Figure \ref{fig:IRAC4/MIPS band ratios}a, therefore
	  we select $\qpah=0.5\%$.
	  From Figure \ref{fig:MIPS band ratios}a,
	  we find $\Umin\approx2.3$ and $\gamma\approx0.012$.

Mark~33:  From Figure \ref{fig:IRAC4/MIPS band ratios}e,
          we estimate $\qpah\approx 3.2\%$.
	  The very high value of $P_{24}=0.46$
	  requires $\Umin\approx 7$ and $\gamma\approx 0.14$ on
	  Figure \ref{fig:MIPS band ratios}e in order to
	  have enough hot dust to reproduce the observed
	  24\um\ flux.
	  The very large value found for $\gamma$ suggests that a large
	  fraction of the dust heating in Mark 33 is taking place in
	  photodissociation regions near OB associations,
	  consistent with the dwarf starburst nature of this galaxy
	  \citep{Hunt+Bianchi+Maiolino_2005}.

NGC~1266: The low value of $P_{7.9}\approx 0.044$ can only
          be reproduced in Figure \ref{fig:IRAC4/MIPS band ratios}
          by models with small values of $\qpah$.  
          Of the values of $\qpah$ for which spectra have been computed,
          $\qpah=0.47\%$ is preferred.
          Turning now to Figure \ref{fig:MIPS band ratios}a, we find that
          $\Umin\approx12.5$ and $\gamma\approx0.029$.
          The values of $\qpah$, $\Umin$, and $\gamma$ found by this
          graphical exercise are in good agreement with the best-fit
          values of $\qpah=0.47\%$, $\Umin=12$, and $\gamma=0.029$
          \citep{Draine+Dale+Bendo_etal_2006}.

NGC~3521: From Figure \ref{fig:IRAC4/MIPS band ratios}g, we find
          $\qpah\approx 4.6\%$.
          From Figure \ref{fig:MIPS band ratios}g, we obtain
	  $\Umin\approx 1.1$ and $\gamma\approx 0.007$.
	  SCUBA fluxes are also available for this galaxy
	  \citep{Dale+Bendo+Engelbracht_etal_2005};
	  when the 850\um\ flux is used to constrain
	  the grain model,
	  \citet{Draine+Dale+Bendo_etal_2006} find that it is difficult
	  to simultaneously reproduce the MIPS and SCUBA photometry.
	  The overall best fit is found for
	  $\Umin=2$, $\gamma=0.01$.

NGC~6822: From
          Figure \ref{fig:IRAC4/MIPS band ratios}b 
          we estimate $\qpah=1.1\%$.
          From Figure \ref{fig:MIPS band ratios}g,
          we estimate $\Umin\approx2.0$ and $\gamma\approx 0.006$.
	  These are in agreement with the best-fit values
	  $\qpah=1.2\%$, $\Umin=2.0$, $\gamma=0.005$ found by
	  \citet{Draine+Dale+Bendo_etal_2006}.


\subsection{Estimating the Dust Mass $\Mdust$}

Consider a galaxy at distance $D$.  If $j_\nu$ is the dust emissivity
per H nucleon, then the flux is
\beq
F_\nu = \frac{M_{\rm H}}{m_{\rm H}}\frac{j_\nu}{D^2} ~~~,
\eeq
and the dust mass
\beq \label{eq:Mdust}
\Mdust 
= \left(\frac{\Mdust}{M_{\rm H}}\right)\,m_{\rm H}\,\frac{F_\nu}{j_\nu}\,D^2
~~~.
\eeq
If we know the values of $(\Mdust/M_{\rm H})$ and $j_\nu$ for a given grain
model, we can estimate $\Mdust$ directly from the distance $D$ and
the observed flux
$F_\nu$.
However, our estimate for $j_\nu$ may depend sensitively on our
estimates for $\qpah$, $\Umin$, and $\gamma$.

A relatively robust approach to dust mass estimation is to note that
for a given dust mixture
the total dust luminosity is proportional to $\Mdust\langle U\rangle$.
Define the quantity
\beq
\Psi(\qpah,\gamma,\Umin) 
\equiv 
\left(\frac{\Mdust}{M_{\rm H}}\right) m_{\rm H}
\frac{\langle U\rangle}
{\left[\langle\nu j_\nu\rangle_{24} +\langle\nu j_\nu\rangle_{71}+
\langle\nu j_\nu\rangle_{160}\right]}
~~~,
\eeq
where $m_{\rm H}$ is the mass of an H atom,
and $\langle U\rangle$ is the mean starlight intensity.
For the distribution function (\ref{eq:dM_d/dU}),
with $\alpha=2$, $\langle U\rangle$ is
\beq \label{eq:<U>}
\langle U\rangle \equiv \frac{\int U d\Mdust}{\int d\Mdust}
= (1-\gamma)\Umin + \frac{\gamma\Umin\ln(\Umax/\Umin)}{1-\Umin/\Umax}
~~~.
\eeq
The dust mass is related to the observed fluxes by
\beq \label{eq:compute Mdust}
\Mdust = \frac{\Psi}{\langle U\rangle} 
\left[\langle\nu F_\nu\rangle_{24} +\langle\nu F_\nu\rangle_{71}+
\langle\nu F_\nu\rangle_{160}\right]D^2
~~~.
\eeq
For a given dust model, we can calculate $\Psi$, which is
shown in Figures \ref{fig:Psi grid}(a-g).
We observe that the values of $\Psi$ in Figures \ref{fig:Psi grid}(a-g)
range only from 0.044 to $\sim$0.066$\g/(\erg\s^{-1})$, with
$\Psi\approx 0.055\g/(\erg\s^{-1})$ as a representative value.
However, if we already have estimates for $\qpah$,
$\Umin$, and $\gamma$, it is straightforward to obtain an accurate value
for $\Psi(\qpah,\Umin,\gamma)$ from Figure \ref{fig:Psi grid}.

To summarize, 
the procedure for estimating the dust mass $\Mdust$ is as follows:
\begin{enumerate}
\item Use $\langle F_\nu^{\rm ns}\rangle_{7.9}$, 
$\langle F_\nu^{\rm ns}\rangle_{24}$,
$\langle F_\nu\rangle_{71}$, and 
$\langle F_\nu\rangle_{160}$ to calculate
$P_{7.9}$, $P_{24}$, and $R_{71}$.
\item Use $R_{71}$ and $P_{7.9}$ to estimate $\qpah$ by finding the value of
$\qpah$ such that
$(R_{71},P_{7.9})$ falls just above the $\gamma=0$ curve in Figure
\ref{fig:IRAC4/MIPS band ratios}.
\item Using this value of $\qpah$,
locate $(R_{71},P_{24})$
on Figure \ref{fig:MIPS band ratios}
to determine $\Umin$ and $\gamma$.
\item Using the values of $\qpah$, $\Umin$, and $\gamma$,
use Figure \ref{fig:Psi grid} to find $\Psi$.
\item Use $\Umin$ and $\gamma$ (and $\Umax=10^6$) to 
calculate $\langle U\rangle$ from eq.\ (\ref{eq:<U>}).
\item Calculate $\Mdust$ using eq.\ (\ref{eq:compute Mdust}).
\end{enumerate}

We have estimated $\langle U\rangle$, $\Psi$, and the resulting dust
mass $\Mdust$ for the 5 sample galaxies, with results given in Table
\ref{tab:examples}.

\section{Discussion
         \label{sec:discussion}}

\subsection{Origin of near-IR Continuum 
Opacity of PAHs\label{sec:pah_continuum}}

The present models assume that the small-particle end 
of the grain size distribution is dominated by PAH particles, 
as assumed by WD01 and Li \& Draine (2001b), 
and the PAH particles are therefore assumed to have 
a ``continuum'' component to $\Cabs$ in order to 
be able to reproduce 2--6$\mum$ emission seen by ISO 
\citep{Lu+Helou+Werner+Dinerstein+Dale_2003}
and emission in the IRAC 4.5$\mum$ filter
\citep{Helou+Roussel+Appleton+Frayer+Stolovy_etal_2004}.
Continuum emission also appears to 
be present at other wavelengths,
e.g., between the
12.6 and 16.45$\mum$ features.
In the 2--6$\mum$ wavelength range, even very small 
carbonaceous grains are arbitrarily taken to have $\Cabs$ equal to
1\% of the continuum that would be produced by free electron absorption
in graphite, added to the discrete absorption bands expected for
PAHs.
The nature of this continuum absorption is unclear.
Perhaps some PAHs, e.g., tubular PAHs with appropriate
chirality 
\citep{Zhou+Sfeir+Zhang_etal_2006}
have zero bandgap as in graphite.

We are assuming that the PAHs are responsible for the 2--5\um\
continuum.  
\citet{An+Sellgren_2003} 
find that the 2\um\ continuum
emission and 3.29\um\ PAH emission have different spatial
distributions in NGC~7023.  
Perhaps the 2\um\ emission is strongest in
regions where
the PAHs have been dehydrogenated, thereby suppressing the 3.29\um\
emission.

The assumption that the continuum emission originates in PAHs
is consistent with the fact that the 9.8$\mum$ and 18$\mum$
silicate features are generally not seen in emission, except from regions
(dusty winds, compact H~II regions, dust near the Trapezium stars in Orion) 
where the starlight is thought to be sufficiently intense to heat
$a\approx0.1\micron$ grains to $T\gtsim 200\K$.
However, 
\citet{Li+Draine_2001a}
showed that the observed 
emission from the diffuse interstellar medium in the Milky Way
would in fact allow larger amounts of ultrasmall silicate grains than had
been estimated previously, because the 9.7$\mum$ silicate emission feature
can be ``hidden'' between the 8.6$\mum$ and 11.3$\mum$ PAH features.

In summary, the origin of the 2--5\um\ continuum emission remains uncertain.

\subsection{Near-IR Absorption of PAH Ions: 
Astrophysical Implications\label{sec:pah_cabs_ir}}

The far-red to near-IR opacity of PAH ions described in
\citet{Li+Draine_2001b} 
was derived from earlier 
experimental data for a small number of small PAH ions 
[see 
\citet{Salama+Bakes+Allamandola+Tielens_1996}
and references therein].
More recent laboratory measurements were carried
out by 
\citet{Mattioda+Hudgins+Allamandola_2005}
for 27 PAH cations and anions ranging in size from
C$_{14}$H$_{10}$ to C$_{50}$H$_{22}$. 
The newly-measured absorption of PAH ions
at 0.77--2.5$\mum$ is considerably higher than 
previously reported.
This further supports our earlier conclusion \citep{Li+Draine_2002b}
that visible and near-IR photons are able to excite 
PAHs to temperatures high enough to emit the mid-IR bands,
in agreement with the detection
of the PAH features in \newtext{interstellar} regions 
lacking UV photons
\citep{Uchida+Sellgren+Werner_1998,
       Pagani+Lequeux+Cesarsky_etal_1999},\footnote{%
  \newtext{
   PAH emission features have also been detected in
   UV-poor dust debris disks around main-sequence stars 
   SAO\,206462 [Sp=F8V, 
   $\Teff\approx 6250\K$; \cite{Coulson+Walther_1995}]
   and HD\,34700 [Sp=G0V,
   $\Teff\approx 5940\K$; \cite{Smith+Clayton+Valencic_2004}].
   However, in an extensive {\it Spitzer} IRS spectroscopic
   survey of 111 T Tauri stars in the Taurus star-forming region,
   \citet{Furlan+Hartmann+Calvet_etal_2006} found that the PAH emission
   bands are not seen in dust disks around T Tauri stars of 
   spectral type later than G1. 
   }
   }
and in agreement with the observation by 
\citet{Sellgren+Luan+Werner_1990}
that the ratio of the IRAS 12$\mum$ 
emission (to which the PAH features are the dominant contributor) 
to the total far-IR surface brightness
is independent of
the effective temperature $T_{\rm eff}$
of the exciting stars, for 24 reflection nebulae
for $5000\K\le T_{\rm eff} \le 33000\K$.

This also relates to our understanding of 
the origin of interstellar PAHs.
At present, the origin and evolution of interstellar PAHs 
are not very clear. Suggested sources for interstellar
PAHs include the injection (into the interstellar medium) 
of PAHs formed in carbon star outflows 
\citep{Jura_1987,
       Latter_1991}.
However, the PAH emission features are commonly not seen
in the mid-IR spectra of C-rich AGB stars. 
The few
C stars which display the PAH features all have a hot companion
which emits UV photons 
\citep{Speck+Barlow_1997,
       Boersma+Hony+Tielens_2006}.
It has been suggested that PAHs are present in all C stars 
but they are simply not excited sufficiently to emit at 
mid-IR due to lack of UV photons.
The new visible-IR cross sections for PAH ions
\citet{Mattioda+Hudgins+Allamandola_2005} suggest that
PAHs can be excited even by the starlight from a C star,
in which case the absence of PAH emission places a
limit on the abundance of small PAHs in these outflows.\footnote{%
  \newtext{ 
  \citet{Jura+Bohac+Sargent_etal_2006} reported detection
  of PAH emission features in HD\,233517, an evolved oxygen-rich
  K2III red giant ($\Teff \approx 4390\K$)
  with circumstellar dust. 
  \citet{Jura_2003} argued that 
  the IR excess around HD\,233517 is unlikely to be produced 
  by a recent outflow in a stellar wind. 
  \citet{Jura+Bohac+Sargent_etal_2006} hypothesized that
  there is a passive, flared disk orbiting HD\,233517
  and the PAH molecules in the orbiting disk 
  may be synthesized in situ as well as having been 
  incorporated from the ISM.
  }
  }

\subsection{\label{sec:6.2um absorption feature}
            Detectability of 6.2\um\ Absorption by Interstellar PAHs}

The dust model appropriate to Milky Way dust is thought to have
$\qpah\approx4.6\%$ of the dust mass in PAHs with $N_{\rm C}<10^3$ C atoms.
The carbonaceous grains with $a < 50\Angstrom$ contain 52 ppm of C per
H nucleon,
divided approximately equally between neutral and ionized PAHs.
For this model, we predict a 6.22\um\ absorption feature
with integrated strength
\beq \label{eq:tau62 model}
\frac{1}{N_{\rm H}}\int \Delta \tau ~d\lambda^{-1} = 
6.7\times10^{-23}
\left(\frac{\rm C_{\rm PAH}/H}{52~\rm ppm}\right)\cm/{\rm H}
~~~.
\eeq
\citet{Chiar+Tielens_2001} 
found an upper limit 
$\int\Delta\tau ~d\lambda^{-1} < 0.8\cm^{-1}$
toward Cyg OB2 No.\,12.  
For the estimated $N_{\rm H}=1.9\times10^{22}\cm^{-2}$, this
gives an upper limit 
\beq \label{eq:tau62 obs}
\frac{1}{N_{\rm H}}\int \Delta \tau ~d\lambda^{-1} < 
4.2\times10^{-23}\cm/{\rm H} ~~~,
\eeq
somewhat smaller than our estimate (\ref{eq:tau62 model}).

While the present dust model has 52 ppm C in PAHs 
with $N_{\rm C}<5\times10^4$, the 6.22\um\ emission is produced mainly
by PAHs with $N_{\rm C}< 500$, which account for only 35 ppm.
The 6.2\um\ absorption contributed by PAHs with $N_{\rm C}< 500$ in our
model is consistent with the \citet{Chiar+Tielens_2001} upper limit.

The absolute PAH band strengths in Table \ref{tab:drude} are
only estimates.
Because the 3--25\um\ emission spectrum for $U=1$ is almost entirely the
result of single-photon heating, the emission spectrum would be almost
unaffected if all the PAH band strengths were uniformly reduced by a
common factor.
However, based on the comparison with theoretical calculations shown
in Figure \ref{fig:Cabs_vs_theory}, the actual band strengths should
be similar in magnitude to the values adopted in Table \ref{tab:drude}.
Spectroscopy of heavily extinguished stars using the IRS instrument
on Spitzer Space Telescope should be carried out to attempt to
detect the predicted 6.22\um\ absorption from interstellar PAHs.

\subsection{Deuterated PAHs\label{sec:dpah}}

Since the aromatic C--D bond zero-point energy is about 30\%
lower than that of the C--H bond, H loss is favored over 
D loss. 
The PAH D/H ratio might 
therefore be substantially larger than the 
\newtext{total} D/H~$\approx2\times10^{-5}$ 
\citep{Allamandola+Tielens+Barker_1989}.
The observed sightline-to-sightline variations of (D/H) in interstellar
gas have been interpreted as
requiring depletion of D by interstellar dust
\citep{Linsky+Draine+Moos_etal_2006}.
\citet{Draine_2004a,Draine_2006b} 
has suggested that this depletion of
D may take via interactions of PAH ions with the gas, and may
imply D/H$\approx0.3$ in interstellar PAHs.
\citet{Peeters+Allamandola+Bauschlicher_etal_2004} 
report tentative
detection of emission at 4.4 and 4.65\um\ from deuterated PAHs from
the Orion Bar and M17 PDRs.
If the emission is in fact due to C-D stretching modes, 
\citet{Peeters+Allamandola+Bauschlicher_etal_2004} 
estimate 
${\rm D/H}\approx0.17\pm0.03$ for the emitting PAHs in the Orion Bar and 
$0.36\pm0.08$ in M17.

Our model for the PAH absorption cross section is based on observed
emission spectra, and some of the modeled absorption may be associated with
C-D bending modes if interstellar PAHs are appreciably deuterated.
Because the observations of 4.4 and 4.65\um\ emission in the Orion Bar
and M17 are tentative, we do not attempt here to include any
absorption and emission that might result from C-D stretching modes 
in the 4.5\um\ region.

\subsection{Ubiquity and Absence of PAHs in Astrophysical Regions: 
Rationale for Variable $q_{\rm PAH}$\label{sec:reason_4_qpah}}

The ISO and Spitzer imaging and spectroscopy have 
revealed that PAHs are a ubiquitous feature of 
both the Milky Way and external galaxies. 
Recent discoveries 
include the detection of PAH emission in a wide range of systems:
        distant Luminous Infrared Galaxies (LIRGs) with redshift $z$ ranging 
        from 0.1 to 1.2 
        \citep{Elbaz+LeFloch+Dole+Marcillac_2005};
        distant Ultraluminous Infrared Galaxies 
         (ULIRGs) with redshift $z\sim$\,2 
         \citep{Yan+Chary+Armus_etal_2005};
          distant luminous submillimeter galaxies at 
          redshift $z\sim$\,2.8 
          \citep{Lutz+Valiante+Sturm_etal_2005};
    elliptical galaxies with a hostile environment
    (containing hot gas of temperature $\sim$\,10$^7\K$) 
    where PAHs can be easily destroyed 
    through sputtering by plasma ions
    \citep{Kaneda+Onaka+Sakon_2005};
    faint tidal dwarf galaxies with metallicity $\sim Z_\odot/3$
    \citep{Higdon+Higdon+Marshall_2006};
     and 
    galaxy halos 
    \citep{Irwin+Madden_2006,
           Engelbracht+Kundurthy+Gordon_etal_2006}.

However, the PAH features are weak or even absent 
in low-metallicity galaxies and AGN:
\begin{itemize}
\item 
Based on an analysis of the mid-IR spectra of 60 galaxies
obtained by ground-based observations, 
\citet{Roche+Aitken+Smith+Ward_1991}
reported the lack of PAH emission features in AGN.
\item 
The Small Magellanic Cloud (SMC), an irregular dwarf galaxy
with a metallicity just $\sim$\,1/10 of solar, exhibits a local 
minimum at $\sim$\,12$\mum$ in its IR emission spectrum, 
suggesting the lack of PAHs 
\citep{Li+Draine_2002c}.
\item 
\citet{Thuan+Sauvage+Madden_1999},
\newtext{\citet{Plante+Sauvage_2002},}
and 
\citet{Houck+Charmandaris+Brandl_etal_2004}
have shown that the PAH features are 
absent in the ISO and Spitzer 
spectra of SBS\,0335-052, a metal-poor ($Z\sim$\,$Z_{\odot}/41$) 
blue compact dwarf galaxy.
\item More recently, 
\citet{Hunt+Bianchi+Maiolino_2005},
\citet{Madden+Galliano+Jones+Sauvage_2006},
and 
\citet{Wu+Charmandaris+Hao_etal_2006}
performed a more systematic investigation 
of the mid-IR spectra of a large number of 
low-metallicity galaxies 
(with $Z/Z_\odot$ ranging from 0.02 to 0.6) 
obtained with ISO and Spitzer.
They found the PAH features are 
substantially suppressed in metal-poor dwarf galaxies.
\item 
\newtext{ 
Using Spitzer IRAC [3.6]--[7.9] colors 
and {\it Sloan Digital Sky Survey} (SDSS) [g--r] 
colors of 313 visible-selected SDSS main sample 
galaxies, \citet{Hogg+Tremonti+Blanton_etal_2005} found that low-luminosity
galaxies show a deficiency in PAH emission, 
with weak evidence for a dependence of 
the PAH-to-stellar radiation ratio on metallicity.
\citet{Rosenberg+Ashby+Salzer+Huang_2006} performed a similar study
for a statistically complete sample of 19 star-forming 
dwarf galaxies, but they found a significant number of 
low-luminosity galaxies with very red [3.6]--[7.9] colors, 
indicating the presence of PAHs (and/or hot dust).
They also found that the 7.9$\mum$ emission is more
strongly correlated with the star formation rate
than it is with the metallicity.
}
\item 
\citet{Engelbracht+Gordon+Rieke_etal_2005}
examined the Spitzer mid-IR colors
of 34 galaxies ranging over 2 orders of magnitude in metallicity.
They found that the 8\,$\mu$m-to-24\,$\mu$m color 
changes abruptly from 
$F_\nu(8\mu {\rm m})$/$F_\nu(24\mu {\rm m})$$\sim$\,0.7
for galaxies with $Z/Z_\odot$\,$>$\,1/3
to $F_\nu(8\mu {\rm m})$/$F_\nu(24\mu {\rm m})$$\sim$\,0.08
for galaxies with $Z/Z_\odot$\,$<$\,1/5--1/3.
They attributed this color shift to 
a decrease in the 7.7$\mum$ PAH feature at low metallicity. 
\item  The silicate-graphite-PAH model described in this
paper has been employed to quantitatively determine the 
fraction $\qpah$ of the dust mass contributed by PAHs
for 61 galaxies in the SINGS survey
\citep{Draine+Dale+Bendo_etal_2006}.
There appears to be a threshold metallicity: galaxies with 
[O/H]$\equiv 12+\log_{10}({\rm O/H})< 8.1$ (i.e., $Z < Z_\odot/3$)
have $\qpah< 1.5\%$.
\end{itemize}

Knowledge of how $q_{\rm PAH}$ varies as 
a function of galaxy parameters (such as metallicity, luminosity,
and morphological type) will 
have a broad impact --
\begin{itemize}
\item It will be essential for properly 
interpreting the IR cosmological surveys which now often
assume a {\it constant} PAH strength for galaxies 
[see, e.g., 
\citet{Lagache+Dole+Puget_etal_2004}].
\item It may provide physical insight into 
the validity of using the line-to-continuum ratio of
the 7.7$\mum$ PAH feature as a discriminator
between starburst and AGN activity in ULIRGs,
i.e., whether the dominant luminosity source of ULIRGs 
is an AGN or a starburst; see 
\citet{Genzel+Lutz+Sturm_etal_1998}.
\item As local analogs of early galaxies at high redshift
which must have formed at very low metallicity, 
nearby low metallicity systems can provide insight into
the early stages of galaxy evolution.
The low PAH abundances observed in these galaxies reflect the
balance of PAH formation and destruction, and thereby is
a valuable probe of the physical conditions following the onset
of star formation in primordial gas. 
\item By comparing with the star-formation rates 
derived from other tracers, it will allow us to 
quantitatively evaluate the validity of using 
the IRAC 8$\mum$ photometry as a reliable tracer for 
star-formation rates. It is generally believed that
the IRAC 8$\mum$ flux is stronger in regions with
stronger star-forming activities; however, observations suggest
that PAHs are destroyed in star-forming regions
with very strong and hard radiation fields
\citep{Contursi+Lequeux+Cesarsky_etal_2000,
       ForsterSchreiber+Roussel+Sauvage+Charmandaris_2004,
       Beirao+Brandl+Devost_etal_2006}.
\end{itemize}

\subsection{Diversity of PAH Emission Spectra\label{sec:diverse}}

The PAH absorption and emission properties presented in 
this series of papers are for ``astronomical'' PAHs
-- we fix the peak wavelength and bandwidth of each PAH band 
to what is observed in most astronomical regions. 
The strength of each PAH band (on a per carbon or hydrogen 
atom basis) is also fixed for a specific charge state 
to be consistent with what is observed in space,
measured in the laboratory, or calculated theoretically.

However, there exist variations in the central wavelength, 
bandwidth and strength of PAH emission bands 
among different astronomical objects 
or among different locations within one object:
\begin{itemize}
\item 
\citet{Cohen+Tielens+Bregman_etal_1989} 
reported that the central wavelength
of the 7.7$\mum$ band varies among different types of objects:
it shifts from shorter wavelengths for objects with heavily
processed PAHs (e.g. HII regions, reflection nebulae)
to longer wavelengths for objects with freshly created PAHs
(e.g. planetary nebulae). This wavelength shift has been 
confirmed by high resolution spectroscopy obtained with 
the ISO Short Wavelength Spectrometer 
\citep{Peeters+Hony+vanKerckhoven_etal_2002}.
Similar shifts have also been seen in other bands 
[e.g. see 
\citet{vanDiedenhoven+Peeters+vanKerckhoven_etal_2004}].
More recently, 
\citet{Bregman+Temi_2005} 
observed a progressive
blueshift in the central wavelength of the 7.7$\mum$ band 
within 3 reflection nebulae from the edge of the nebulae
to places closer to the exciting stars.
\item 
\citet{Uchida+Sellgren+Werner+Houdashelt_2000} 
reported a progressive
broadening of the FWHM of the 7.7$\mum$ band 
moving toward the exciting star
in the reflection nebula vdB 17.
This broadening has also been reported
for the reflection nebula Ced 21
\citep{Cesarsky+Lequeux+Ryter+Gerin_2000}. 
But such a broadening has not been seen 
in other PAH bands. 
\item 
\citet{Roelfsema+Cox+Tielens_etal_1996} 
reported that 
the 8.6$\mum$ band of the compact HII region 
IRAS 18434-0242 is almost twice as strong as
the 7.7$\mum$ band, while for most objects
the 7.7$\mum$ band is stronger than the 8.6$\mum$ band
by a factor of $\sim$\,5--10.
\item 
\newtext{
\citet{Reach+Boulanger+Contursi+Lequeux_2000} reported that 
the 11.3$\mum$/7.7$\mum$ band ratio
of the PAH emission spectrum of 
the quiescent molecular cloud SMC B1 No.\,1 
in the SMC is much higher than that of 
any other Galactic or extragalactic objects.
(see Figs.\,16 and 17 of \citet{Draine+Li_2001}).
}
\end{itemize}

While the band-ratio variations 
can presumably be accounted for by the combined effects of 
variations in the radiation exciting the PAHs and 
variations in PAH sizes and ionization state 
(small PAHs emit more strongly 
at the 6.2, 7.7, 8.6$\mum$ bands, 
while relatively large PAHs emit more strongly 
at the 11.3$\mum$ band;
neutral PAHs are much stronger 3.3$\mum$ 
and 11.3$\mum$ emitters 
and weaker 6.2, 7.7 and 8.6$\mum$ emitters 
compared to their charged counterparts),
we do not intend to confront our PAH model
with extreme spectral variations, such as the observed spectrum
of IRAS 18434-024 (with the 8.6$\mum$ band almost
twice as strong as the 7.7$\mum$ band) which,
according to \citet{Roelfsema+Cox+Tielens_etal_1996}, 
is possibly rich in non-compact PAHs.
Because of the way we design the ``astronomical'' PAH model, 
the present model is not able to account for
the central wavelength shifts and the 7.7$\mum$ band
broadening, although experimental investigations
have been carried out [e.g. see \citet{Hudgins+Allamandola_1999}].

\citet{Hudgins+Allamandola_2005} discuss recent work toward PAH identification,
with attention to the effects of N substitution.
Future, more realistic modeling will require more extensive knowledge
of PAH properties than is now available from
laboratory studies [e.g., 
\citet{Hudgins+Allamandola_1999,
       Mattioda+Hudgins+Allamandola_2005,
       Mattioda+Allamandola+Hudgins_2005}]
and theoretical calculations [e.g.,
\citet{Bakes+Tielens+Bauschlicher_2001,
       Bakes+Tielens+Bauschlicher_etal_2001,
       Malloci+Mulas+Joblin_2004,
       Mulas+Malloci+Joblin+Toublanc_2006,
       Malloci+Joblin+Mulas_2006}].

\subsection{Other Grain Models\label{sec:others}}

The dust model used in this study is, of course, provisional.
The model assumes the amorphous silicate and carbonaceous material
to be in physically separate grains.  While this appears to be
consistent with observations, including the observed absence of
polarization in the 3.4\um\ absorption feature
\citep{Chiar+Adamson+Whittet_etal_2006},
it is by no means observationally established.

Our models have also used the dielectric tensor of crystalline graphite
to estimate the infrared absorption by carbonaceous grains with
$a\gtsim0.01\micron$.  Some form of amorphous carbon may provide
a better approximation for interstellar carbonaceous material that
may be formed from agglomeration of PAHs followed by cross-linking
due to ultraviolet photolysis and cosmic ray exposure.
The main effect on the present models would be to do away with the
broad absorption feature near 30\um\ that appears in the calculated
$\Cabs(\lambda)$ for graphitic material, as seen in Figure
\ref{fig:Cabs for carbonaceous grains}, and the corresponding broad
emission feature near 30\um\ when the starlight intensity is
high enough to heat $a\gtsim0.01\micron$ carbonaceous grains to
$T\gtsim10^2\K$.

In this work we focus on the silicate-graphite-PAH model.
For other dust models, if they are tailored to reproduce 
the Milky Way interstellar extinction
{\it and} albedo in the optical/UV 
{\it and} meanwhile satisfy plausible interstellar 
abundance constraints, the derived parameters 
$\Mdust\langle U\rangle$ will be nearly the same as
derived from the silicate-graphite-PAH model,
simply from energy balance considerations: 
the energy emitted in 
the IR must be absorbed by the same dust in the optical/UV.
If the dust model reproduces the observed far-infrared emission spectrum
of Milky Way dust for $U\approx 1$, then the temperatures of the larger
dust grains will scale with changes in $U$ in a manner similar to the
present model, and therefore the inferred properties of the
radiation field ($\Umin$, $\gamma$) will presumably be similar.

However, the composite model will probably have a lower optical/UV 
albedo than observed 
\citep{Dwek_1997}. 
Because the extinction per H atom is fixed by observations, the lower
albedo will result in more total far-infrared emission per H nucleon.
Given the uncertainties regarding both the intensity of interstellar
starlight and the far infrared emission from dust,
composite grain models must be considered to be viable.
Additional observational tests to distinguish between dust models are needed.

\section{Summary
         \label{sec:summary}}

The principal results of this study are as follows:
\begin{enumerate}

\item The wavelength dependence of absorption by interstellar PAHs has
      been revised to better reproduce oberved emission spectra.
      The adopted $C_{\rm abs}(\lambda)$ are generally consistent with
      measured or calculated absorption cross sections for PAHs
      (Figure \ref{fig:Cabs_vs_theory}).

\item Temperature distribution functions have been calculated for PAHs
      of different sizes, heated by starlight with the spectrum estimated
      for the local interstellar radiation field.
      These temperature distribution functions have been used
      to compute the time-averaged emission spectra for PAHs of different
      sizes (Figure \ref{fig:PAHi and PAHn emission spectra}).

\item Figures \ref{fig:band efficiencies} and \ref{fig:band efficiencies, dism}
      show the
      efficiency with which PAHs radiate absorbed energy into 
      different emission features.  The strong 7.7$\mum$ feature is
      radiated efficiently only by PAHs with $N_{\rm C}\ltsim 10^3$ carbon
      atoms -- larger PAHs radiate primarily at longer wavelengths.

\item Emission spectra have been calculated for dust models that
      reproduce the average Milky Way extinction, using size distributions
      with different abundances of small PAHs
      (see Fig.\ \ref{fig:variousqpah_spec}).

\item Emission spectra have been calculated for dust mixtures heated by
      different starlight intensities 
      (see Fig.\ \ref{fig:variousU_spec}).

\item The dust model used in this paper is found to be in good agreement
      with observed
      IRAC~3.6\um/ IRAC~7.9\um\ and IRAC~5.7\um/ IRAC~7.9\um\ band ratios
      (see Fig.\ \ref{fig:irac band ratios}), but underpredicts the 
      observed IRAC~4.5\um/ IRAC~7.9\um\ ratio by about a factor 1.5.

\item The far-infrared and submm emission spectrum calculated for
      the present model appears to be in good agreement with
      the observed emission from dust in the local Milky Way
      (see Figure \ref{fig:firspec}).

\item A prescription (eq.\ \ref{fig:TIR estimator}) is found that allows the
      total IR emission to be estimated from observed fluxes
      in the IRAC~7.9\um\ band and the 3 MIPS bands.
      For dust heated by starlight intensities $0.1 < U \ltsim 10^2$,
      eq.\ (\ref{fig:TIR estimator}) is accurate to within $\sim10\%$.

\item Emission spectra have been calculated for dust mixtures heated
      by power-law distributions of starlight intensities
      (see Fig.\ \ref{fig:spec for Udists}).

\item A distribution function (eq.\ \ref{eq:dM_d/dU}) is proposed for
      representing the distribution of starlight intensities heating
      dust in a galaxy, or region within a galaxy.

\item Graphical procedures are presented that allow dust model parameters
      to be found using 3 ratios ($P_{7.9}$, $P_{24}$, and $R_{71}$)
      constructed from fluxes measured in IRAC band 4 and
      MIPS bands 1--3.  The graphs provided here 
      (Figs. \ref{fig:IRAC4/MIPS band ratios} and
      \ref{fig:MIPS band ratios})
      allow model parameters $q_{\rm PAH}$, $\Umin$, and $\gamma$
      to be estimated.
\item The fraction $f_{\rm PDR}$ of the dust infrared emission that is
      contributed by dust in photodissociation regions with $U > 10^2$
      can be estimated from IRAC and MIPS photometry using Figure
      \ref{fig:fpdr}.
\item Using a coefficient $\Psi$ obtained from Figure \ref{fig:Psi grid},
      the total dust mass $\Mdust$ can be estimated from fluxes in the
      3 MIPS bands using eq.\ (\ref{eq:compute Mdust}).
\end{enumerate}

\acknowledgements
We are grateful to G.~Malloci for making available to us theoretical
absorption cross sections for a number of PAHs in advance of publication;
to E.~Peeters for providing SWS spectra of the Orion Bar, M17, and
NGC~7027; 
\newtext{
to K.~Sellgren for providing IRS spectra of NGC~7023;
to T.~Onaka for providing response functions for Akari;
and to A.~Poglitsch and M.~Griffin 
for providing response functions for PACS and SPIRE.
We thank the referee, G.~Clayton, for helpful comments.
}
BTD was supported in part by NSF grant AST-0406883.
AL is supported in part by the University of Missouri
Summer Research Fellowship, the University of Missouri
Research Board, and NASA/Spitzer Theory Programs.
We are grateful to R.H. Lupton for availability of the SM graphics
program, used extensively in preparation of this paper.

\bibliography{btdrefs}
\end{document}

%% file: table4.dat
 0.5 & 0.47 & 5.66e-28 & 1.54e-28 & 1.31e-27 & 4.05e-27 & 4.86e-27 & 5.91e-27 & 5.83e-27 & 5.04e-26 & 1.25e-25\cr
 0.5 & 1.12 & 1.55e-27 & 4.30e-28 & 3.53e-27 & 9.17e-27 & 6.41e-27 & 6.27e-27 & 6.05e-27 & 4.76e-26 & 1.21e-25\cr
 0.5 & 1.77 & 2.56e-27 & 7.09e-28 & 5.77e-27 & 1.43e-26 & 7.97e-27 & 6.62e-27 & 6.26e-27 & 4.54e-26 & 1.19e-25\cr
 0.5 & 3.19 & 4.78e-27 & 1.34e-27 & 1.07e-26 & 2.60e-26 & 1.24e-26 & 8.87e-27 & 8.16e-27 & 4.11e-26 & 1.12e-25\cr
 0.5 & 4.58 & 7.07e-27 & 1.98e-27 & 1.58e-26 & 3.81e-26 & 1.73e-26 & 1.14e-26 & 1.04e-26 & 3.66e-26 & 1.05e-25\cr
 1.0 & 0.47 & 1.13e-27 & 3.08e-28 & 2.63e-27 & 8.10e-27 & 9.84e-27 & 1.21e-26 & 1.20e-26 & 1.35e-25 & 2.27e-25\cr
 1.0 & 1.12 & 3.11e-27 & 8.59e-28 & 7.06e-27 & 1.84e-26 & 1.29e-26 & 1.29e-26 & 1.24e-26 & 1.28e-25 & 2.21e-25\cr
 1.0 & 1.77 & 5.11e-27 & 1.42e-27 & 1.15e-26 & 2.87e-26 & 1.60e-26 & 1.35e-26 & 1.28e-26 & 1.22e-25 & 2.18e-25\cr
 1.0 & 3.19 & 9.55e-27 & 2.68e-27 & 2.14e-26 & 5.20e-26 & 2.50e-26 & 1.80e-26 & 1.66e-26 & 1.11e-25 & 2.06e-25\cr
 1.0 & 4.58 & 1.41e-26 & 3.96e-27 & 3.15e-26 & 7.62e-26 & 3.47e-26 & 2.32e-26 & 2.11e-26 & 9.89e-26 & 1.94e-25\cr
 3.0 & 0.47 & 3.39e-27 & 9.24e-28 & 7.90e-27 & 2.44e-26 & 3.07e-26 & 3.95e-26 & 3.92e-26 & 5.80e-25 & 5.19e-25\cr
 3.0 & 1.12 & 9.32e-27 & 2.58e-27 & 2.12e-26 & 5.51e-26 & 3.99e-26 & 4.14e-26 & 4.03e-26 & 5.53e-25 & 5.06e-25\cr
 3.0 & 1.77 & 1.53e-26 & 4.25e-27 & 3.47e-26 & 8.62e-26 & 4.91e-26 & 4.32e-26 & 4.12e-26 & 5.35e-25 & 5.01e-25\cr
 3.0 & 3.19 & 2.86e-26 & 8.05e-27 & 6.41e-26 & 1.56e-25 & 7.58e-26 & 5.66e-26 & 5.25e-26 & 4.89e-25 & 4.77e-25\cr
 3.0 & 4.58 & 4.24e-26 & 1.19e-26 & 9.46e-26 & 2.29e-25 & 1.05e-25 & 7.22e-26 & 6.60e-26 & 4.41e-25 & 4.57e-25\cr
 5.0 & 0.47 & 5.66e-27 & 1.54e-27 & 1.32e-26 & 4.07e-26 & 5.28e-26 & 6.97e-26 & 6.95e-26 & 1.08e-24 & 7.30e-25\cr
 5.0 & 1.12 & 1.55e-26 & 4.29e-27 & 3.53e-26 & 9.19e-26 & 6.80e-26 & 7.27e-26 & 7.10e-26 & 1.03e-24 & 7.13e-25\cr
 5.0 & 1.77 & 2.56e-26 & 7.08e-27 & 5.78e-26 & 1.44e-25 & 8.32e-26 & 7.54e-26 & 7.23e-26 & 1.00e-24 & 7.06e-25\cr
 5.0 & 3.19 & 4.77e-26 & 1.34e-26 & 1.07e-25 & 2.60e-25 & 1.28e-25 & 9.76e-26 & 9.10e-26 & 9.20e-25 & 6.76e-25\cr
 5.0 & 4.58 & 7.06e-26 & 1.98e-26 & 1.58e-25 & 3.81e-25 & 1.77e-25 & 1.24e-25 & 1.14e-25 & 8.36e-25 & 6.50e-25\cr
10.0 & 0.47 & 1.13e-26 & 3.08e-27 & 2.63e-26 & 8.17e-26 & 1.12e-25 & 1.54e-25 & 1.54e-25 & 2.37e-24 & 1.13e-24\cr
10.0 & 1.12 & 3.11e-26 & 8.59e-27 & 7.07e-26 & 1.84e-25 & 1.42e-25 & 1.59e-25 & 1.56e-25 & 2.27e-24 & 1.11e-24\cr
10.0 & 1.77 & 5.11e-26 & 1.42e-26 & 1.15e-25 & 2.88e-25 & 1.72e-25 & 1.63e-25 & 1.57e-25 & 2.21e-24 & 1.10e-24\cr
10.0 & 3.19 & 9.55e-26 & 2.68e-26 & 2.14e-25 & 5.21e-25 & 2.61e-25 & 2.07e-25 & 1.94e-25 & 2.05e-24 & 1.06e-24\cr
10.0 & 4.58 & 1.41e-25 & 3.96e-26 & 3.15e-25 & 7.63e-25 & 3.59e-25 & 2.59e-25 & 2.39e-25 & 1.88e-24 & 1.02e-24\cr
100.0 & 0.47 & 1.13e-25 & 3.09e-26 & 2.67e-25 & 8.67e-25 & 1.68e-24 & 2.80e-24 & 2.89e-24 & 2.16e-23 & 3.72e-24\cr
100.0 & 1.12 & 3.11e-25 & 8.62e-26 & 7.12e-25 & 1.89e-24 & 1.94e-24 & 2.78e-24 & 2.83e-24 & 2.09e-23 & 3.66e-24\cr
100.0 & 1.77 & 5.12e-25 & 1.42e-25 & 1.16e-24 & 2.93e-24 & 2.20e-24 & 2.75e-24 & 2.77e-24 & 2.06e-23 & 3.64e-24\cr
100.0 & 3.19 & 9.57e-25 & 2.69e-25 & 2.15e-24 & 5.26e-24 & 3.07e-24 & 3.09e-24 & 3.03e-24 & 1.95e-23 & 3.55e-24\cr
100.0 & 4.58 & 1.41e-24 & 3.97e-25 & 3.17e-24 & 7.70e-24 & 4.07e-24 & 3.56e-24 & 3.41e-24 & 1.85e-23 & 3.48e-24\cr

%% file: table5.dat
 0.5 & 0.47 & 3.75e-28 & 1.68e-28 & 4.37e-27 & 3.55e-27 & 3.49e-27 & 5.13e-27 & 5.48e-27 & 5.88e-27 & 4.46e-26 & 8.43e-26 & 1.32e-25 & 1.26e-25\cr
 0.5 & 1.12 & 1.03e-27 & 4.67e-28 & 1.03e-26 & 7.54e-27 & 6.36e-27 & 6.71e-27 & 6.38e-27 & 6.18e-27 & 4.20e-26 & 8.02e-26 & 1.28e-25 & 1.22e-25\cr
 0.5 & 1.77 & 1.69e-27 & 7.70e-28 & 1.62e-26 & 1.16e-26 & 9.26e-27 & 8.30e-27 & 7.28e-27 & 6.47e-27 & 3.99e-26 & 7.73e-26 & 1.25e-25 & 1.20e-25\cr
 0.5 & 3.19 & 3.16e-27 & 1.46e-27 & 2.95e-26 & 2.09e-26 & 1.63e-26 & 1.29e-26 & 1.06e-26 & 8.57e-27 & 3.62e-26 & 7.04e-26 & 1.18e-25 & 1.13e-25\cr
 0.5 & 4.58 & 4.68e-27 & 2.15e-27 & 4.33e-26 & 3.05e-26 & 2.36e-26 & 1.79e-26 & 1.42e-26 & 1.10e-26 & 3.25e-26 & 6.31e-26 & 1.10e-25 & 1.06e-25\cr
 1.0 & 0.47 & 7.50e-28 & 3.35e-28 & 8.75e-27 & 7.12e-27 & 7.01e-27 & 1.04e-26 & 1.12e-26 & 1.21e-26 & 1.23e-25 & 2.03e-25 & 2.48e-25 & 2.25e-25\cr
 1.0 & 1.12 & 2.05e-27 & 9.33e-28 & 2.06e-26 & 1.51e-26 & 1.27e-26 & 1.35e-26 & 1.30e-26 & 1.27e-26 & 1.16e-25 & 1.93e-25 & 2.40e-25 & 2.19e-25\cr
 1.0 & 1.77 & 3.37e-27 & 1.54e-27 & 3.25e-26 & 2.31e-26 & 1.85e-26 & 1.67e-26 & 1.47e-26 & 1.32e-26 & 1.11e-25 & 1.87e-25 & 2.36e-25 & 2.16e-25\cr
 1.0 & 3.19 & 6.32e-27 & 2.91e-27 & 5.90e-26 & 4.18e-26 & 3.26e-26 & 2.59e-26 & 2.13e-26 & 1.74e-26 & 1.01e-25 & 1.71e-25 & 2.22e-25 & 2.04e-25\cr
 1.0 & 4.58 & 9.34e-27 & 4.30e-27 & 8.66e-26 & 6.10e-26 & 4.72e-26 & 3.60e-26 & 2.86e-26 & 2.23e-26 & 8.97e-26 & 1.55e-25 & 2.09e-25 & 1.93e-25\cr
 3.0 & 0.47 & 2.25e-27 & 1.01e-27 & 2.63e-26 & 2.14e-26 & 2.12e-26 & 3.24e-26 & 3.56e-26 & 3.93e-26 & 5.66e-25 & 7.27e-25 & 5.93e-25 & 5.01e-25\cr
 3.0 & 1.12 & 6.16e-27 & 2.80e-27 & 6.17e-26 & 4.54e-26 & 3.84e-26 & 4.18e-26 & 4.08e-26 & 4.08e-26 & 5.38e-25 & 6.98e-25 & 5.78e-25 & 4.89e-25\cr
 3.0 & 1.77 & 1.01e-26 & 4.62e-27 & 9.76e-26 & 6.95e-26 & 5.58e-26 & 5.12e-26 & 4.60e-26 & 4.23e-26 & 5.19e-25 & 6.80e-25 & 5.71e-25 & 4.84e-25\cr
 3.0 & 3.19 & 1.89e-26 & 8.73e-27 & 1.77e-25 & 1.25e-25 & 9.80e-26 & 7.88e-26 & 6.56e-26 & 5.48e-26 & 4.73e-25 & 6.28e-25 & 5.43e-25 & 4.61e-25\cr
 3.0 & 4.58 & 2.80e-26 & 1.29e-26 & 2.60e-25 & 1.83e-25 & 1.42e-25 & 1.09e-25 & 8.77e-26 & 6.95e-26 & 4.24e-25 & 5.76e-25 & 5.17e-25 & 4.42e-25\cr
 5.0 & 0.47 & 3.75e-27 & 1.68e-27 & 4.39e-26 & 3.59e-26 & 3.57e-26 & 5.58e-26 & 6.20e-26 & 6.93e-26 & 1.09e-24 & 1.25e-24 & 8.52e-25 & 6.96e-25\cr
 5.0 & 1.12 & 1.03e-26 & 4.67e-27 & 1.03e-25 & 7.57e-26 & 6.43e-26 & 7.12e-26 & 7.05e-26 & 7.16e-26 & 1.04e-24 & 1.20e-24 & 8.32e-25 & 6.80e-25\cr
 5.0 & 1.77 & 1.69e-26 & 7.70e-27 & 1.63e-25 & 1.16e-25 & 9.33e-26 & 8.68e-26 & 7.90e-26 & 7.37e-26 & 1.00e-24 & 1.17e-24 & 8.23e-25 & 6.74e-25\cr
 5.0 & 3.19 & 3.16e-26 & 1.45e-26 & 2.95e-25 & 2.09e-25 & 1.64e-25 & 1.33e-25 & 1.12e-25 & 9.45e-26 & 9.18e-25 & 1.09e-24 & 7.85e-25 & 6.45e-25\cr
 5.0 & 4.58 & 4.67e-26 & 2.15e-26 & 4.33e-25 & 3.05e-25 & 2.37e-25 & 1.83e-25 & 1.48e-25 & 1.19e-25 & 8.29e-25 & 1.01e-24 & 7.53e-25 & 6.22e-25\cr
10.0 & 0.47 & 7.50e-27 & 3.35e-27 & 8.82e-26 & 7.22e-26 & 7.25e-26 & 1.18e-25 & 1.34e-25 & 1.52e-25 & 2.48e-24 & 2.50e-24 & 1.36e-24 & 1.06e-24\cr
10.0 & 1.12 & 2.05e-26 & 9.33e-27 & 2.06e-25 & 1.52e-25 & 1.30e-25 & 1.48e-25 & 1.50e-25 & 1.55e-25 & 2.37e-24 & 2.41e-24 & 1.33e-24 & 1.04e-24\cr
10.0 & 1.77 & 3.37e-26 & 1.54e-26 & 3.25e-25 & 2.32e-25 & 1.87e-25 & 1.79e-25 & 1.66e-25 & 1.59e-25 & 2.30e-24 & 2.36e-24 & 1.32e-24 & 1.04e-24\cr
10.0 & 3.19 & 6.32e-26 & 2.91e-26 & 5.90e-25 & 4.18e-25 & 3.28e-25 & 2.71e-25 & 2.31e-25 & 2.00e-25 & 2.12e-24 & 2.21e-24 & 1.26e-24 & 9.96e-25\cr
10.0 & 4.58 & 9.34e-26 & 4.30e-26 & 8.67e-25 & 6.11e-25 & 4.74e-25 & 3.72e-25 & 3.05e-25 & 2.50e-25 & 1.93e-24 & 2.06e-24 & 1.22e-24 & 9.65e-25\cr
100.0 & 0.47 & 7.51e-26 & 3.37e-26 & 9.26e-25 & 7.97e-25 & 8.76e-25 & 1.72e-24 & 2.11e-24 & 2.63e-24 & 2.55e-23 & 1.77e-23 & 4.84e-24 & 3.36e-24\cr
100.0 & 1.12 & 2.05e-25 & 9.37e-26 & 2.11e-24 & 1.59e-24 & 1.44e-24 & 1.99e-24 & 2.22e-24 & 2.60e-24 & 2.47e-23 & 1.72e-23 & 4.75e-24 & 3.30e-24\cr
100.0 & 1.77 & 3.38e-25 & 1.54e-25 & 3.30e-24 & 2.39e-24 & 2.00e-24 & 2.25e-24 & 2.33e-24 & 2.57e-24 & 2.43e-23 & 1.70e-23 & 4.73e-24 & 3.29e-24\cr
100.0 & 3.19 & 6.33e-25 & 2.92e-25 & 5.96e-24 & 4.26e-24 & 3.41e-24 & 3.16e-24 & 2.95e-24 & 2.91e-24 & 2.29e-23 & 1.62e-23 & 4.60e-24 & 3.20e-24\cr
100.0 & 4.58 & 9.36e-25 & 4.31e-25 & 8.74e-24 & 6.19e-24 & 4.88e-24 & 4.20e-24 & 3.69e-24 & 3.36e-24 & 2.15e-23 & 1.55e-23 & 4.51e-24 & 3.15e-24\cr

%% file: table6.dat
 0.5 & 0.47 & 4.21e-26 & 7.92e-26 & 9.61e-26 & 5.71e-26 & 2.05e-26 & 5.80e-27\cr
 0.5 & 1.12 & 3.97e-26 & 7.57e-26 & 9.31e-26 & 5.58e-26 & 2.01e-26 & 5.70e-27\cr
 0.5 & 1.77 & 3.78e-26 & 7.34e-26 & 9.17e-26 & 5.53e-26 & 2.00e-26 & 5.68e-27\cr
 0.5 & 3.19 & 3.43e-26 & 6.71e-26 & 8.61e-26 & 5.28e-26 & 1.92e-26 & 5.50e-27\cr
 0.5 & 4.58 & 3.05e-26 & 6.05e-26 & 8.09e-26 & 5.07e-26 & 1.87e-26 & 5.39e-27\cr
 1.0 & 0.47 & 1.13e-25 & 1.78e-25 & 1.73e-25 & 8.74e-26 & 2.83e-26 & 7.50e-27\cr
 1.0 & 1.12 & 1.08e-25 & 1.71e-25 & 1.68e-25 & 8.55e-26 & 2.77e-26 & 7.38e-27\cr
 1.0 & 1.77 & 1.03e-25 & 1.67e-25 & 1.66e-25 & 8.48e-26 & 2.76e-26 & 7.35e-27\cr
 1.0 & 3.19 & 9.33e-26 & 1.53e-25 & 1.57e-25 & 8.13e-26 & 2.66e-26 & 7.13e-27\cr
 1.0 & 4.58 & 8.32e-26 & 1.40e-25 & 1.48e-25 & 7.85e-26 & 2.60e-26 & 7.01e-27\cr
 3.0 & 0.47 & 4.96e-25 & 5.69e-25 & 3.94e-25 & 1.56e-25 & 4.40e-26 & 1.07e-26\cr
 3.0 & 1.12 & 4.73e-25 & 5.49e-25 & 3.84e-25 & 1.53e-25 & 4.32e-26 & 1.05e-26\cr
 3.0 & 1.77 & 4.57e-25 & 5.37e-25 & 3.80e-25 & 1.52e-25 & 4.30e-26 & 1.05e-26\cr
 3.0 & 3.19 & 4.17e-25 & 5.00e-25 & 3.62e-25 & 1.47e-25 & 4.17e-26 & 1.02e-26\cr
 3.0 & 4.58 & 3.76e-25 & 4.64e-25 & 3.47e-25 & 1.43e-25 & 4.09e-26 & 1.01e-26\cr
 5.0 & 0.47 & 9.28e-25 & 9.23e-25 & 5.54e-25 & 1.99e-25 & 5.29e-26 & 1.25e-26\cr
 5.0 & 1.12 & 8.87e-25 & 8.92e-25 & 5.41e-25 & 1.95e-25 & 5.20e-26 & 1.23e-26\cr
 5.0 & 1.77 & 8.59e-25 & 8.75e-25 & 5.36e-25 & 1.94e-25 & 5.18e-26 & 1.22e-26\cr
 5.0 & 3.19 & 7.89e-25 & 8.20e-25 & 5.12e-25 & 1.88e-25 & 5.03e-26 & 1.19e-26\cr
 5.0 & 4.58 & 7.17e-25 & 7.67e-25 & 4.93e-25 & 1.83e-25 & 4.94e-26 & 1.18e-26\cr
10.0 & 0.47 & 2.04e-24 & 1.71e-24 & 8.59e-25 & 2.73e-25 & 6.77e-26 & 1.53e-26\cr
10.0 & 1.12 & 1.96e-24 & 1.66e-24 & 8.41e-25 & 2.68e-25 & 6.67e-26 & 1.51e-26\cr
10.0 & 1.77 & 1.91e-24 & 1.63e-24 & 8.35e-25 & 2.67e-25 & 6.64e-26 & 1.50e-26\cr
10.0 & 3.19 & 1.76e-24 & 1.54e-24 & 8.02e-25 & 2.59e-25 & 6.47e-26 & 1.47e-26\cr
10.0 & 4.58 & 1.62e-24 & 1.45e-24 & 7.77e-25 & 2.54e-25 & 6.36e-26 & 1.45e-26\cr
100.0 & 0.47 & 1.88e-23 & 9.29e-24 & 2.84e-24 & 6.49e-25 & 1.35e-25 & 2.75e-26\cr
100.0 & 1.12 & 1.82e-23 & 9.08e-24 & 2.79e-24 & 6.40e-25 & 1.33e-25 & 2.71e-26\cr
100.0 & 1.77 & 1.79e-23 & 9.00e-24 & 2.78e-24 & 6.38e-25 & 1.33e-25 & 2.71e-26\cr
100.0 & 3.19 & 1.70e-23 & 8.66e-24 & 2.71e-24 & 6.24e-25 & 1.30e-25 & 2.66e-26\cr
100.0 & 4.58 & 1.61e-23 & 8.39e-24 & 2.66e-24 & 6.17e-25 & 1.29e-25 & 2.64e-26\cr